\documentclass[iop,apjl,numberedappendix,appendixfloats,floatfix]{emulateapj}
\usepackage{lineno}
\usepackage{color}
\usepackage{amsmath}
\usepackage{url}
\usepackage{yfonts}
\usepackage[T1]{fontenc}

\newcommand{\tco}{$^{12}$CO}
\newcommand{\ttco}{$^{13}$CO}
\newcommand{\ceto}{C$^{18}$O}

\newcommand{\kms}{\,km\,s$^{-1}$}
\newcommand{\degree}{$^{\circ}$}
\newcommand{\fdeg}{$^{\circ}$\hspace{-1mm}.}

\newcommand{\td}{$T_{\rm dust}$}

\newcommand{\nhtwo}{$N_{\rm H_2}$}
\newcommand{\nco}{$N_{\rm ^{12}CO}$}

\def\lapp{\ifmmode\stackrel{<}{_{\sim}}\else$\stackrel{<}{_{\sim}}$\fi}
\def\gapp{\ifmmode\stackrel{>}{_{\sim}}\else$\stackrel{>}{_{\sim}}$\fi}
\shorttitle{Magnetic Fields in the western $\eta$ Car GMC}
\shortauthors{Barnes et al.}

\begin{document}

\title{The Magnetic Keys to Massive Star Formation: \\  The Western $\eta$ Carinae Giant Molecular Cloud}

\author{Peter J. Barnes\altaffilmark{1}, Stuart D. Ryder\altaffilmark{2,3}, Giles Novak\altaffilmark{4,5}, and Laura M. Fissel\altaffilmark{6}}
\affiliation{$^{1}$Space Science Institute, 4765 Walnut St. Suite B, Boulder, CO 80301, USA}
\affiliation{$^{2}$School of Mathematical \& Physical Sciences, Macquarie University, NSW 2109, Australia}
\affiliation{$^{3}$Astrophysics and Space Technologies Research Centre, Macquarie University, NSW 2109, Australia}
\affiliation{$^{4}$Center for Interdisciplinary Exploration \& Research in Astrophysics, 1800 Sherman Ave., Evanston, IL 60201, USA} 
\affiliation{$^{5}$Dept.\ of Physics and Astronomy, Northwestern University, 2145 Sheridan Rd., Evanston, IL 60208, USA}
\affiliation{$^{6}$Dept.\ of Physics, Engineering Physics \& Astronomy, Queen's University, 64 Bader Lane, Kingston, ON, K7L 3N6, Canada}

%\altaffiliation{7}{Society of Fellows, Harvard University}
%\altaffiliation{8}{Patron, Alonso's Bar and Grill}
%\altaffiliation{9}{All-round Mensch}

\email{pbarnes@spacescience.org}

\begin{abstract}
\hspace{3mm}We present SOFIA/HAWC+ continuum polarisation data on the magnetic fields threading 17 pc-scale massive molecular clumps at the western end of the $\eta$ Car GMC (Region 9 of CHaMP, representing all stages of star formation from pre-stellar to dispersing via feedback), revealing important details about the field morphology and role in the gas structures of this clump sample. %~61 words

\hspace{3mm}We performed Davis-Chandrasekhar-Fermi and Histogram of Relative Orientation analyses tracing column densities 25.0 $<$ log($N$/m$^{-2}$) $<$ 27.2.  With HRO, magnetic fields 
change from mostly parallel to column density structures to mostly perpendicular at a threshold $N_{\rm crit}$ = (3.7$\pm$0.6)$\times$10$^{26}$\,m$^{-2}$, %$n_{\rm crit}$ = 2.5$\times$10$^{11}$\,m$^{-3}$??, and $B_{\rm crit}$ = 42$\pm$7\,nT??. %~96 words 
indicating that gravitational forces exceed magnetic forces above this value.  The same analysis in 10 individual clumps gives similar results, with the same clear trend in field alignments and a threshold $N_{\rm crit}$ = (1.9$^{+1.5}_{-0.8}$)$\times$10$^{26}$\,m$^{-2}$.  In the other 7 clumps, the alignment trend with $N$ is much flatter or even reversed, %~144 words 
inconsistent with the usual HRO pattern.  Instead, these clumps' fields reflect external environmental forces, such as %radiation %or overpressure 
from the nearby HII region NGC\,3324.

\hspace{3mm}%A computation of 
DCF analysis reveals field strengths somewhat higher than typical of nearby clouds, with the $Bn$ data lying mostly above the \cite{cru12} relation. The mass:flux ratio $\lambda$ across all clumps has a gaussian distribution, with log$\lambda_{\rm DCF}$ = %~201 words
--0.75$\pm$0.45 (mean$\pm\sigma$):\ only small areas are dominated by gravity.  However, a significant trend of rising log$\lambda$ with falling \td\ parallels \cite{p19}'s result: \td\ falls as \nhtwo\ rises towards clump centres.  In this massive clump sample, magnetic fields provide enough support against gravity to explain their overall low star formation rate. % 250 words
\end{abstract}

%% The macro also takes an optional argument in parentheses in cases where the 
%% data center identification differs from what is to be printed in the paper.

%%%%%%%%%
%                       %
%    Section 1    %
%                       %
%%%%%%%%%
\section{Introduction}

Magnetic fields (hereafter $B$ fields) likely play an important role in the evolution of the interstellar medium (ISM).  Particularly in molecular clouds where stars form, the impact of $B$ fields is a long-standing question \citep{mo07,cru12}, stimulated further in the last few years by new models and data.  Far-IR instruments (e.g., {\em Planck}, SOFIA's HAWC+ camera, BLASTpol) have made sensitive, higher resolution, and/or wide-field maps of the plane-of-sky component $B_{\perp}$ via linear polarisation measurements \citep[e.g.,][]{pc16}.  Polarised radiation is thought to arise from non-spherical dust grains being oriented by radiative alignment torques (RAT) to the $B$ field.  While not all possible alignment mechanisms are magnetic, non-magnetic mechanisms are probably not dominant \citep{L07}.

Observationally, $B$ field measurements rely on accurate values for the polarised contributions to emission or absorption (i.e., the Stokes parameters $Q$, $U$, $V$), which are usually much weaker than the total intensity $I$; and then interpreting the data in terms of particular physical polarisation mechanisms, whether from alignment of dust grains or atomic/molecular effects (Zeeman, etc.), as explained by \citet{cru12} or \citet{BL15}.  This is compounded by challenges of making high-quality Stokes measurements in large cloud samples at high spatial dynamic range (SDR), and relating these to the clouds' other physical conditions.  

If the dust polarisation alignment is magnetic, statistical methods can convert turbulent variations in field orientation $\theta_{B_{\perp}}$ to estimates of  $|B_{\perp}|$ \citep[][hereafter DCF]{d51,cf53}.  Another analysis method, the histogram of relative orientations (HRO), compares the $B_{\perp}$ field alignment to the orientation of dense gas structures to infer the threshold density $n_0$ where gravity appears to dominate $B$ field pressure support \citep{faa16,saa17,LY18}.  Although approximate, both methods have been effectively used from cloud (10\,pc) to core (0.1\,pc) scales to meaningfully constrain the importance of $B$ fields in different situations \citep{mg91,BL15,faa19,ppvii,b23}.

%%%%%%
%   Fig.1  %   FIR-mm cartography
%%%%%%
\begin{figure*}[ht]
\vspace{0mm}
\centerline{\includegraphics[angle=0,scale=0.9]{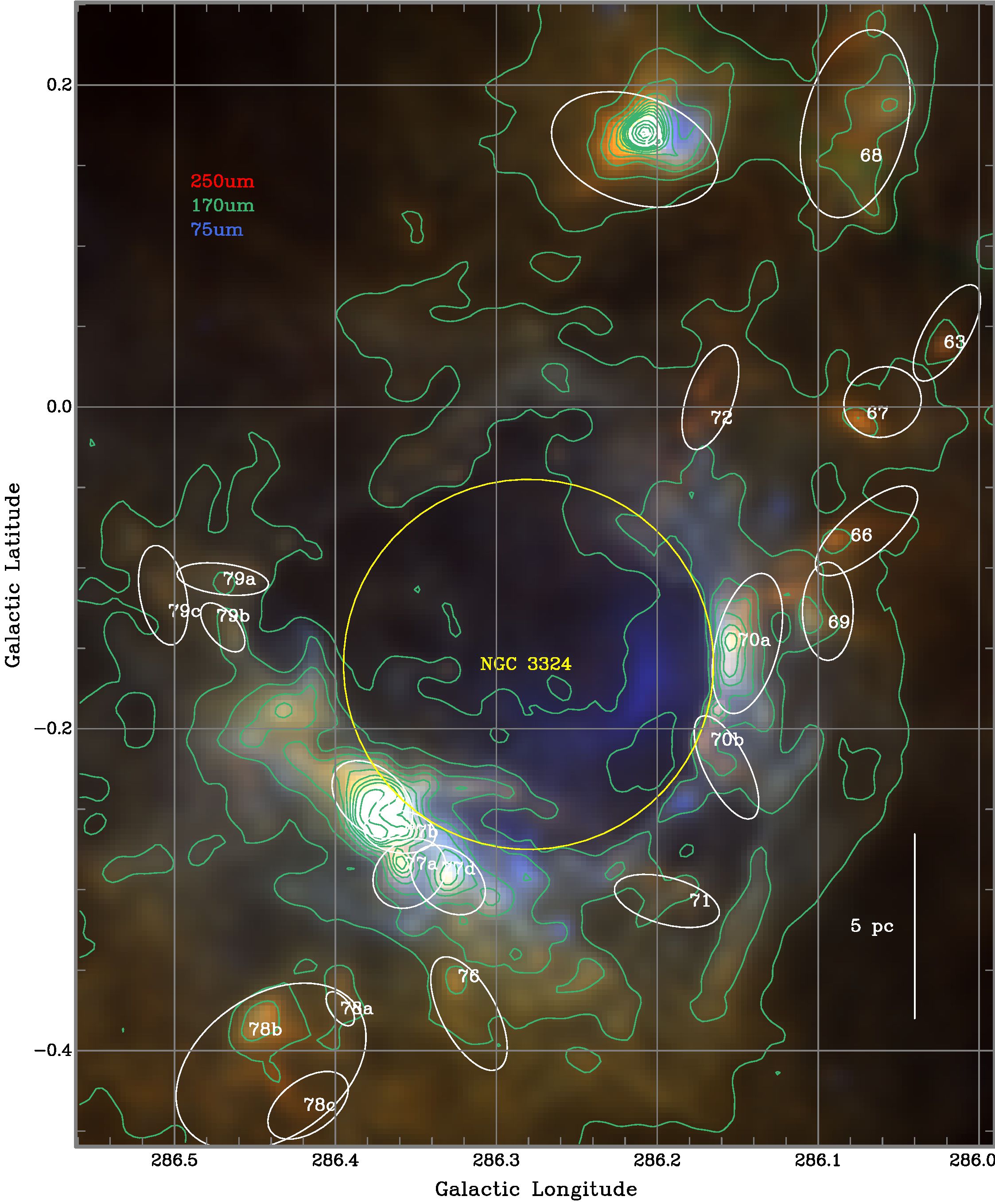}}
% All overlays included in figure using annotations file(s).  MUCH simpler!
\caption{%\footnotesize % uncomment for aasj
Composite RGB image as labelled from {\em Herschel}-PACS and -SPIRE data of the western end of the $\eta$ Car GMC \citep{p19}.  This shows the general molecular cloud cartography in a roughly 0\fdeg7$\times$0\fdeg5 area surrounding the classic HII region NGC\,3324 at a distance of 2.5\,kpc \citep[][giving the scale bar as shown]{s21} and encompassing most of Region 9 from the CHaMP project \citep{b18}.  The FIR colour contrast was maximised in order to make clear the variations corresponding to the SED-fitted dust temperature \td: redder colours indicate cooler dust \lapp10\,K where the 250\,$\mu$m emission is more dominant, bluer colours show warmer dust up to $\sim$40\,K where 75\,$\mu$m emission dominates.  All the images are somewhat saturated for the brightest clumps in order to bring out details of the fainter FIR emission.  Green contours are overlaid for the PACS 170$\mu$m data at levels of 15(15)135 and 160(100)560 mJy/arcsec$^2$, where x(y)z denotes ``from x in steps of y to z''.  Also overlaid as white ellipses are the approximate half-power sizes of the CHaMP clumps' \tco\ emission \citep{b16}, each labelled by their BYF number at the \tco\ emission peak.
}\label{higal}\vspace{0mm}
\end{figure*}

%A third approach uses the Zeeman effect, observed through circular polarisation effects.  This is the only practical technique to measure the line-of-sight component of the field, $B_{||}$.  
Studies using the Zeeman effect, which measures the line-of-sight component of the field $B_{||}$, have shown that below $n_0$ $\approx$ 300\,cm$^{-3}$, $B$ fields can support gas against gravity and have fairly uniform strength.  Above this level, the line-of-sight component increases with density, $B_{||} \propto n^\kappa$ where $\kappa$ = 0.65, and the ratio of magnetic to gravitational forces is close to critical \citep{cru12}.  A more recent study by \cite{wh24} suggests that the change in slope $\kappa$ at $n_0$ is more muted, and that $n_0$ itself is larger.  Tracking local variations in the transition density $n_0$ and density dependence $\kappa$ are both important to star formation (SF) theory, since SF is only observed in higher-density gas and a variable transition could change the SF efficiency and/or initial mass function.  
% In Vela C, for example, the alignment of $\theta_{B_{\perp}}$ with dense structures changes from parallel to perpendicular near the same threshold $n_0$ as in the Zeeman data \citep{faa19}.

Catching massive protostars in the act of formation, however, is more difficult than for low-mass protostars, because of their greater distances, accelerated timescales, and rapid alteration of initial conditions.  Thus, data on massive cluster-scale clumps, where most massive protostars likely also form, are very sparse.  We need to precisely measure both $|B|$ and $n$ in a wider variety of clouds and environments to test these ideas.

Region 9 of CHaMP \citep[the {\em Galactic Census of High- and Medium-mass Protostars};][]{b18} is a promising area for study, covering the western $\sim$third of the 120\,pc-long $\eta$ Carinae GMC ({\color{red}Fig.\,\ref{higal}}).  It comprises 21 massive parsec-scale clumps across a 30\,pc field, and includes the 9\,pc-wide classic HII region/open cluster NGC\,3324 at a distance of 2.50$\pm$0.27\,kpc \citep{s21}.  These clumps \citep[denoted individually by their catalogued BYF number from CHaMP;][]{b11} form a representative sample of molecular clouds at all significant stages of star formation, from starless or pre-stellar (BYF\,68, 71, 72, 76, 79a--c) to early-protostellar (BYF\,63, 66, 67, 69, 78a--c) to more advanced massive protostellar stages (BYF\,73, 77a--d) to post-SF feedback affected clumps (BYF\,70a--b).

Importantly, all these clouds are physically and kinematically associated with each other at one distance, that of NGC\,3324; see \cite{p19}, \cite{p21}, and references therein for information on a wide range of physical properties of these clouds.  Intercomparing properties among {\em these} clouds avoids complications arising from samples at widely disparate distances observed with a single angular resolution.  The project described herein was therefore conceived to examine, at a fixed physical resolution of 0.16\,pc, the role of magnetic fields in star formation via polarisation mapping of this varied but representative sample.

Among these clumps, we have previously published observational studies of BYF\,73 using SOFIA and other facilities.  \cite{p18} reported SOFIA/FIFI-LS, {\em Gemini}, {\em Spitzer}, and ATCA data on the status of the most massive protostellar object in the cloud, MIR\,2, and its less massive protostellar neighbours.  \cite{b23} described SOFIA/HAWC+ and ALMA polarisation data on the cloud's detailed physical conditions, including $B$ fields, gas dynamics, and energetics.  We now turn to the wider sample of clouds near BYF\,73 in CHaMP Region 9, in order to put their overall and individual magnetic properties into a wider context.

In this paper, we describe the observational and data reduction procedures in \S\ref{observ}.  In \S\ref{analysis} we use two standard statistical methods to analyse our polarisation data and obtain constraints on the role of $B$ fields in these clouds.  We discuss all these results in \S\ref{disc} in order to highlight new insights from the data as well as their limitations.  We present our conclusions in \S\ref{concl}.

%%%%%%
%   Fig.2  %	Global HAWC+ data
%%%%%%
\begin{figure*}[ht]
\vspace{0mm}\centerline{
\includegraphics[angle=0,scale=0.88]{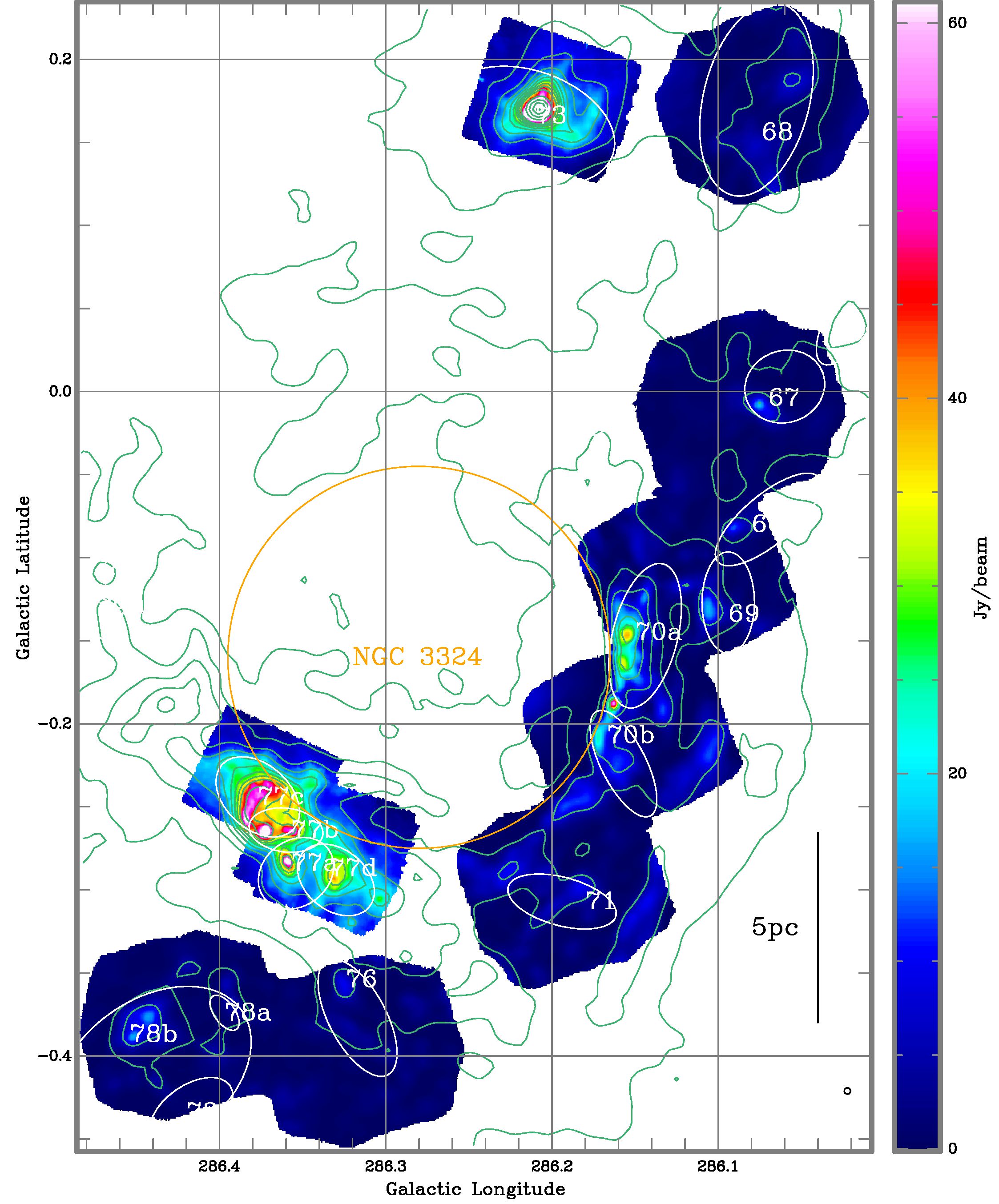}}  % option B

%%%{\footnotesize % uncomment for aasj version
\vspace{-1mm}\caption{%\renewcommand{\baselinestretch}{0.5}\footnotesize % uncomment for aasj
Overview of all SOFIA HAWC+ band D (154$\mu$m) total intensity (Stokes $I$) maps of Region\,9 from both Cycle 7 (2019) and Cycle 9 (2022) observations, covering almost the same area as Fig.\,\ref{higal}.  The band D angular resolution, 13\farcs6, is shown in the bottom right corner as a very small black circle.  The three Cycle 7 fields show up as smallish rectangles, reflecting the instantaneous field of view of the HAWC+ camera and the standard nod-chop (NMC) imaging procedure with fixed blank-sky reference locations.  The seven Cycle 9 fields appear as slightly larger irregular octagons, reflecting the coverage of Lissajous tracks on the sky for the more efficient on-the-fly (OTF) imaging procedure available then.
The $I$ image is overlaid by the same white \tco\ ellipses and green PACS contours as in Fig.\,\ref{higal}; the latter can be converted to the HAWC+ data scale by a factor 210\,arcsec$^2$/bm, and both instruments' maps are seen to conform well to each other.
The peak $I$ intensity is highly variable across the different clumps visible here.  Compact sources in BYF\,73 (north of NGC\,3324) and 77 (on the SE edge of NGC\,3324) reach peak flux densities of 478 and 90\,Jy/bm (respectively) and so are saturated on the displayed $I$ scale (colour bar on the right), but the extended emission under 60\,Jy/bm is well-reflected in the image.  In either observing mode, the typical rms $I$ error in the interior of each field has a central minimum around 0.10--0.13\,Jy/bm, rising to 2--5$\times$ these values around the field boundaries; the S/N in the various emission features then runs from a few 100s to $>$5000.  
}\label{hawcmaps}\vspace{0mm}
\end{figure*}

%%%%%%%%%
%                      %
%   Section 2    %
%                      %
%%%%%%%%%
\section{HAWC+ Observations and Data Reduction}\label{observ}

%%%%%%%%%
%   Section 2.1  %
%%%%%%%%%
%\subsection{SOFIA/HAWC+}

This project was executed with HAWC+'s band D ($\lambda$154\,$\mu$m) filter in two stages, with first maps in Cycle 7 (one flight in July 2019)\footnote{See the HAWC+ description at https://irsa.ipac.caltech.edu/d ata/SOFIA/docs/instruments/hawc, its Data Handbook at https:/ /irsa.ipac.caltech.edu/data/SOFIA/docs/instruments/handbooks/ HAWC\_Handbook\_for\_Archive\_Users\_Ver1.0.pdf, and the Cycle 7 Observer's Handbook at https://irsa.ipac.caltech.edu/data/S OFIA/docs/sites/default/files/Other/Documents/OH-Cycle7.pdf for details of the observing modes.} 
of clumps BYF\,73 and 77a--d, followed up by more extended mapping in Cycle 9 (four flights in July 2022)\footnote{See the Cycle 9 Observer's Handbook at https://irsa.ipac. caltech.edu/data/SOFIA/docs/sites/default/files/2022-12/oh-cycl e9.pdf for details of the observing modes.} 
of 11 more clumps.  The project was also favourably reviewed for Cycle 10, but the sudden defunding of the observatory prevented us from completing the planned coverage of the other 5 clumps and achieving our overall sensitivity target.

During both Cycles we carefully planned to orient the fields at various position angles in the equatorial coordinate frame so as to optimise the camera's coverage onto the brightest emission, and produce a relatively seamless mosaic.  Unfortunately however, different software bugs in each Cycle prevented this optimisation during the observing runs, resulting in the fields being oriented in unexpected ways with respect to each other, as described further below.

The Cycle 7 mapping was done in the standard NMC nod-chop mode for that Cycle, which produces rectangular maps of angular size $\sim$4$'$ corresponding approximately to the HAWC+ camera's instantaneous sky footprint.  Chopping and nodding were done asymmetrically due to the nearby FIR emission to the Galactic west and south.  The total on-source integration times were 784.4\,s for BYF\,73 (1 field) and 1007.7\,s for the BYF\,77 complex (2 fields).  Pipeline processing with HAWC-DRP produced final Level 4 quality Stokes image products which were downloaded from the SOFIA archive.  This processing produces data that has all known instrumental and atmospheric effects removed, giving an absolute Stokes $I$ calibration uncertainty of 20\%, a relative polarisation uncertainty of 0.3\% in flux and 3\degree\ in angle, and astrometry which should be accurate to better than 3$''$ \citep{hrd18}.  However, we found the HAWC+ L4 astrometry for BYF\,73 was still consistently offset $\sim$2$''$ to the Galactic south compared to the other maps described by \cite{b23}.  Similarly, we found that the BYF\,77 map was offset $\sim$4$''$ to the Galactic south compared to the {\em Herschel} 70\,$\mu$m map.  We inserted both of these corrections by hand into the HAWC+ data files.

%%%%%%
%   Fig.3  %	Region 9 North
%%%%%%
\begin{figure*}[ht]
\vspace{0mm}\centerline{
\includegraphics[angle=0,scale=0.65]{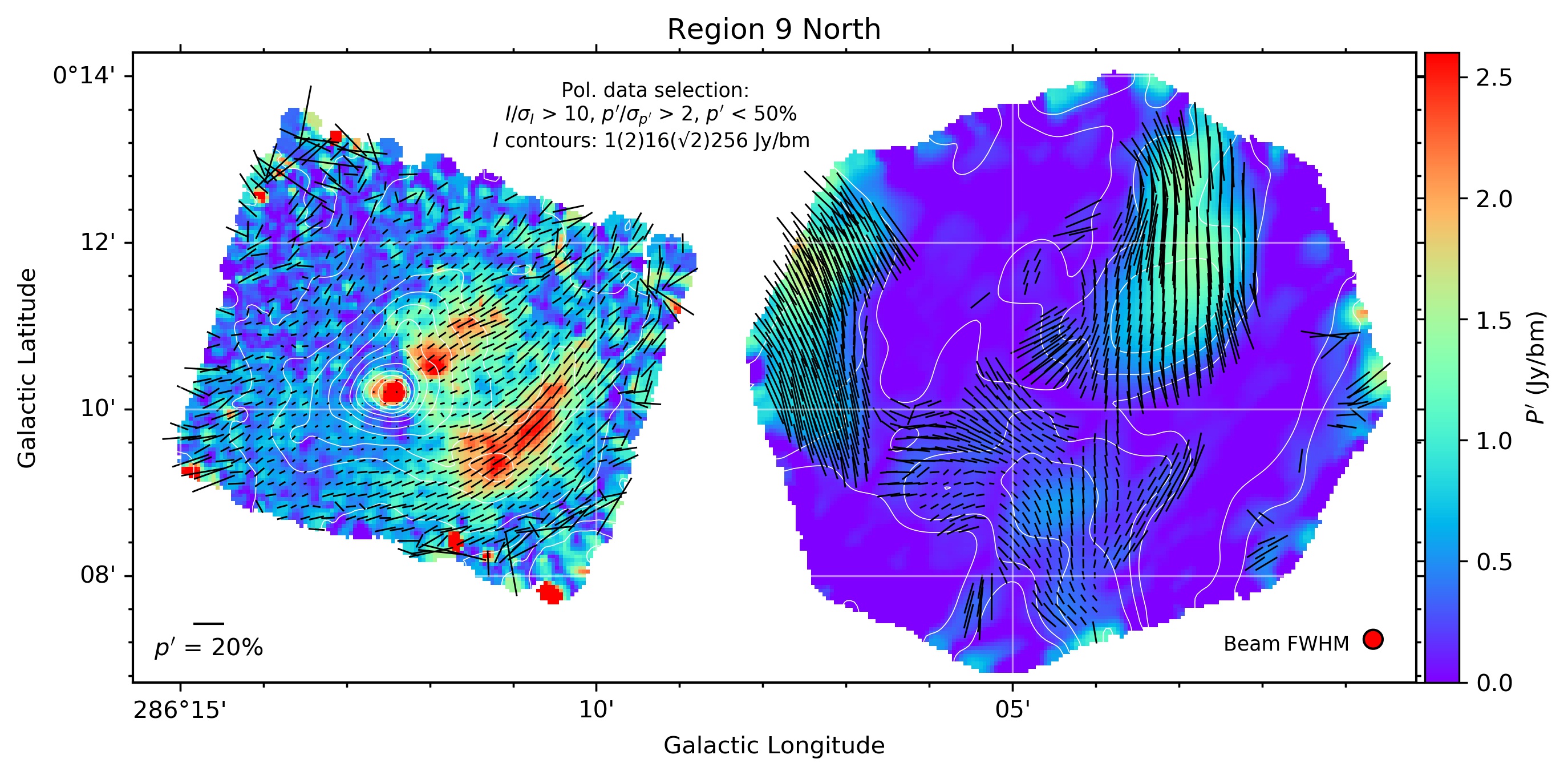}}
\vspace{-3mm}\caption{
Zoom in to the northern portion of Fig.\,\ref{hawcmaps} showing fields for the clumps BYF\,73 (left) and 68 (right).  The background image is now the HAWC+ debiased polarised flux density $P'$, overlaid here by white HAWC+ $I$ contours as labelled.  At every 3rd pixel (0.6 beam) satisfying the indicated selection criteria, we also display black ``vectors'' showing the debiased polarisation percentage ($p'$, with a scale bar shown in the bottom left corner) at a position angle (with the usual $\pm\pi$ degeneracy) of the plane-of-sky $B$ field component (i.e., rotated 90\degree\ from the observed polarisation direction).  The respective BYF\,73/68 peak $P'$ values are 5.2/1.7 Jy/bm, with central uncertainties 0.12/0.05 Jy/bm rising slowly to each field's edge, and S/N peaking at 33/20.  The percentage polarisation $p'$ has very similar S/N to $P'$, while uncertainties in the position angle/inferred $B$ field orientation are typically a few degrees, except for the noisier pixels at the edge of the BYF\,73 field.  Consult Fig.\,\ref{quality} to see how the polarisation data quality varies across all the HAWC+ fields.
%Above $I$ = 16 Jy/bm for the HAWC+ data, %(or the 76\,mJy/arcsec$^2$ PACS contour), 
%nearly all $p'$ vectors here and in subsequent Figures have S/N ranging from $\sim$5--25; for 8$<$$I$$<$16 Jy/bm, displayed vectors have S/N$\sim$2--6. %and are not displayed if their S/N$<$2 or where $I$$<$0.25 Jy/pixel.  
}\label{r9north}\vspace{-1mm}
\end{figure*}

The Cycle 9 mapping utilised the newly-available on-the-fly (OTF) mode, in order to improve the observing efficiency \& sensitivity and produce slightly larger ($\sim$6$'$) \& more-easily mergeable fields, albeit with a somewhat variable noise level across each field due to the integrated sky coverage of the camera using a Lissajous scanning pattern.  The total on-source integration time across the 7 OTF fields was 3388\,s for an average of 484\,s per field.  In addition to the normal calibrations as above, all 7 fields were merged in the HAWC-DRP to produce the final L4 Stokes mosaics.  The astrometry in this case was found to be internally consistent across the mosaic, but offset by $\sim$3$''$ to the Galactic west compared to the {\em Herschel} 70\,$\mu$m map, a correction for which was again inserted by hand.

For the data analysis described in the following sections, the Level 4 data for all fields from both Cycles were mosaicked together onto a consistent J2000 equatorial grid for each Stokes parameter using the {\sc Miriad} package \citep{st95} in a set of custom unix shellscripts, then transformed to Galactic coordinates to simplify comparisons with other data.  Supplementary analysis and display also included use of Jupyter notebooks from sample python scripts in the online HAWC+ handbooks, the {\em karma} package \citep{g97}, and the SuperMongo package \citep{lm00}.

%%%%%%
%   Fig.4  %   Region 9 West
%%%%%%
\begin{figure*}[t]
\hspace{-4mm}
\centerline{\includegraphics[angle=0,scale=1.18]{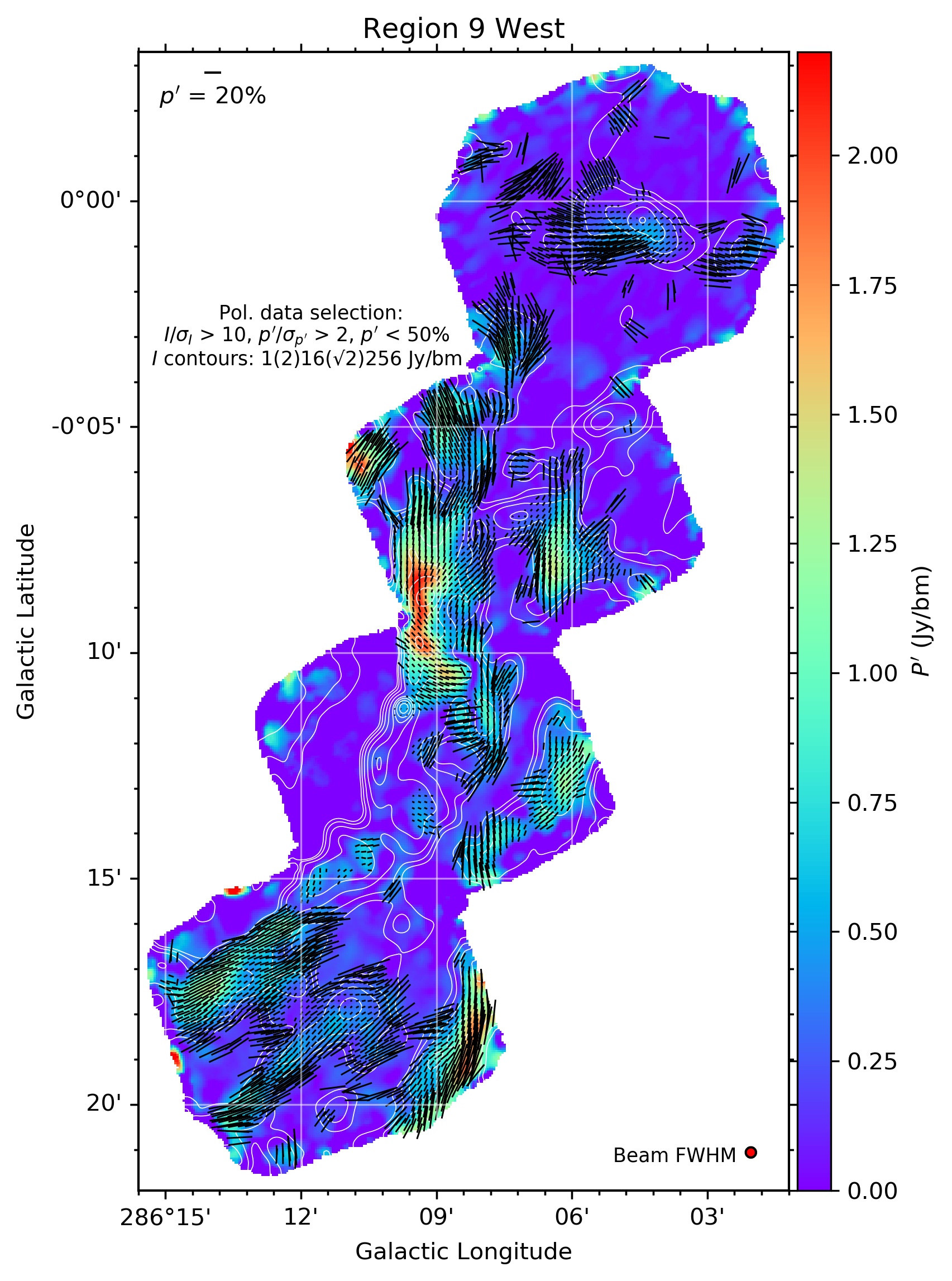}}
\vspace{-3mm}\caption{ %\footnotesize % uncomment for aasj
Similar zoom in to Fig.\,\ref{r9north} but here for the western portion of Region 9 (clumps BYF\,66, 67, 69, 70a--b, and 71).  The background image is again of $P'$ with a peak of 2.2\,Jy/bm, uncertainty minima 0.03, 0.06, 0.10, and 0.06\,Jy/bm in the 4 fields from north to south, and peak S/N of 18.  The vector selection criteria are as before.
}\label{r9west}\vspace{0mm}
\end{figure*}

%%%%%%
%   Fig.5  %   Region 9 South
%%%%%%
\begin{figure*}[t]
\hspace{-2mm}
\centerline{\includegraphics[angle=0,scale=0.76]{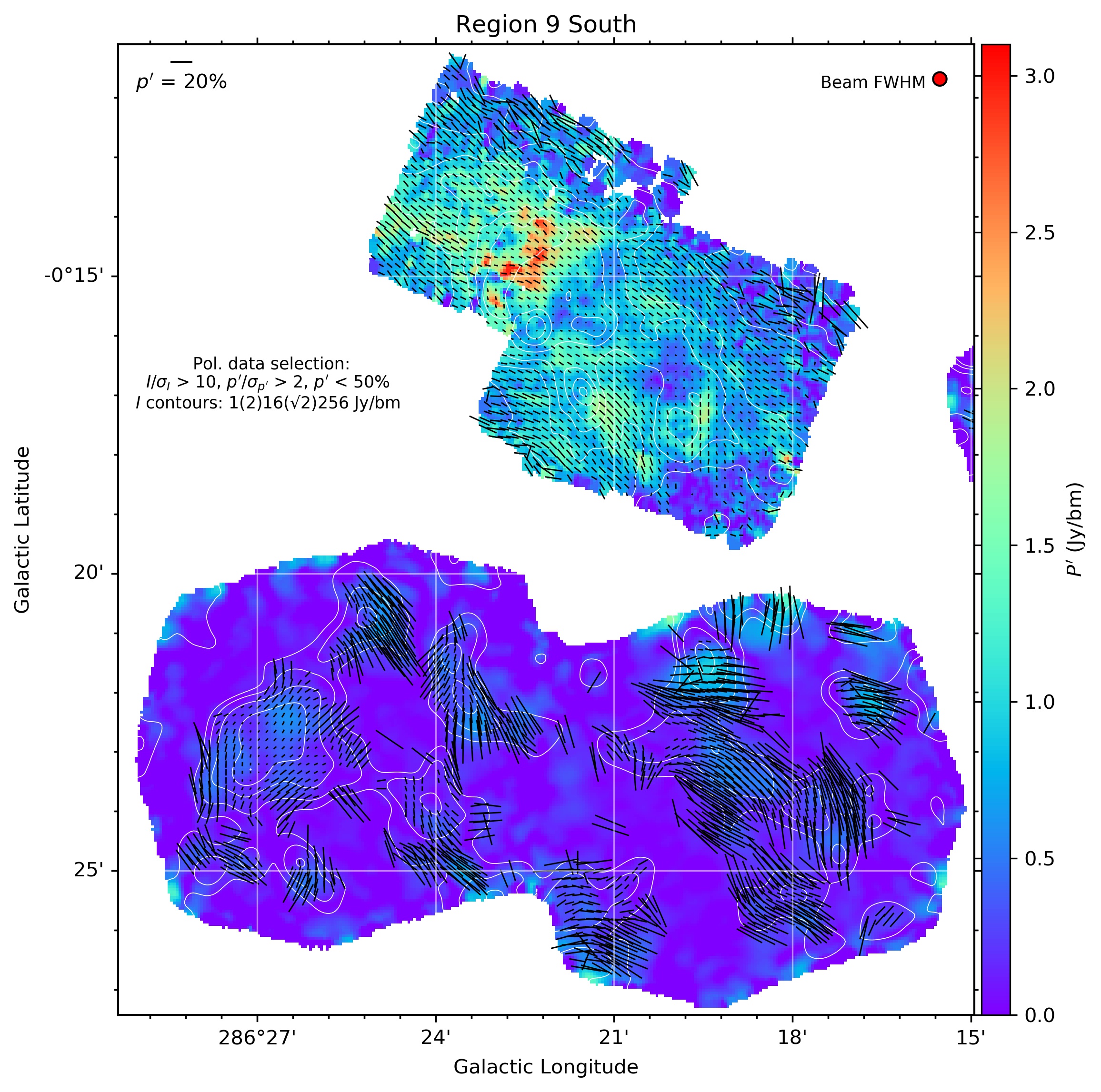}}
\vspace{-3mm}\caption{ %\footnotesize % uncomment for aasj
Zoom in to the southern portion of Region 9 (clumps BYF\,76, 77a--d, and 78a--c).  The background $P'$ image has a peak of 3.1\,Jy/bm, uncertainty minima $\sim$0.2, 0.08, 0.05, \& 0.04\,Jy/bm in the 4 fields from top-left to bottom-right, and peak S/N of 26.  The vector selection criteria are as before.
}\label{r9south}\vspace{-1mm}
\end{figure*}

%%%%%%%%%
%                      %
%   Section 3    %
%                      %
%%%%%%%%%
\section{Data Analysis}\label{analysis}

At a distance of 2.50$\pm$0.27\,kpc for NGC\,3324 \citep{s21}, the scale in Region 9 is 36$''$ = 0.44\,pc, or 0.1\,pc = 8\farcs25.  Thus, HAWC+ band D gives us a useful spatial dynamic range from 0.08\,pc to the maximum scale in the mosaic, $\sim$30\,pc, a linear factor of about 360 and exceeding 10$^4$ resolution elements in area. 

The final Stokes $I$ mosaic is shown in {\color{red}Figure \ref{hawcmaps}}, overlaid with the same features as in Figure \ref{higal}.  Poor overlaps of adjacent fields due to the observing software PA bugs, especially for the Cycle 9 data, are also visible.

%%%%%%
%   Fig.6  %   Data quality
%%%%%%
\begin{figure}[t]
\hspace{-1mm}{\includegraphics[angle=0,scale=0.208]{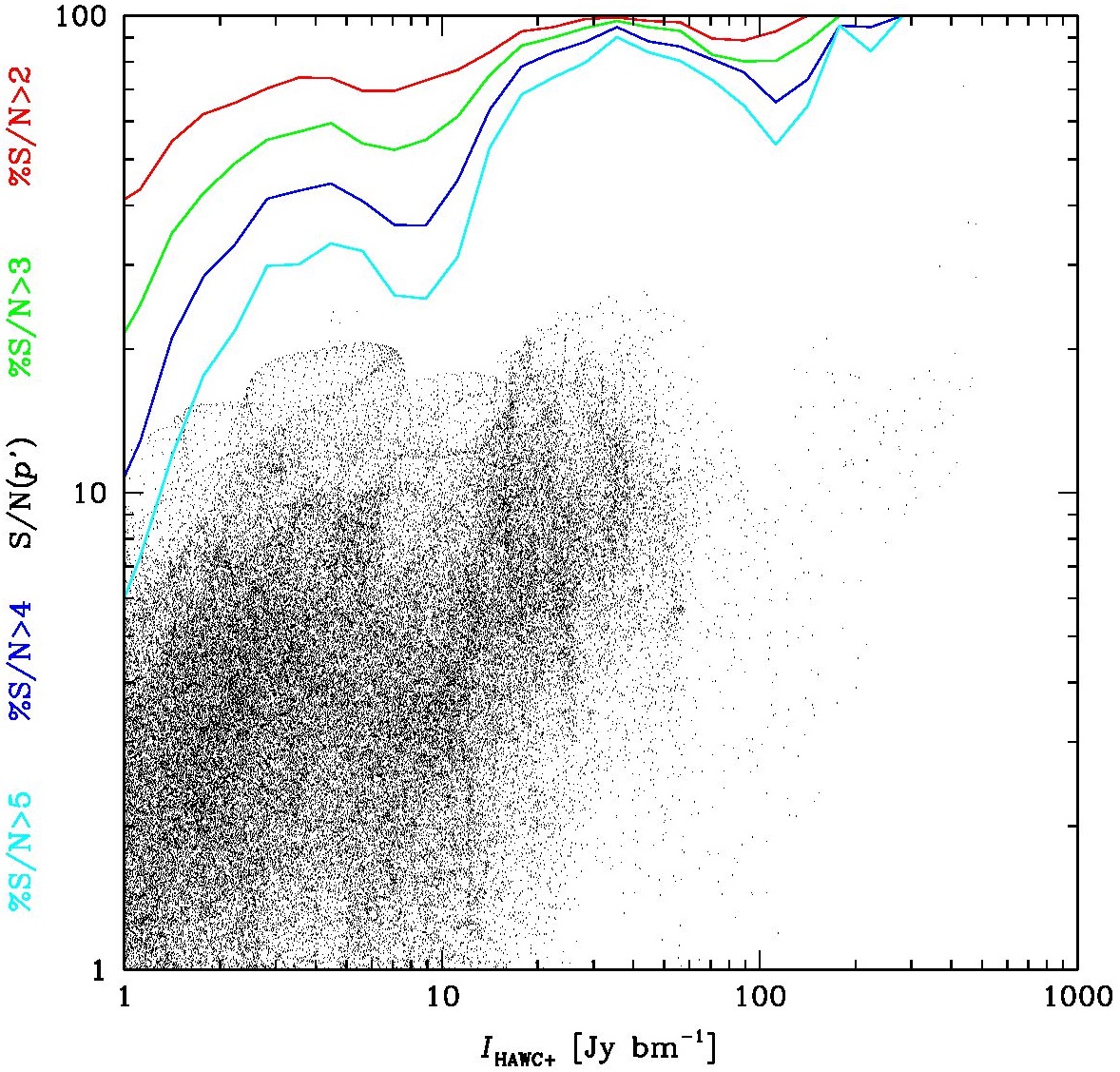}}
\vspace{-2mm}\caption{ %\footnotesize % uncomment for aasj
Summary of data quality metrics for all polarisation vectors shown in Figs.\,\ref{r9north}--\ref{r9south}.  The data points show the S/N in $p'$ at each pixel as a function of the pixel's HAWC+ $I$ flux (black axis labels).  The maximum S/N is 37, and the y-scale goes down to S/N = 1 for simplicity.  The coloured lines and coloured y-axis labels show the fraction of these points, as a percentage, that have S/N $>$ the indicated values as a function of $I$.
}\label{quality}\vspace{-1mm}
\end{figure}

To display the details of the HAWC+ polarisation data more clearly, we show zoomed-in displays of the northern ({\color{red}Fig.\,\ref{r9north}}), western ({\color{red}Fig.\,\ref{r9west}}), and southern ({\color{red}Fig.\,\ref{r9south}}) HAWC+ fields, all on a common printed scale.  In each field, the background image is the debiased Stokes $P'$, % = $\sqrt{Q^2+U^2-n_{\rm rms}^2}$, where $n_{\rm rms}$ is the combined instrumental and sky noise
overlaid by white contours of the Stokes $I$ (the same data as the Fig.\,\ref{hawcmaps} background image) plus ``vectors'' of the debiased polarisation percentage $p'$.  These data products can be directly analysed by standard techniques.

A summary plot of the data quality is given in {\color{red}Figure \ref{quality}}.  The fraction of pixels with polarisation S/N above any quality level goes up as the HAWC+ $I$ flux rises.  For example, at $I$ \gapp\ 2.5\,Jy\,bm$^{-1}$ (slightly lower than the lowest PACS contour in Figs.\,\ref{higal} and \ref{hawcmaps}), $>$50\% of the polarisation data points have S/N$>$3 in $p'$.  At $I$ \gapp\ 15\,Jy\,bm$^{-1}$, $>$80\% of the vectors have S/N$>$3 and $>$50\% have S/N$>$5.  Allowing for the oversampling of the HAWC+ beam in Figs.\,\ref{r9north}--\ref{r9south}, there are roughly 9,000 independent polarisation measurements with S/N$>$3 in the combined Cycle 7+9 data sets.

In what follows, we use the same general analysis approach (DCF and HRO methods) as for our prior detailed BYF\,73 study \citep{b23}.  This will facilitate comparisons in the respective results.

%%%%%%%%%
%   Section 3.1  %
%%%%%%%%%
\subsection{Davis-Chandrasekhar-Fermi (DCF) Method}\label{DCFstuff}

DCF analysis \citep{d51,cf53} is a widely-used approach that yields $B$ field estimates from plane-of-sky polarisation data; however, it is rather approximate \citep[see reviews by][]{cru12,liu22}.  Its appeal is to directly link the dispersions in polarisation angle $\delta\theta$ = $s$ (tracing variations in the $B$ field orientation) to physical parameters in the ISM, assuming that $s$ arises from the propagation of a transverse MHD wave in a turbulent plasma.  The degree to which this is true gives rise to the uncertainties.

Under this assumption, the dispersion $s$ is related to three other physical parameters that simply and naturally describe the MHD wave: the gas density $n$, the line-of-sight velocity dispersion $\delta$$V$, and the plane-of-sky magnetic field strength $B_{\perp}$.  That is, $s$ will increase as (1) $B$ decreases, since then the magnetic restoring forces are reduced; (2) $n$ increases, since then the medium's inertia to the MHD wave is greater; or (3) $\delta$$V$ increases, since that describes the strength of the MHD wave.  

Following \cite{BL15}, we use the SI formulation (1\,nT = 10\,$\mu$G) of the DCF relation: %According to \cite{cn04}, with appropriate SI unit conversions the projected $B$ field strength %(in picotesla) is
%\begin{displaymath} 
%	B_{\rm pos} = Q~\sqrt{4\pi\rho}~\delta V/s~;
%\end{displaymath}
% With rho = mu.m.n, delta-V in km/s, s in degrees, and leaving m,n in cgs, B =
% sqrt [pi * (mu=2.34871884058) * 1.67353283716x10-24 g * n cm-3 / 8ln2] * 10^5 cm/s * (180/pi) = 85.5x10-7 G = 8.55 uG = 0.855012530777 nT, or 0.55790077393 nT without mu.  More precise!
% Now I get 0.83252692784561 nT and 0.54322878395873 nT
% For equivalent Crutcher et al 2004 mu value, * sqrt(2.8/2.34871884058) = 0.933286334173 nT.
% Changing n to m-3, we have an extra sqrt(10-6) and nT --> pT:
\begin{equation} % EQUATION ONE
	B_{\rm \perp,DCF} = 0.558\,{\rm pT}~\sqrt{\mu n}~(\Delta V/s)~,
\end{equation}
where %$\rho$\,=\,$\mu\,m_{\rm H}n$ as before 
% Although, I still wonder about the lack of g-->kg conversion: sqrt(10-3 kg) => 1/31.6x smaller.  Also, where is sqrt(mu0)?  => 3160 bigger???  A factor of 100 is missing to cancel these two out...  bloody cgs!!!
$n$ (m$^{-3}$) is the gas density with mean molecular weight $\mu$=2.35, $\Delta V$ = $\sqrt{8{\rm ln}2}~\delta V$ is the velocity FWHM (\kms) in the cloud, and $s$ is measured in degrees.  Included in the constant is a numerical factor $Q$\,=\,0.5 from \cite{cn04} to correct for various smoothing effects \citep[e.g., see][]{osg01}, although recent work summarised by \cite{liu22} suggests $Q$ can vary somewhat ($\sim$0.3--0.6) in various circumstances.  

However, several modifications to the classical DCF method have also been proposed \citep[see][]{ppvii}.  Among these is that of \cite{st21}, who point out that the original DCF approach was for incompressible Alfv\'en waves; according to them, compressible MHD waves should dominate in the ISM.  They propose
\begin{equation} % EQUATION TWO
	B_{\rm \perp,ST} = 0.1042\,{\rm pT}~\sqrt{\mu n/s}~(\Delta V)~,
\end{equation}
where we have converted their relation to the same SI units as Eq.\,(1).  The general effect of Eq.\,(2) is to give, with the same inputs, slightly smaller $B_{\perp}$ values than the classical DCF method.  Other alternatives have also been proposed, as described by \cite{ppvii}, but the differences are subtle, and we are content to consider the above two approaches as representative.

Either way, with estimates, or even better, {\em maps} of the quantities $n$, $\Delta$$V$, and $s$, we can compute maps of the estimated line-of-sight component $B_{\perp}$ (subject to the uncertainties that go into Eqs.\,1 and 2).

We start by deriving $s$ maps from our HAWC+ data.  The somewhat involved procedure is described in {\color{red}Appendix \ref{DCFdetails}}; the final result is shown in {\color{red}Appendix \ref{parmaps}}.

\begin{figure*}[t]
\centerline{(a)\vspace{-2mm}
	\includegraphics[angle=-90,scale=0.53]{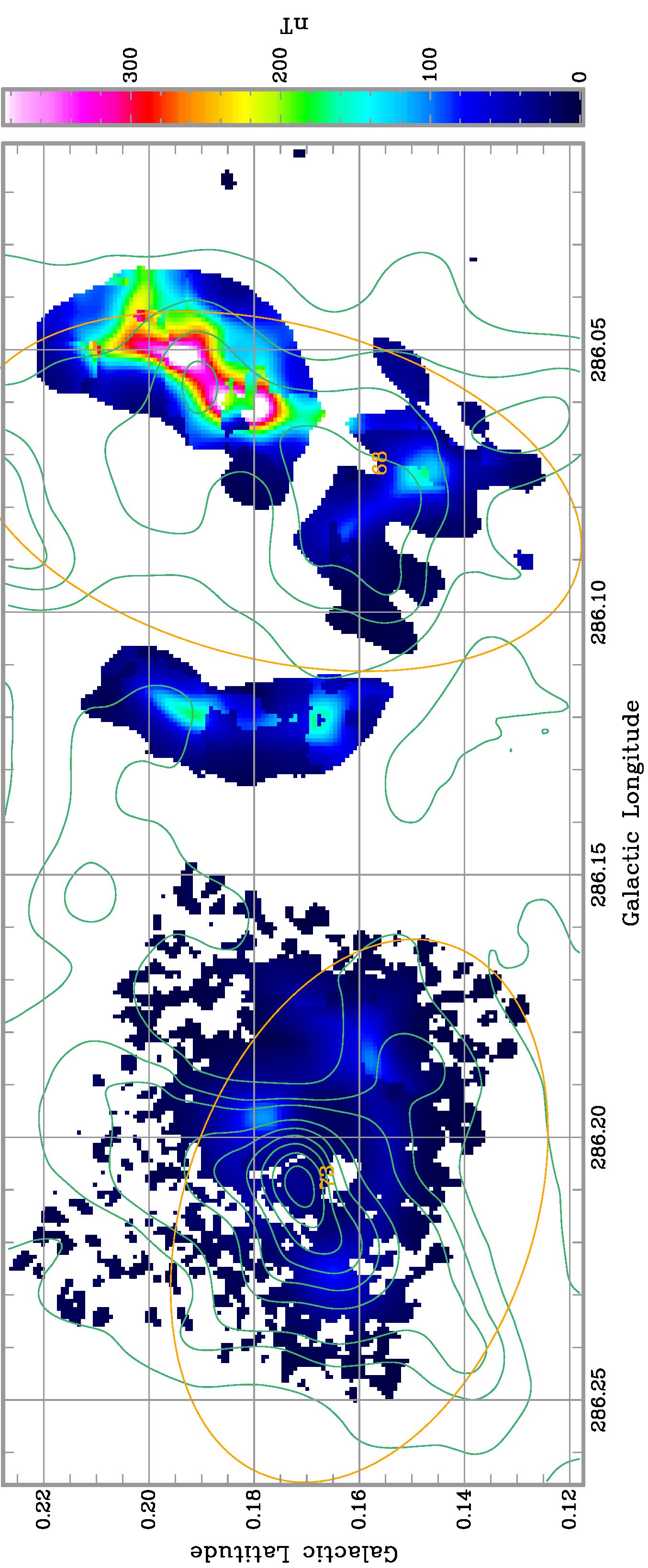}} %kvis ctrpix453,823 on(b)frame
\vspace{-4mm}
\centerline{\hspace{-0.4mm}(b)\vspace{-1mm}
	\includegraphics[angle=-90,scale=0.53]{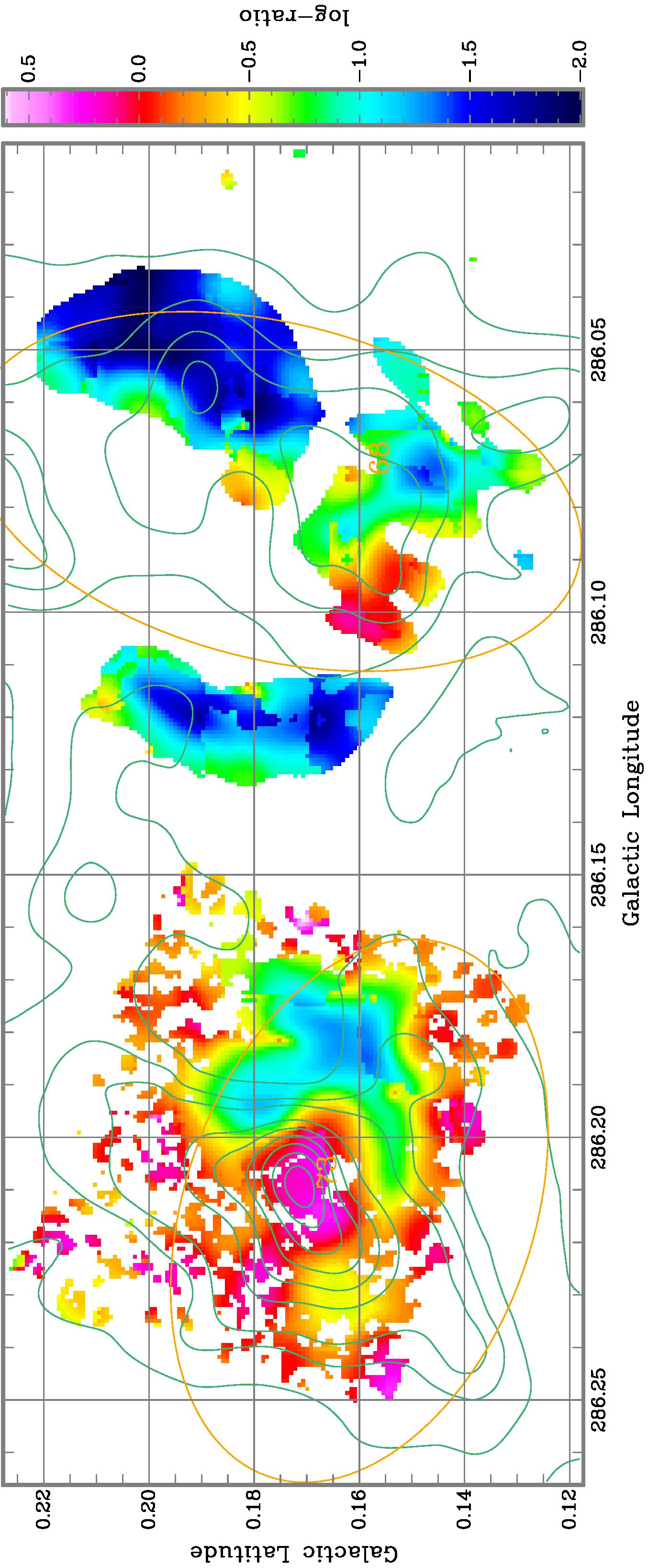}} % kvis ctr pix 453,823
\vspace{-1mm}
\caption{ %\footnotesize  % uncomment for aasj
Maps of (a) $B_{\rm \perp}$ from DCF computation of Eq.\,(1), and (b) log$_{10}\lambda$ from Eq.\,(4), in Region 9 North.  Clump sizes of BYF\,73 and 68 are shown by ellipses as in Figs.\,\ref{higal} and \ref{hawcmaps}, while contours are of column density \nhtwo\ from \cite{p19}, at levels 0.5(0.5)2 and 4(2)14 $\times$10$^{26}$ molecules\,m$^{-2}$.  In (a), the colour scale is slightly saturated to show weaker $B_{\rm \perp}$ features; the peak $B_{\rm \perp}$ value in BYF\,68 is 469\,nT.  In (b), the colour scale spans the full range of log$\lambda$ values. %The conversion from SI to cgs units is 100\,nT = 1\,mG. %(1\,$\mu$T = 10\,mG).
}\label{northB}\vspace{-1mm}
\end{figure*}

\begin{figure*}[t]
\centerline{\includegraphics[angle=0,scale=0.73]{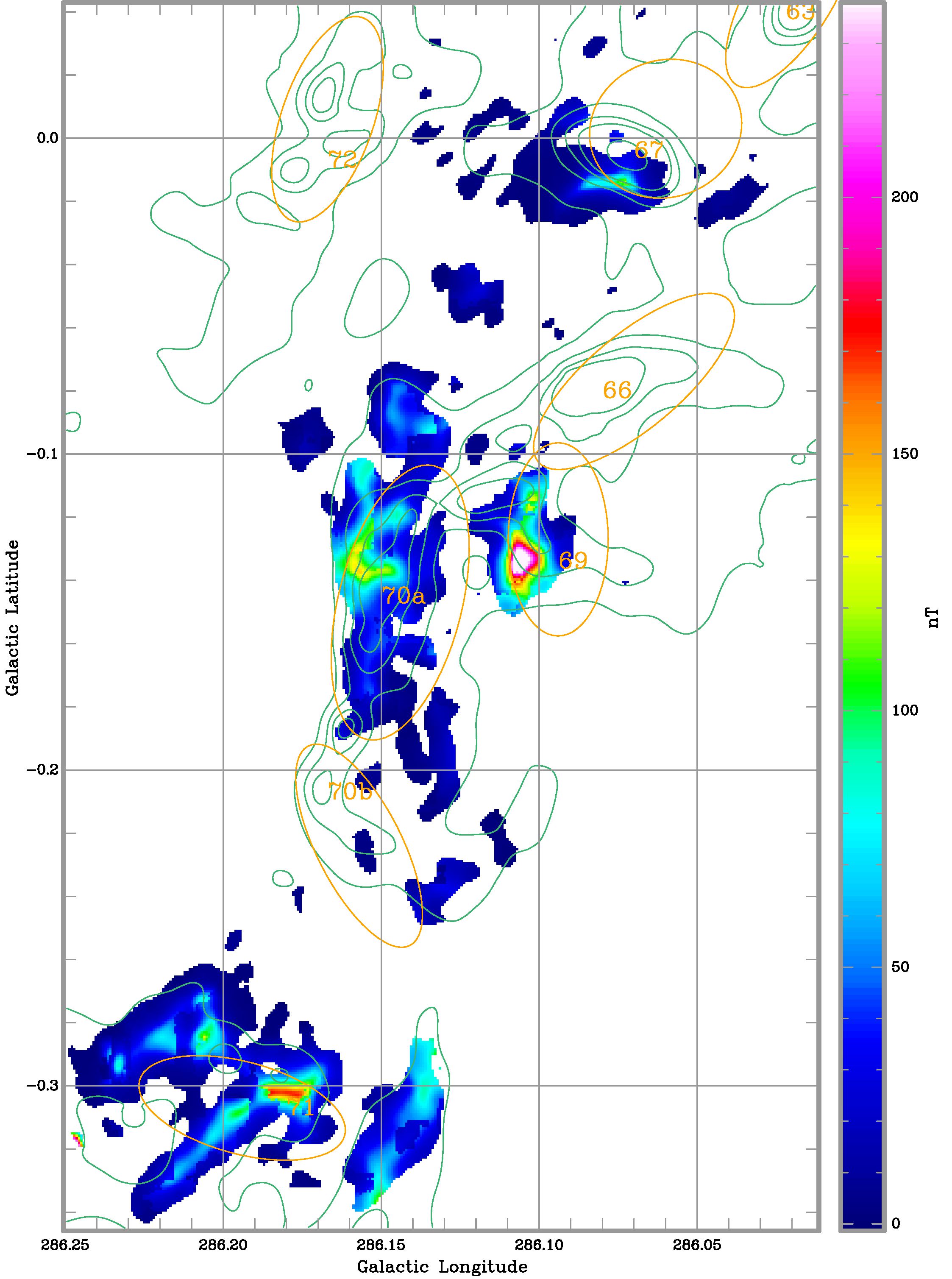}} % kvis ctr pix 463,399
\vspace{-2mm}
\caption{ %\footnotesize  % uncomment for aasj
(a) Map of $B_{\rm \perp}$ as in Fig.\,\ref{northB}a but for Region 9 West (BYF\,66, 67, 69, 70a--b, 71).  The printed scale is the same, as are the overlaid $N$(H$_2$) contours and \tco\ ellipses.  The colour scale is again slightly saturated; the peak $B_{\rm \perp}$ value in BYF\,69 is 324\,nT.
%\centerline{\hspace{2mm}\includegraphics[angle=0,scale=0.72]{reg9w-loglamcl.jpg}} % kvis ctr pix 463,399
%Map of log$_{10}\lambda$ as in Fig.\,\ref{northB}b but for Region 9 West.  The printed scale is the same, as are the overlaid green $N$(H$_2$) contours and red \tco\ ellipses.
}\label{westB}
\vspace{-2mm}
\end{figure*}

\begin{figure*}[t]
\centerline{\includegraphics[angle=0,scale=0.73]{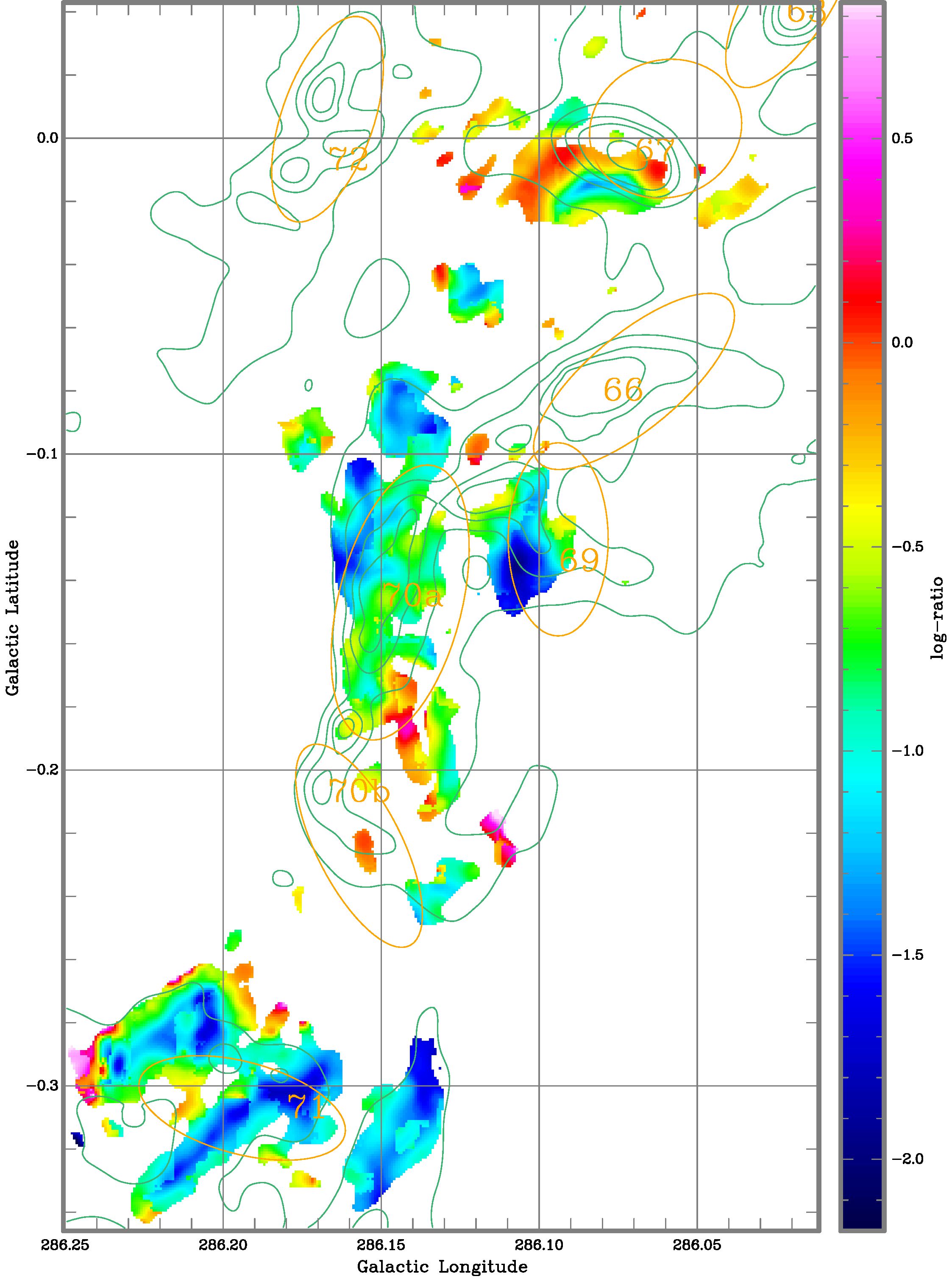}} % kvis ctr pix 463,399
\vspace{-2mm}
\caption{ %\footnotesize  % uncomment for aasj
(b) Map of log$_{10}\lambda$ as in Fig.\,\ref{northB}b but for Region 9 West (BYF\,66, 67, 69, 70a--b, 71).  The printed scale is the same, as are the overlaid $N$(H$_2$) contours and \tco\ ellipses.  The colour scale spans the full range of log$\lambda$ values.
}\label{Wlamb}
\end{figure*}

\begin{figure*}[t]
\centerline{(a)\includegraphics[angle=0,scale=0.565]{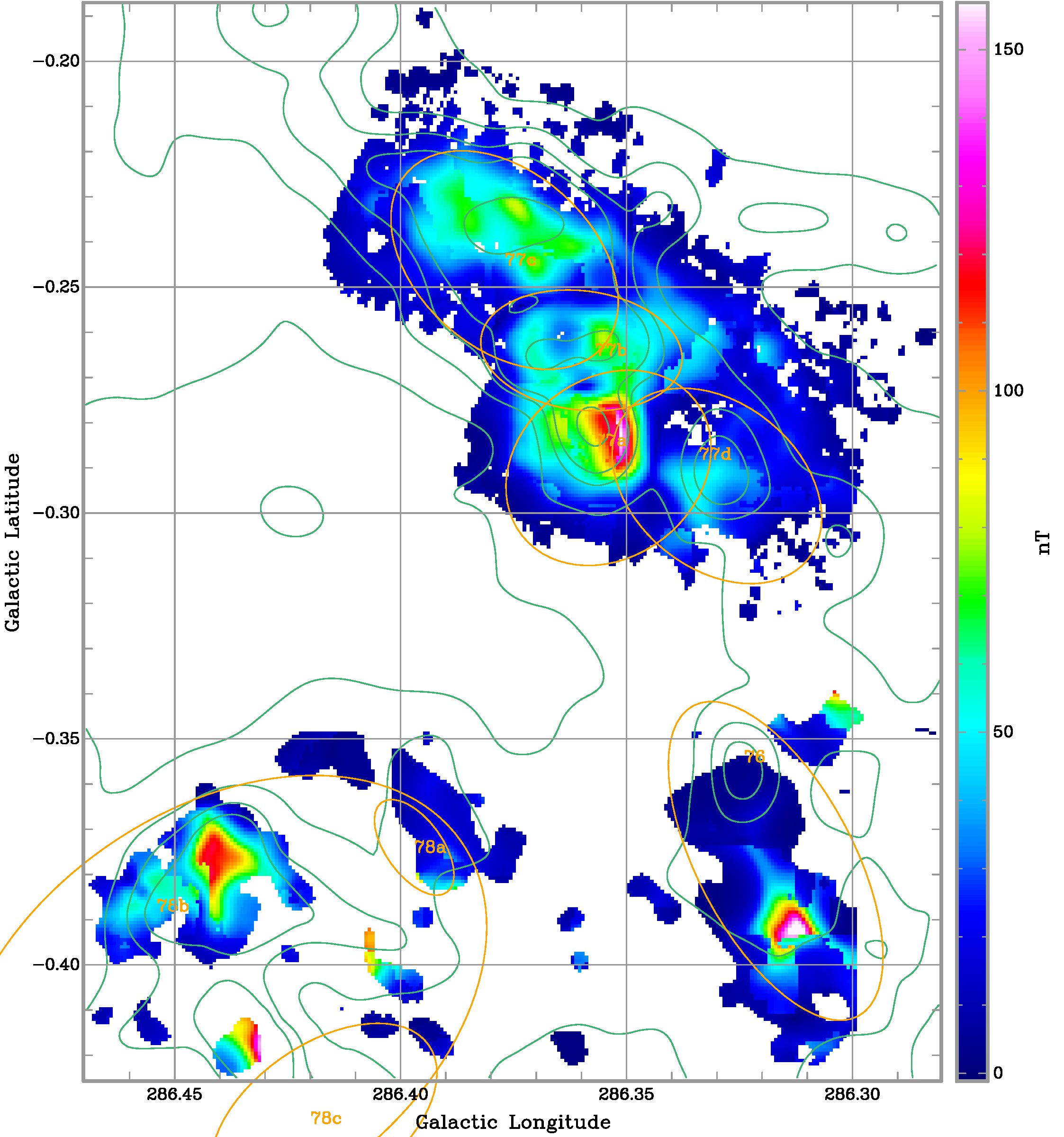}} % kvis ctr pix (156,190)
\vspace{-4mm}
\centerline{\hspace{-0.5mm}(b)\includegraphics[angle=0,scale=0.565]{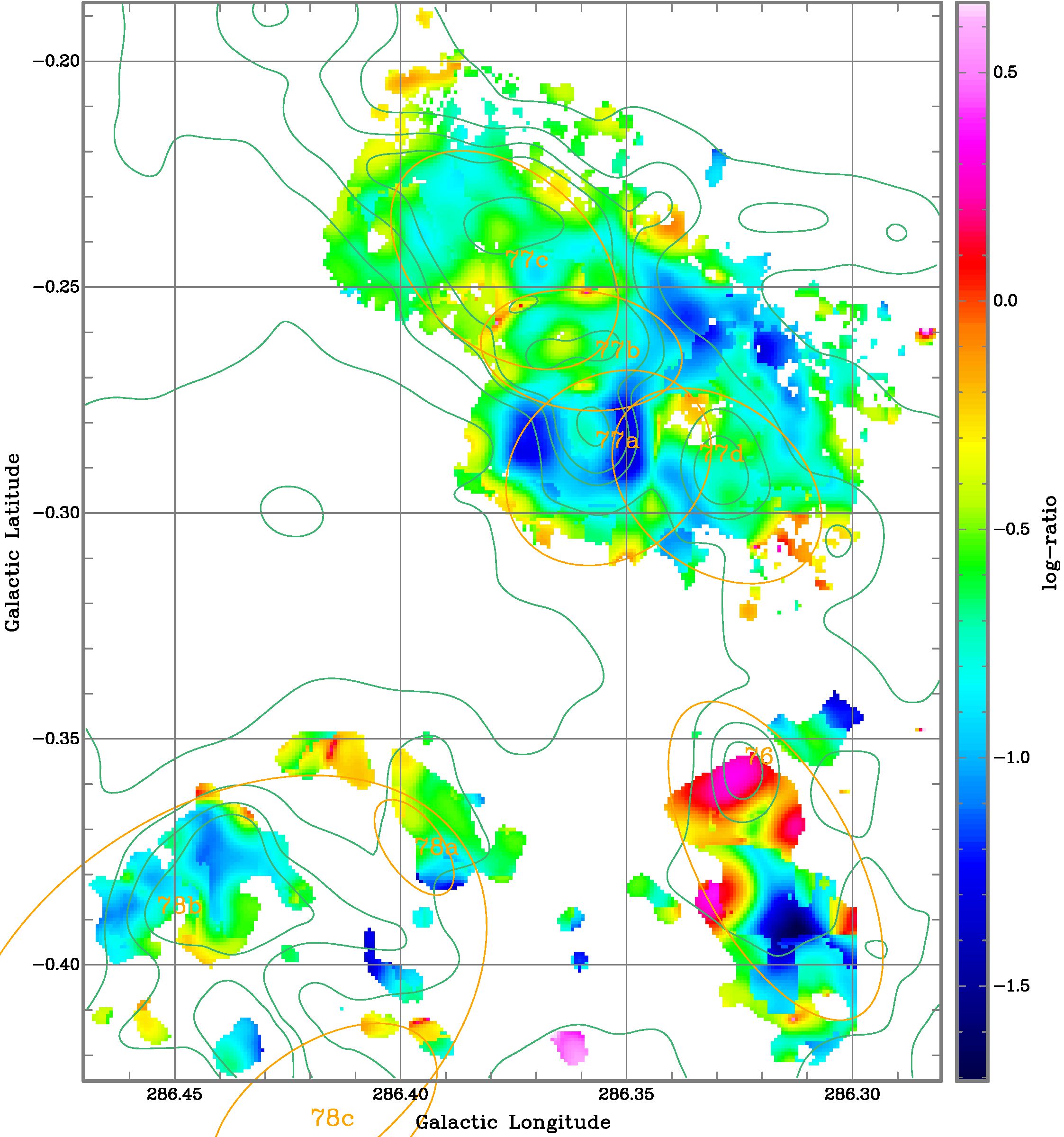}} % kvis ctr pix (156,190) 143,196
\vspace{-1mm}
\caption{ %\footnotesize  % uncomment for aasj
Maps of (a) $B_{\rm \perp}$ as in Figs.\,\ref{northB}a \& \ref{westB}a, and (b) log$_{10}\lambda$ as in Figs.\,\ref{northB}b \& \ref{Wlamb}b, but for Region 9 South (BYF\,76, 77a--d, 78a--c).  The printed scale is the same, as are the overlaid $N$(H$_2$) contours and \tco\ ellipses.  In (a), the colour scale is again slightly saturated; the peak $B_{\rm \perp}$ value in BYF\,76 is 177\,nT.  In (b), the colour scale spans the full range of log$\lambda$ values.
}\label{southB}
\end{figure*}

We next consider the density.  We can't measure $n$ directly, but estimate it instead from two other parameters.  These are: (1) the \nhtwo\ from SED fitting of {\em Herschel} data \citep[shown as green contours in Figs.\,\ref{northB}--\ref{southB}, from][]{p19}, which is likely the best type of estimate for the total gas column density in the cold ISM; and (2) a simple but appropriate estimate for the ({\em a priori} unknown) line-of-sight depth $R$ of the clumps across Region 9, to convert from column to volume density.

We could use various structure-finding approaches to evaluate the sizes of the more obvious clumps in the FIR or spectral-line data (such as the Mopra ellipses displayed in these figures), and then assume depths for these structures commensurate with their projected sizes.  But this approach is somewhat unsatisfying, since it uses only some of the structural information available in the maps.  We choose instead an estimate for $R$ which is structure-agnostic. 

Consider the gradient $\nabla$\nhtwo\ of the column density map.  Molecular clouds have structure on scales all the way from a map's size to the resolution limit, often with local maxima embedded in the more extended structure.  Thus, gradients like $\nabla$\nhtwo\ generally indicate where values rise towards each local peak.  Given this, an appropriate size scale or depth for the more extended cloud envelope will be larger than the $\sim$beam-sized scale/depth for the smaller structures.  In other words, the shallower the gradient, the larger the size and inferred depth.  Conversely, the steeper the gradient, the smaller the size/depth.  So, we initially estimate the appropriate depth across our maps as the inverse of the relative gradient (hereafter, IRG) in the \nhtwo\ map, i.e., \nhtwo/$\nabla$\nhtwo\ = IRG = $R$ (with the relevant unit conversions).

However, this estimate will tend to fail where the gradient, assumed to be smoothly increasing towards each isolated clump peak, becomes very small around local maxima, since a very small $\nabla$\nhtwo\ means a very large $R$, making the derived density very small.  Away from the actual clump peaks this is of little consequence, since there, the derived densities are already small (and there is less likelihood of polarisation data anyway).  But near the smaller clump peaks where $N$ reaches significant maxima and the polarisation S/N is also optimal, the gradients will systematically approach zero, leading to very large $R$ and very small $n$, just where we expect $n$ to peak.  This appears as prominent, unphysical ``doughnut holes'' in the $n$ maps at the $N$ peaks.

To cure this, we also compute the Laplacian (i.e., curvature) of the column density map, $\nabla^2$\nhtwo, which would be positive everywhere that the gradient keep increasing.  Where the gradient decreases --- and especially around significant peaks, goes to zero --- the Laplacian becomes negative.  Where this happens, we use this threshold as a discriminant to substitute a maximum $R$ scale for the IRG, namely the projected beamsize.  The resulting maps for $R$ \& $n$ appear in Appendix \ref{parmaps}. 

In summary, we use
\begin{eqnarray} % EQUATION THREE
	n &=& \frac{N_{H_2}}{R} = \frac{N_{H_2}}{\rm IRG} = \frac{N_{H_2}}{N_{H_2}/\nabla N_{H_2}} \nonumber \\
	&=& \nabla N_{H_2}~,~~~~~~~~~{\rm where}~\nabla^2N_{H_2} > 0 \\
	&=& N_{H_2}/R_{\rm beam}~,~~{\rm where}~\nabla^2N_{H_2} \leq 0 . \nonumber \end{eqnarray}

Finally, we need a velocity dispersion map to compute the DCF-based $B_{\rm \perp}$.  We take the \nco-derived $\sigma_V$ for Region 9 from \cite{b18}, which has the advantage over any single-species $\sigma_V$ (such as from \tco\ or \ttco) of being intrinsically corrected for the line opacity in either species (see Appx.\,\ref{parmaps}).  Combining these maps of $s$, $n$, and $\Delta$$V$ according to Eq.\,(1), we show maps of $B_{\rm \perp}$ in {\color{red}Figures \ref{northB}a--\ref{southB}a}.

Having computed a map of $B_{\rm \perp}$, we can add another important data product as a bonus.  As a measure of the relative importance of gravity (compression of clumps or cores) to $B$ fields (pressure support), the mass-to-flux ratio $M$/$\Phi$ has been widely used to highlight structures that are susceptible or resistant to star formation \citep[e.g.,][]{cn04}.  From \cite{BL15} we can write
\begin{equation} % EQUATION FOUR
	\lambda = \frac{(M/\Phi)_{\rm obs}}{(M/\Phi)_{\rm crit}} = 6.38~\frac{N_{\rm H_2}/10^{26}{\rm m}^{-2}}{B_{\rm TOT}/{\rm nT}}~~
% CNWK04: 2.8mH 2pi*sqrt(G) N/B = 2.8*1.67e-24*6.28*2.583e-4/1e-6 Ncm-2/Bug = 7.601347e-21 ug.cm2
% pseudo-SI: (2.34871884058* 1.67353283716x10-24 g*cm-2/10-6G) *2pi*sqrt(6.6732x10-8) = 63.7987557306e-22 ug.cm2 = 63.7987557306e-27 nT.m2 = 6.37987557306 nT/(10e26cm-2)  Precise!
\end{equation}
and assume that $B_{\rm TOT}$ = 2$B_{\rm \perp,DCF}$ on average.  Then, wherever $\lambda$$\gg$1 (or equivalently, log$_{10}\lambda$ \gapp\ 0.5), gravitational forces are stronger than magnetic forces and the likelihood of SF is enhanced; such a structure is considered to be ``supercritical.''  Conversely, where $\lambda$$\ll$1 (or log$_{10}\lambda$ \lapp\ --0.5), SF is inhibited (subcritical); $\lambda$$=$1 suggests that these forces are close to equilibrium.  Thus, a map of $\lambda$ can be compared to other tracers and provide useful insights into the dominant processes in a given cloud.  With the $B$ and $N$ maps already in hand, we get the log$_{10}\lambda$ maps shown in {\color{red}Figures \ref{northB}b--\ref{southB}b}.

%%%%%%
%   Fig.7  %   Region 9N DCF Bpos map
%%%%%%
%%%%%%%
% Fig.12not %   Region 9N lambda map as a separate figure
%%%%%%%
%\begin{figure*}[t]
%\centerline{\includegraphics[angle=-90,scale=0.53]{reg9n-loglamcl.jpg}} %kvis ctrpix (460,823) 453,823
%\caption{ %\footnotesize  % uncomment for aasj
%Map of log$_{10}\lambda$ from Eq.\,(4) in Region 9 North, with clumps BYF\,73 and 68 shown by red ellipses plus green contours of column density \nhtwo\ from \cite{p19} (levels 0.5(0.5)2 and 4(2)14 $\times$10$^{26}$ molecules\,m$^{-2}$), as in Fig.\,\ref{northB}.
%}\label{Nlamb}\vspace{0mm}
%\end{figure*}

These log$\lambda$ maps are suggestive in places, but not necessarily definitive.  For example, in the North (Fig.\,\ref{northB}), the massive collapsing core MIR\,2 in BYF\,73 \citep[a known massive YSO;][]{b23} is surrounded by a dense core with log$\lambda$ systematically $>$0.2, as one might expect where apparently gravitationally-driven inflows are overwhelming any means of support.  Meanwhile, the cloud's adjacent compact HII region transitions sharply to log$\lambda$ $<$ --0.7, again as expected for an overpressured bubble.  Nearby to the west, the quiescent, cold, starless clump BYF\,68 has the strongest $B$ field measured in Region 9.  Its log$\lambda$ is $<$--0.5 almost everywhere across it, except for a small patch peaking around 0.1, consistent with the $B$ field alone being capable of preventing most SF.

In the West and South subregions, the $B$ fields are somewhat strong in places, but otherwise rather moderate (i.e., $\sim$50--200\,nT).  Reflecting this, the log$\lambda$ values are generally somewhat negative.  The largest patches that approach or exceed log$\lambda$ = 0.2 are on the edges of BYF\,67 ($\sim$0.1) and 76 ($\sim$0.3).  Even where some YSOs are evident (BYF\,70a, 77ab), log$\lambda$ is still $<$0, and one is left to speculate that SF is only being triggered in these clouds because of the strong feedback from NGC\,3324.

Overall in Region 9, the log$\lambda$ distribution is very close to gaussian, with a very small extra tail to positive values (see \S\ref{disc}).  The mean$\pm$sigma values are log$\lambda$ = --0.75$\pm$0.45.  Above the 2$\sigma$ level (log$\lambda$$>$0.15) we find 2.2\% of the pixels; above 3$\sigma$ (log$\lambda$$>$0.6), we find 0.22\%.  Both fractions are about 50\% higher than expected for a truly gaussian tail.  Thus in this sample, it is rare even for massive molecular clumps to be supercritical, but the exceptions to this may be important --- see \S\ref{disc} for more details.

%%%%%%%%%
%   Section 3.2  %
%%%%%%%%%
\subsection{Histogram of Relative Orientations (HRO)}\label{HROstuff}

In this section we similarly follow our prior analysis procedures for BYF\,73 \citep{b23}.

The HRO method of analysing $B$ field orientations in star-forming gas is another widely-used, standard technique \citep[e.g.,][and references therein]{saa17}.  It allows us to examine how the $B$ field orientation changes with column density.  Usually, at lower molecular gas column densities $\sim$10$^{26}$\,m$^{-2}$, the $B$ field direction tends to be mostly parallel, or not show any preferred direction, relative to gas structures.  In contrast, the $B$ field is mostly perpendicular to higher column density structures $\sim$10$^{27}$\,m$^{-2}$.  This generally confirms a series of results by the lower resolution ($\sim$10$'$) Planck Collaboration \citep{pc16} over a wider range of molecular gas column densities.

This is widely attributed to a transition from subcritical gas at lower densities, where the flow is at least guided to some extent by the $B$ field, to near-critical or slightly supercritical gas at higher densities, where gravity is capable of overwhelming the magnetic pressure, allowing stars to form.  Does Region 9 conform to this picture?

%%%%%%
%   Fig.8  %   Region 9W DCF Bpos map; original scale=0.73
%%%%%%

\begin{figure}[t]
\centerline{\includegraphics[angle=0,scale=0.215]{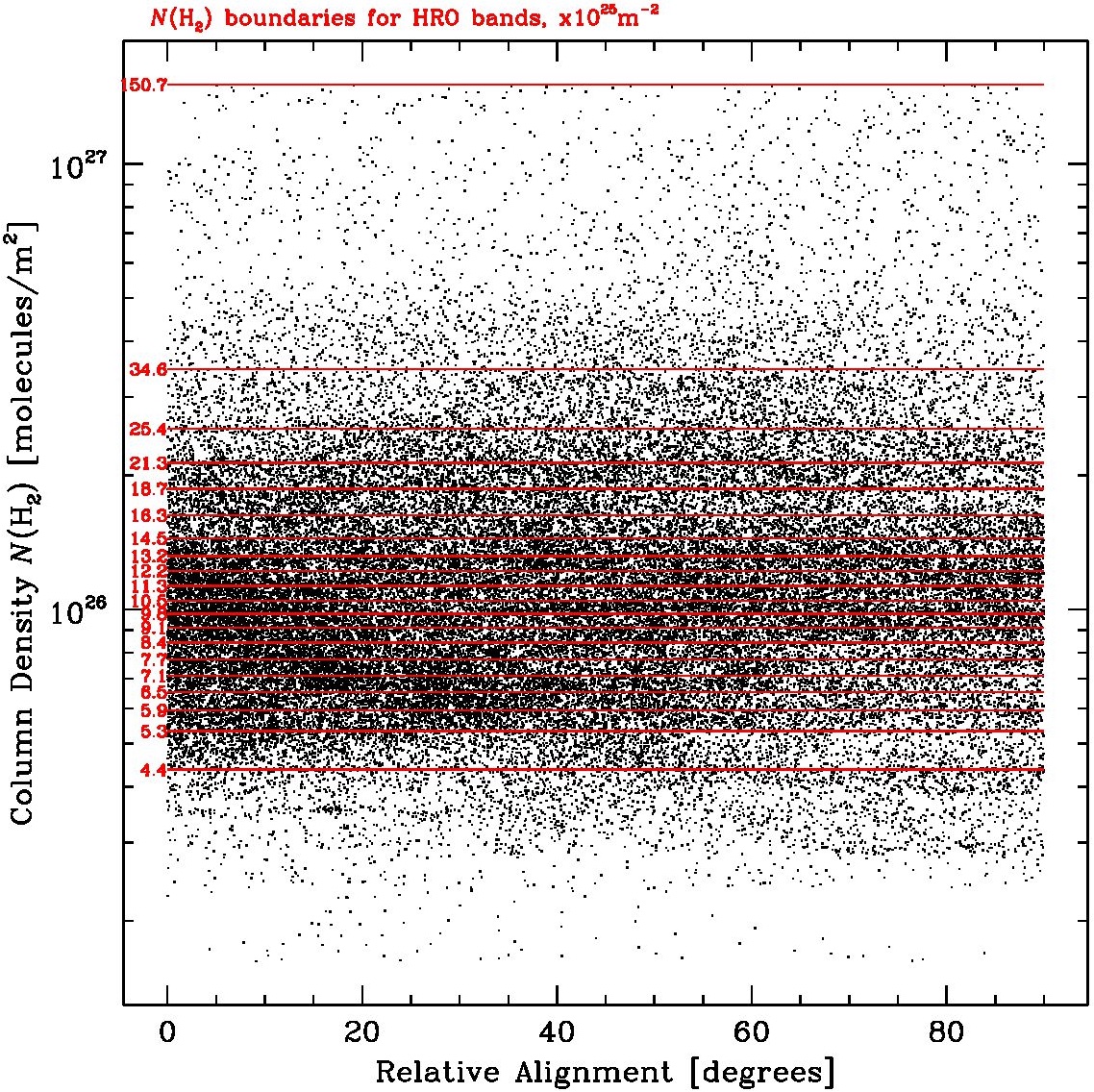}}
\vspace{0mm}\caption{ %\footnotesize % uncomment for aasj
Relative alignment between polarization position angle $\theta_{B_{\perp}}$ and the tangent to iso-column density $N$ contours, as a function of $N$ across the HAWC+ field.  An angle of 0\degr\ means the $B$ field is oriented along the iso-$N$ contours, while at 90\degr\ the field is perpendicular to the contours and aligned with the gradient $\nabla$$N$.  Also shown as red lines and labelled in $N$, in units of 10$^{25}$m$^{-2}$, are the boundaries of the separate bands in $N$ for which each histogram in Figure \ref{hawcHRObins} was computed.  The boundaries were chosen to ensure histogram equalisation, i.e., to divide all $N$ data into 20 equally-populated bins with comparable statistical noise in each $N$-bin.
}\label{hawcNtheta}\vspace{0mm}
\end{figure}

We show first in {\color{red}Figure \ref{hawcNtheta}} the per-pixel relative alignment of $B$ field vectors for all the HAWC+ data in Figures \ref{r9north}--\ref{r9south}, stratified by the SED-fit column density $N$ \citep{p19} which we use as a proxy for ``structure'' in the molecular gas.  That is, where the rotated polarisation vectors $\theta_{B_{\perp}}$ are aligned with the tangent to the iso-$N$ contours, the relative angle is close to 0\degr\ and the field is considered to be ``parallel'' to the gas structures.  Where $\theta_{B_{\perp}}$ is perpendicular to the contours and aligned with the column density gradient $\nabla$$N$, the relative angle is close to 90\degr\ and the field is considered to be ``perpendicular'' to the gas structures.  

This approach has the advantage of not imposing any preconceived interpretation of whether the gas structures represent ``clumps,'' ``cores,'' ``filaments,'' or any other potentially subjective term \citep[see][]{pc16,saa17}.  

\begin{figure}[t]
\centerline{\includegraphics[angle=0,scale=0.22]{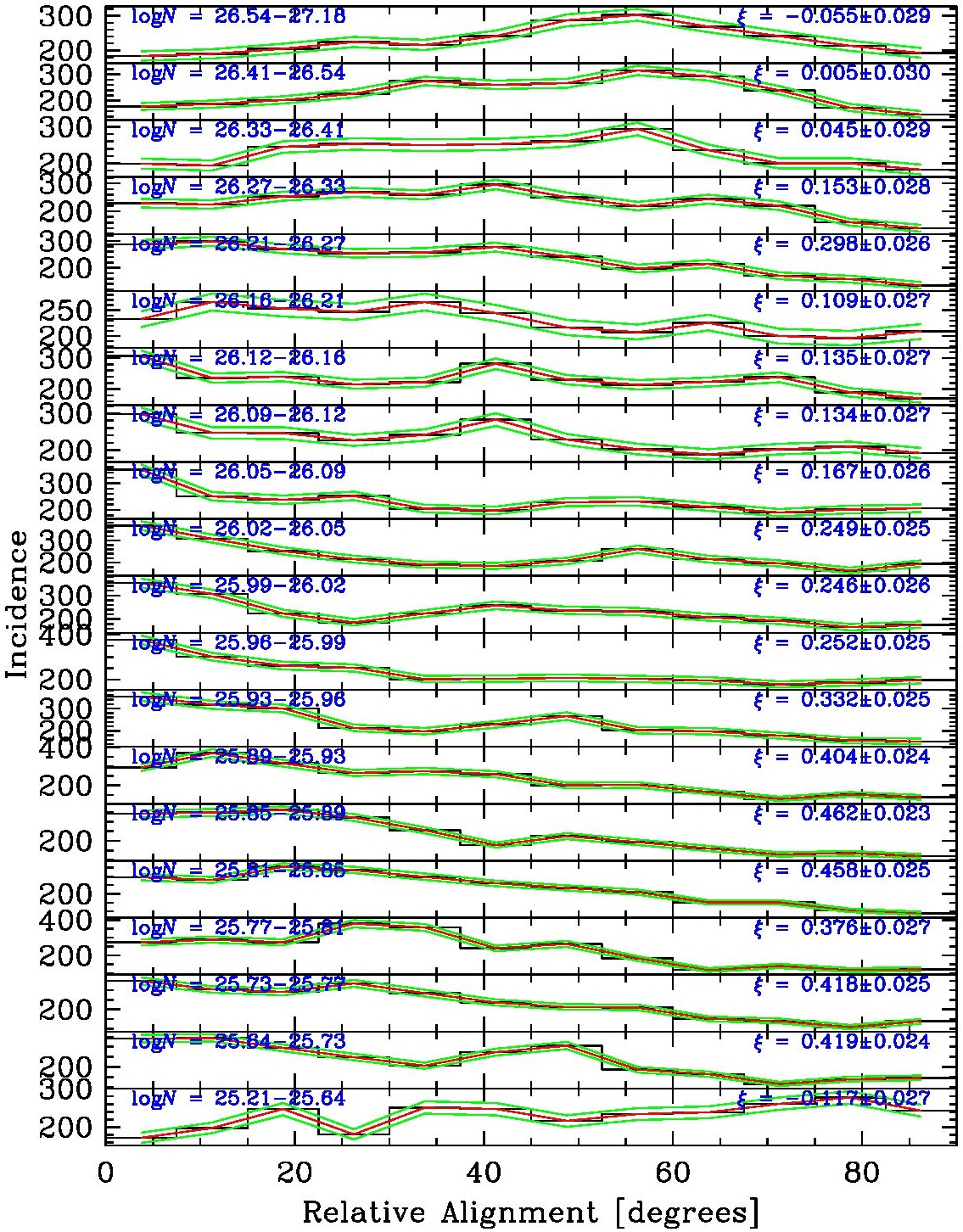}}
\vspace{0mm}\caption{ %\footnotesize % uncomment for aasj
HRO plots in each of the $N$-bins shown in Figure \ref{hawcNtheta}.  Each window is labelled by the range of column density $N$ in that bin and its fitted HRO shape parameter $\xi$ $\pm$ uncertainty, as defined by \citet{saa17}.
}\label{hawcHRObins}\vspace{0mm}
\end{figure}

The angle distribution is then quantified by computing histograms on each $N$-bin separately as in {\color{red}Figure \ref{hawcHRObins}}, including the HRO shape parameter --1 $<$ $\xi$ $<$ 1 computed on each $N$ bin's HRO, as described by \cite{saa17}.  This parameter objectively indicates whether there is a preponderance of parallel ($\xi$$>$0) or perpendicular ($\xi$$<$0) alignments in the data, and can be plotted as a function of $N$ ({\color{red}Fig.\,\ref{hawcHROxi}}) to reveal any trends via linear regression,
\begin{equation} % EQUATION FIVE
	\xi = C_{\rm HRO}~({\rm log}N-X_{\rm HRO})~.
\end{equation}
Already in Figure \ref{hawcNtheta} we can see that the distribution of relative alignments has definite patterns in various column density ranges.  These observations are reflected numerically in Figures \ref{hawcHRObins} and \ref{hawcHROxi}.  

%%%%%%
%   Fig.9  %   Region 9W lambda map as a separate figure; original scale=0.73
%%%%%%   Really want this to be Fig.8b

First, we discount the lowest $N$ bin because those data have low S/N in the $\theta_{B_{\perp}}$ values, hence the relative PA distribution is not so reliable (uncertainties $\sim$20\degr\ or more).  But in the next 16 low-$N$ bins, there is a strong overabundance of parallel alignments (relative PA \lapp\ 20--40\degr) between the inferred $B$ field orientation $\theta_{B_{\perp}}$ and the iso-$N$ contours.  Not only do each of these 16 bins have their shape parameter $\xi$ $>$ 0 to high significance (4--15$\sigma$), but there is a strong downward trend in $\xi$ as $N$ rises.  In the top three $N$ ranges, the downward trend in $\xi$ continues to $\sim$0, with PAs more broadly distributed ($\sim$20--60\degr) in those bins.  Unlike the BYF\,73-only plots, though, which included higher-resolution ALMA data and reached higher $N$ in that clump, in Region 9 generally we do not clearly progress to a column density regime where the alignments are more perpendicular (i.e., $\xi$ $<$ 0 or, say, PA \gapp\ 50\degr).

%%%%%%%
%   Fig.10   %   Region 9S DCF Bpos map
%%%%%%%
%%%%%%%
% Fig.14not %   Region 9S lambda map as a separate figure
%%%%%%%
%\begin{figure*}[t]
%\centerline{\includegraphics[angle=0,scale=0.60]{reg9s-loglamcl.jpg}} % kvis ctr pix (156,190) 143,196
%\caption{ %\footnotesize  % uncomment for aasj
%Map of log$_{10}\lambda$ as in Figs.\,\ref{Nlamb} \& \ref{Wlamb} but for Region 9 South (BYF\,76, 77a--d, 78a--c).  The printed scale is the same, as are the overlaid green $N$(H$_2$) contours and red \tco\ ellipses.
%}\label{Slamb}\vspace{0mm}
%\end{figure*}

Nevertheless, the negative trend of $\xi$ with $N$ is clear, especially for the red fit in Fig.\,\ref{hawcHROxi}, with a slope $C_{\rm HRO}$ = --0.51.  This is of similar magnitude and significance (8$\sigma$) to the BYF\,73-only results \citep{b23}.  Furthermore, the transition value of log$N$, where $\xi$ goes through 0 for Region 9 as a whole, is $X_{\rm HRO}$ = 26.56 (red value in Fig.\,\ref{hawcHROxi}), within 3$\sigma$ of the BYF\,73-only value as well \citep{b23} and similarly sharp (small uncertainty).  Thus, our HAWC+ data and HRO analysis suggest that we are possibly tracing a common relationship in these massive clumps between their $B$ fields and their star-forming structure.  We examine this idea in more detail next.

%%%%%%%%%
%   Section 3.3  %
%%%%%%%%%
\subsection{Comparison Among Region 9 Clumps, \\ and with Other Studies}\label{Bcomps}

It is instructive to break down our HRO analysis by smaller areas, to see if these parameters are everywhere the same or if the above results are simply an average of some wider environmental effect(s).  Indeed, having already defined the ROIs for the DCF analysis, we can re-use them for this purpose.  The results for each are shown in {\color{red}Appendix \ref{HROdetails}}, and a summary plot of the trends appears in {\color{red}Figure \ref{HROsummary}}.

Among these HRO $\xi$-$N$ regression fits, we see that they clearly split into two groups plus 1 outlier.  First, there are 10 ROIs with clearly-defined negative slopes $C_{\rm HRO}$ (average slope/error $\sim$ 4) and log$N$ intercepts covering a relatively narrow range ($X_{\rm HRO}$ = 26.3$\pm$0.4) with small individual log$N$ uncertainties (average $\approx$ 0.13).  This group is comprised of BYF\,67, 68c, 68e, 68n, 71w, 73Hs, 76, 77abc, 77d, and 78b; we dub it the ``nominal'' group.

%%%%%%%
%   Fig.11   %   HAWC+ N vs theta
%%%%%%%

The fits for the other 7 ROIs are quite different, in that their slopes $C$ are positive or flat but poorly-defined, and the $X$ values vary widely with large uncertainties.  Six of these have slopes $C_{\rm HRO}$ consistent with 0 (average slope/error $\sim$ 1) and log$N$ intercepts covering two orders of magnitude in $N$ ($X_{\rm HRO}$ = 26.4$\pm$1) with large individual log$N$ uncertainties ($\sim$ 0.5--12).  We call this group of BYF\,69, 70a, 70aS, 71, 73Hn, and 78a the ``flat'' group.

%%%%%%%
%   Fig.12   %   HAWC+ HRO in N bins
%%%%%%%

%%%%%%%
%   Fig.13   %   HAWC+ HRO xi vs N
%%%%%%%
\begin{figure}[h]
\centerline{\includegraphics[angle=0,scale=0.215]{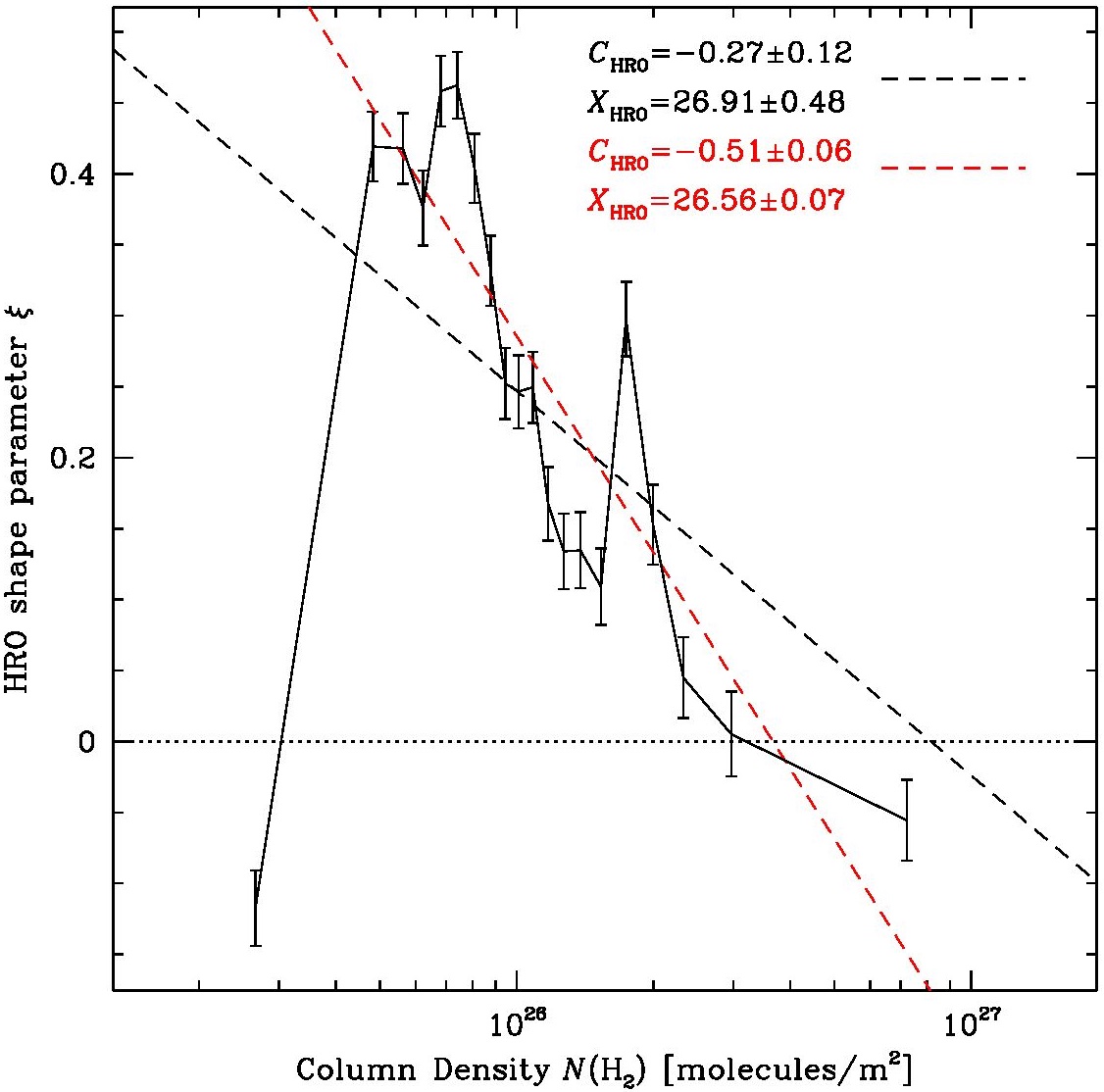}}
\vspace{0mm}\caption{ %\footnotesize % uncomment for aasj
HRO shape parameter $\xi$ as a function of column density $N$ as fitted in Figure \ref{hawcHRObins}.  The black labels and dashed line are solutions to the parameters $C$ (the slope) and $X$ (log of the $N$-axis intercept) of a linear regression to all the $\xi$ data, while the red labels and dashed line are for a fit to all bins except the first (log$N$ $<$ 25.64), which includes the lowest-S/N $\theta_{B_{\perp}}$ data.
}\label{hawcHROxi}\vspace{0mm}
\end{figure}

The last ROI, BYF\,70aN, is an outlier from both the above groups, with a very large positive slope $C_{\rm HRO}$ (both the slope and slope/error are $\sim$8) but unremarkable log$N$ intercept ($X_{\rm HRO}$ = 25.75).

For comparison, the overall Region 9 HRO result is shown in Figure \ref{HROsummary} in black.  From this we can see that the nominal group is fairly representative of the whole Region, in that the log$N$ intercepts are all reasonably close to the Figure \ref{hawcHROxi} value, while the slopes are all clearly negative, although the individual ROIs do have slopes {\em more} negative than the Region-wide average (respectively, --1 to --4 compared to --0.5).  To some extent, one can attribute this trend to a reduction in statistical uncertainty: i.e., as the number of points in a ROI goes up, the slopes become less negative and less uncertain.  Nevertheless, the average slope within the nominal group is --2.0.  Evidently, the contribution of the statistics from the other 7 ROIs seems to push the Region-wide slope to its less negative value, but without erasing its clearly negative (8$\sigma$) signal.

%%%%%%%
%   Fig.14   %   HAWC+ HRO-CX summary plot
%%%%%%%
\begin{figure}[t]
\centerline{\includegraphics[angle=0,scale=0.215]{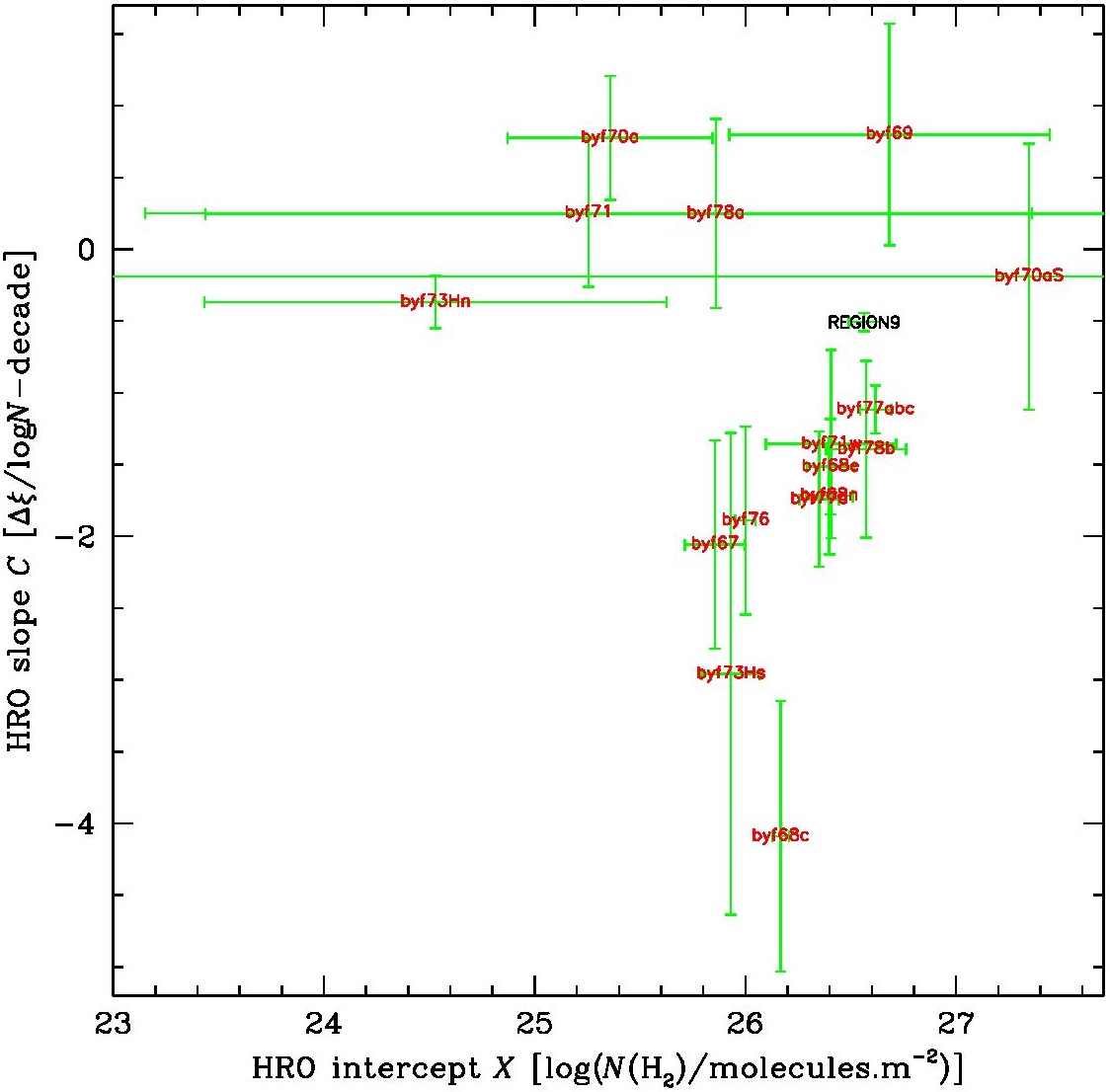}}
\vspace{-1mm}\caption{ %\footnotesize % uncomment for aasj
Results of HRO regression analysis in each ROI shown in Appendix \ref{HROdetails}, except for BYF\,70aN which lies far outside the plot window at ($X$,$C$) = (25.75,7.56).  These are shown with the clump's name in red at the fitted ($X$,$C$) coordinates, each with green error bars from the formal fit uncertainties.  In addition, we show in black the overall Region 9 result from Fig.\,\ref{hawcHROxi}. }\label{HROsummary}\vspace{0mm}
\end{figure}

\begin{figure}[t]
\centerline{\includegraphics[angle=0,scale=0.215]{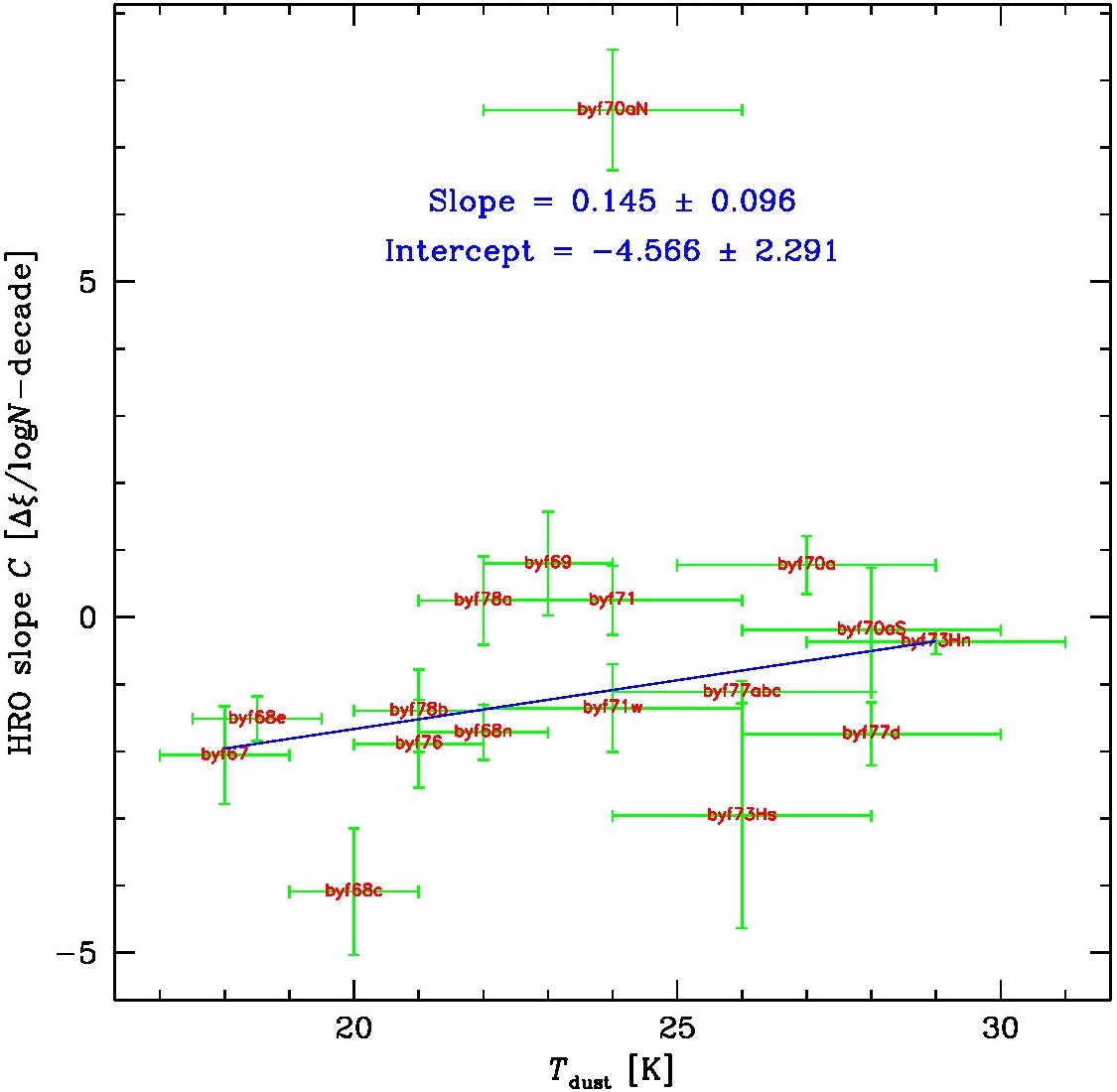}}
\vspace{-1mm}\caption{ %\footnotesize % uncomment for aasj
Comparison of HRO $C_{\rm HRO}$ slope parameter vs.\ mean \td\ across each of the 17 ROIs.  Also shown is a linear fit in this space (blue) to the 16 ROIs excluding BYF\,70aN, with correlation coefficient $r$ = 0.37 and other parameters as labelled.
}\label{HRO-CTd}\vspace{0mm}
\end{figure}

Otherwise, there doesn't appear to be an obvious physical reason for some clumps/ROIs to give a clearly negative slope $C_{\rm HRO}$, and for others to be flat or positive.  For example, we can easily estimate an average \td\ across each ROI from the SED fits of \cite{p19}, and plot this vs.\ the $C_{\rm HRO}$ slope with a linear fit as shown in {\color{red}Figure \ref{HRO-CTd}}.  This hints at a trend, in the sense of steeper slopes $C_{\rm HRO}$ as \td\ falls, i.e., a sharper transition to criticality in colder gas.  This would make physical sense, in that gas that is warmer is expected to be more disturbed from its cooler, more SF-prone state, where gravity is primarily only resisted by $B$ fields.  But the correlation shown here is not very strong, and would need better evidence to be considered reliable.  However, we return to this idea in \S\ref{disc}.

Instead, inspection of the individual ROI ${B_{\perp}}$-vector maps is a little more informative (Figs.\,\ref{r9north}--\ref{r9south}).  Since any anomalies may be most obvious in the outlier or flat ROIs, we consider them first.

{\em BYF\,70aN, outlier.}  In \cite{p19}'s SED-fit \nhtwo\ map, this ROI lies across an area of relatively flat $N$ $\sim$ 0.5$\times$10$^{26}$\,molec/m$^2$, but which nevertheless peaks along a ridge which is well-aligned with most of the $\theta_{B_{\perp}}$ values, +20\degr$\pm$10\degr.  This accounts for the strongly positive $\xi$ values at the highest-$N$ bins in its matching HRO plot (App.\ \ref{HROdetails}), while the alignment tends to perpendicularity going down the ridgeline ($\xi$$<$0 at low-$N$).  Because of the apparently ``crushed'' $B$ field configuration (possibly due to the expanding edge of the NGC\,3324 HII region), the nominal $\xi$ trend is reversed.

{\em BYF\,69, flat.}  The $\theta_{B_{\perp}}$ distribution here is strongly peaked again at --20\degr$\pm$20\degr, although projected to be about 3\,pc away from it.  The polarisation signal is also displaced from the peak $N$ structure, so the angular correlation is poor, tending to perpendicularity everywhere and leading to a weakly positive $\xi$-$N$ trend.

{\em BYF\,70a, flat.}  This ROI is centred on the sharpest $N$ gradient at the edge of the HII region, and most ${B_{\perp}}$ vectors are aligned with its strong ionisation front, so there is an overall dominance of parallel alignments and another weakly positive HRO.

{\em BYF\,70aS, flat.}  Just south of the previous ROI, the data here straddle what appears to be a breakout in the ionisation front, judging by the anticorrelation of the $N$ map with $\theta_{B_{\perp}}$.  That is, where $N$ is high, we see a dominantly parallel/north-south $B$ field alignment, such as would be produced by the HII region's overpressure.  Where there is a gap in $N$ along the front, the field lines turn sharply east-west as if to trace an escape route for the HII.  The $\xi$-$N$ trend is therefore very flat.

%%%%%%%
%   Fig.15   %   HAWC+ HRO-CTd summary plot
%%%%%%%

The above four ROIs seem to be strongly affected by the radiation and dynamical overpressure from the NGC\,3324 HII region.  One surmises that the kind of delicate, pre-stellar interplay between $B$ fields and gravity has been swamped here by post-SF feedback.

{\em BYF\,71, flat.}  Once again, the $\theta_{B_{\perp}}$ distribution is completely uniform at --60\degr$\pm$20\degr, and again parallel to the edge of the HII region despite being projected $\sim$1--4\,pc away from it.  Its rising-then-falling HRO plot, however, seems to be strongly affected by the 2 lowest-$N$ bins showing perpendicular alignments off the NW edge of the $N$ map.  Otherwise, this ROI may actually follow the nominal $\xi$-$N$ pattern, including the high-density peaks inside the clump's NW edge.

{\em BYF\,73Hn, flat.}  Despite being part of BYF\,73's compact HII region, and therefore sampling mostly post-SF gas, the HRO trend is vaguely nominal, although statistically weak (slope/error = 2).

{\em BYF\,78a, flat.}  This ROI has 3 distinct $\theta_{B_{\perp}}$ correlation patches within it, the 2 smaller of which straddle a higher-density feature and align with its contours, at least in part, in the nominal way.  The third patch has $\theta_{B_{\perp}}$ perpendicular to the low-$N$ contours going away from the higher-density feature, so generate a positive $\xi$-$N$ trend there.  The area seems to be somewhat dominated by its surroundings.

In the 10 nominal ROIs, their $B$ fields do not obviously appear to be influenced by NGC\,3324 or its ionisation front.  Their negative $\xi$-$N$ slopes suggest that those clump areas conform to the presumption of the HRO method, i.e., that the balance between $B$ fields and gravity shifts from magnetic at low columns to gravitic at high columns.  One might then argue that the functional reason for a clump exhibiting a nominal $\xi$-$N$ plot vs.\ a flat or reversed $\xi$-$N$ is whether the gas is dominated by slow internal forces (nominal) or more dynamic external forces.  Brief remarks on the 10 ROIs follow.

{\em BYF\,67, nominal.}  This clump has a distinctly elliptical shape in \nhtwo, peaking at 5$\times$10$^{26}$\,molec/m$^2$, with a low-ish average $T_d$ = 18$\pm$1\,K.  However, its polarisation signal is concentrated on its eastern side and has highest $p'$ off the $N$ peak.  Thus, most of its ${B_{\perp}}$ vectors are aligned approximately radially to the $N$ distribution, although at the lowest $N$ contours, the alignment becomes more oblique.

{\em BYF\,68, nominal.}  This clump's appearance in $N$ is rather amorphous with somewhat cool $T_d$ = 20$\pm$2\,K, but its HAWC+ data self-partition into three ROIs.  In the central ROI, BYF\,68c contains a similar mixture of radial and circumferential $p'$ vectors to BYF\,67, with more of the latter at lower $N$ and vice versa.  The eastern BYF\,68e, however, has a strongly aligned $p'$ field across its elongated shape, in such a way as to produce a consistently nominal HRO diagram.  Last, the northern BYF\,68n has a distribution of $p'$ vectors intermediate between c and e.

{\em BYF\,71w, nominal.}  West of the bulk of BYF\,71, this ROI consists of a strongly polarised ridge with its $p'$ vectors mostly aligned along it, similar to the elongated parts of BYF\,71 but without the confusion at the low-$N$ contours.

{\em BYF\,73Hs, nominal.}  Again part of BYF\,73's compact HII region sampling post-SF gas, the $\xi$-$N$ trend is still statistically weak (slope/error = 2) but more negative than in BYF\,73Hn.

{\em BYF\,76, nominal.}  This clump is again somewhat amorphous in $N$ but has a well-ordered $p'$ field.  The net result is a clearly negative $\xi$-$N$ slope but with some variance.

{\em BYF\,77abc, nominal.}  These 3 clumps are nestled close to each other near the southern boundary of NGC\,3324, in a complex area of elevated $T_d$ = 25--30\,K and \nhtwo\ up to $\sim$5$\times$10$^{26}$\,m$^{-2}$.  They are compact ($\sim$beam-sized) at the Mopra resolution (37$''$; indeed, BYF\,77b is itself a double) but BYF\,77b \& c resolve into a $\sim$1.6\,pc-wide shell at the higher HAWC+ resolution (14$''$), with BYF\,77a projected to be 1\,pc outside this shell to the southwest.  They are considered together because the HAWC+ polarisation vectors are remarkably well-aligned ($\theta_{B_{\perp}}$ = 55\degr$\pm$30\degr) across all these features, and in the same direction as the maximum extent of the complex, from a to c.  As such, they give a strongly nominal $\xi$-$N$ plot once the lowest-$N$ bin (with poor S/N) is discounted.

{\em BYF\,77d, nominal.}  This 4th clump in the BYF\,77 complex is far enough away (3\,pc projected to the SW) to be clearly separated in both the Mopra and HAWC+ maps, but it is also distinguished by having a different $\theta_{B_{\perp}}$ distribution (20\degr$\pm$25\degr) than in a--c.  This gives a distinctly nominal HRO plot, but conversely with a slightly rising $\xi$ trend at the higher $N$ bins.

{\em BYF\,78b, nominal.}  The negative $\xi$-$N$ trend in this cooler ($T_d$ = 21$\pm$1\,K) clump is clear, but a little confused by a mixture of alignments at the higher $N$ levels.

Besides these intercomparisons, we also briefly consider some similar HRO studies in other SF regions, as discussed by \cite{b23}.  These employed {\em Planck}, {\em Herschel}, BLASTpol, and HAWC+ data to investigate various Gould Belt, Vela-C, and L1688 clouds \citep{pc16,saa17,zss20,lbc21}.  The other instruments had lower angular resolution than HAWC+, but since these clouds are all closer to us than Region 9/$\eta$ Car, most of these studies obtained a similar sub-parsec physical resolution as herein.  However, the clouds studied have a typical column density from much- to somewhat-lower than the average in Region 9.

The typical threshold column densities $X$ for this HRO work were about 0.6 (Gould), 3 (Vela-C), and 0.6 (L1688) $\times$10$^{26}$\,m$^{-2}$.  Our all-Region 9 value from Figure \ref{HROsummary} is $X$ = 3.6$\times$10$^{26}$\,m$^{-2}$.  Thus, Region 9 seems to contain clumps that not only have higher column and volume densities in general than a number of local ($d$$<$1\,kpc) clouds and complexes, but also have criticality (i.e., the threshold for balancing gravity with magnetic pressure support) displaced to higher densities as well.

%%%%%%%
%   Fig.16   %   all-pixel B-n plot
%%%%%%%
\begin{figure*}[t]
\centerline{\includegraphics[angle=0,scale=0.23]{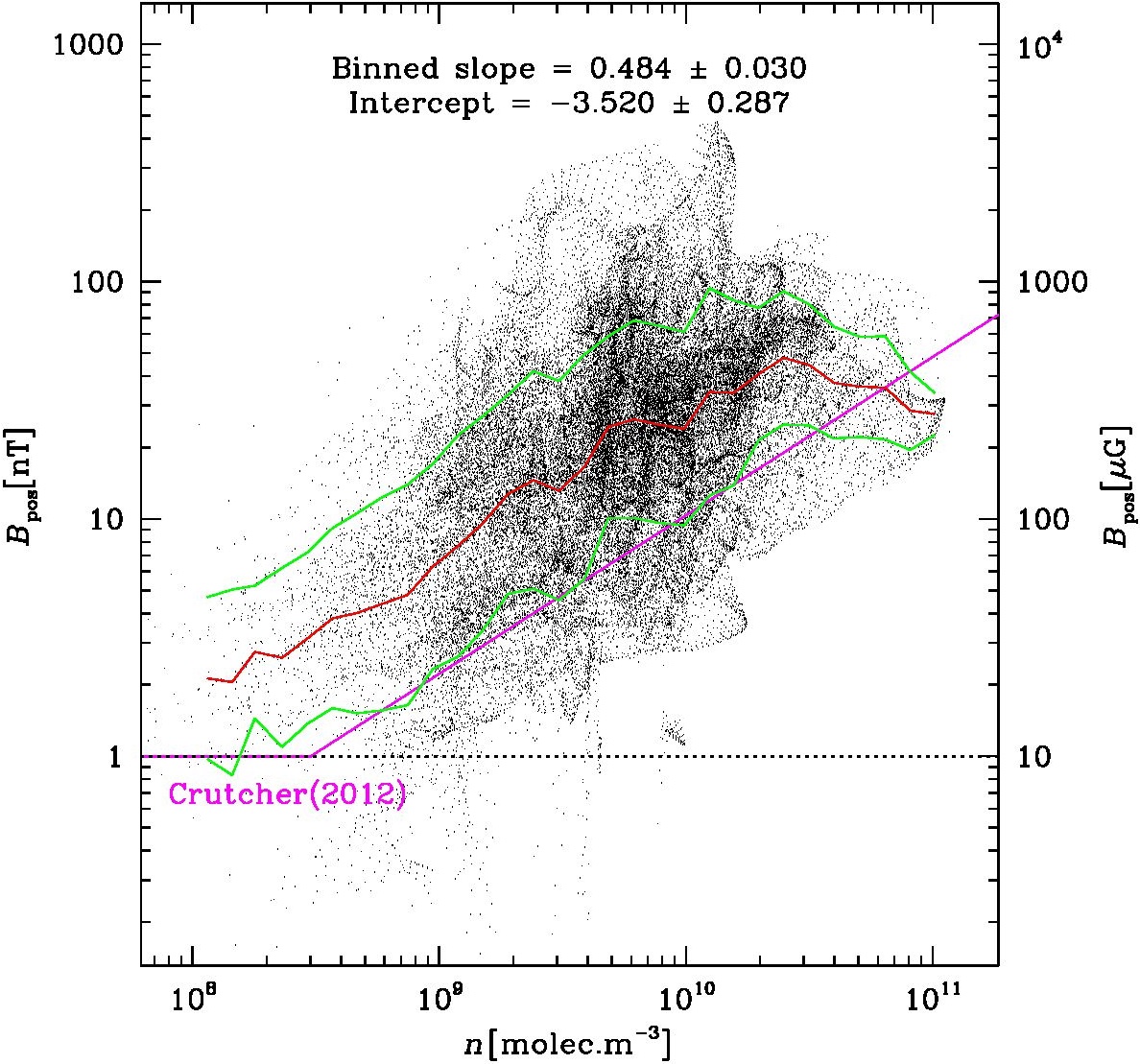}\hspace{-11mm}\includegraphics[angle=0,scale=0.23]{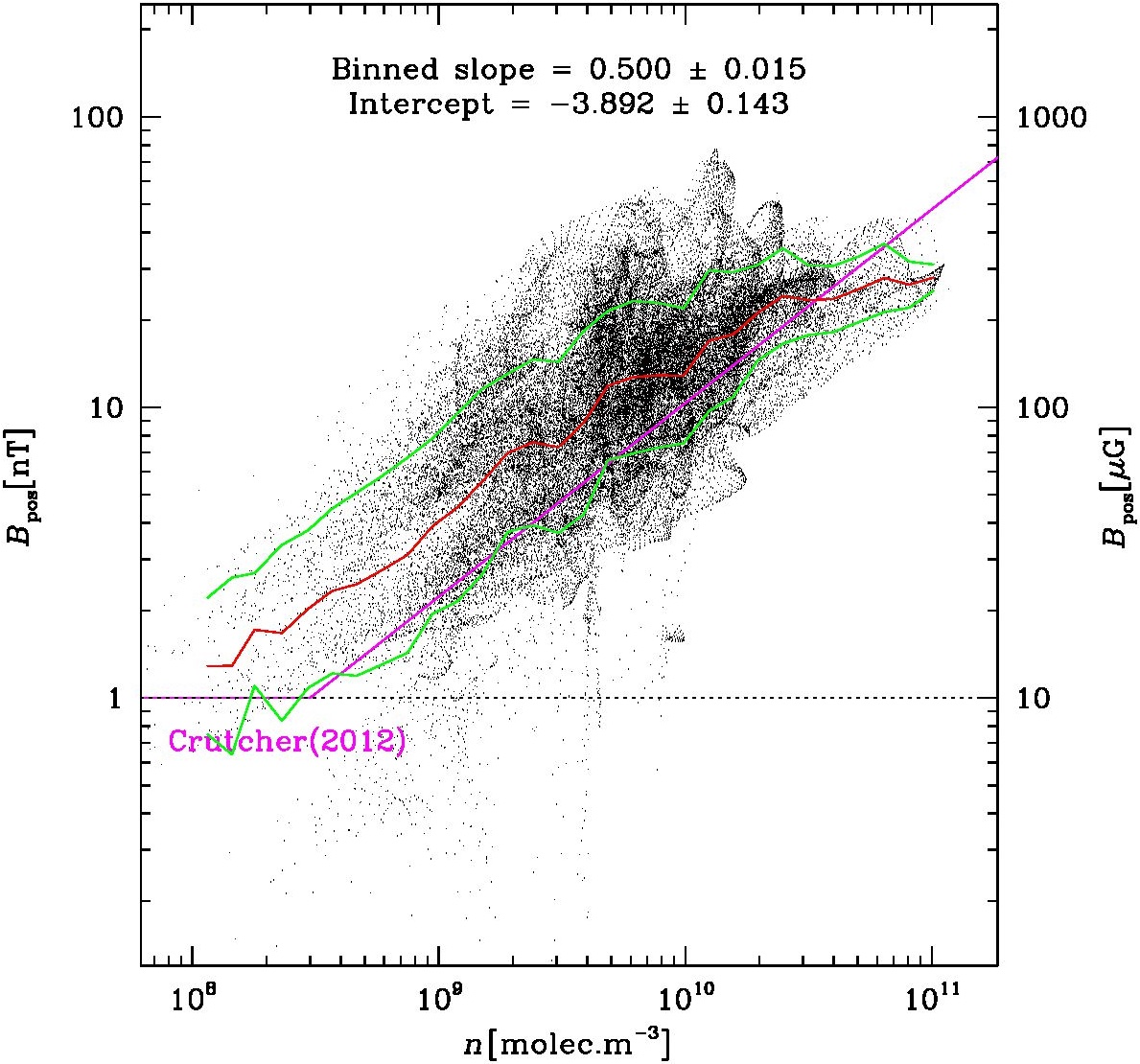}\hspace{-181mm}(a)\hspace{81mm}(b)\hspace{91mm}}
\vspace{0mm}\caption{ %\footnotesize % uncomment for aasj
({\em a}) Comparison of pixel values for $B$ and $n$ maps across all of Region 9 using the classical DCF calculation \citep{ppvii}, overlaid by the \cite{cru12} relationship in magenta.  The log$B$ values were also averaged in each of 30 bins in log\,$n$, and the bins' mean ($\pm$1$\sigma$) log$B$ values are overlaid in red (green).  The label at the top describes a linear fit to the binned means (i.e., a power-law fit to the linear data), with correlation coefficient $r$ = 0.95 and $\chi^{2}$ = 0.59.  A simple linear fit to all the log$B$-log\,$n$ data at once gives a similar result, with slope $\kappa$ = 0.524$\pm$0.004, intercept --3.80$\pm$0.04, and $r$ = 0.50. %and $\chi^{2}$ = 9196.
Below a density threshold of 2.5$\times$10$^{10}$m$^{-3}$, however, the binned means clearly trend close to the \cite{cru12} slope of $\tfrac{2}{3}$, but displaced above it by a factor of $\sim$2.5, although the S/N in the $B$ values tends to be low for $B$$<$10\,nT.  
({\em b}) Similar to (a) but using the \cite{st21} calculation for $B$.  The distribution of points and fits are very similar, except for a smaller scatter in the ordinate due to the weaker dependence on $s$.  Thus, the trend in low-$n$ data is only a factor $\sim$1.5 above the \cite{cru12} line, and the high-$n$ turnover is noticeably softer.  A cgs scale is also shown on the far right. %for convenience.
}\label{bnd}\vspace{2mm}
\end{figure*}

We can also compare the respective equivalent $B_{\rm crit}$ field strengths derived in the nearby clouds with the values obtained here.  From Eq.\,(4) we can write
\begin{equation} % EQUATION SIX		% 1.56742868814
	B_{\rm crit}~({\rm nT}) = N_{\rm H_2}/(1.57 \times 10^{25}{\rm m}^{-2})~~
\end{equation}
where $\lambda$ = 1, i.e., where \nhtwo\ = $X_{\rm HRO}$.  Then, $B_{\rm crit}$ $\sim$ 4\,nT for the Gould Belt \citep{pc16}, $\sim$20\,nT for Vela-C \citep{saa17}, and $\sim$4\,nT for L1688 \citep{lbc21}.  Similarly, for Region 9 as a whole, $B_{\rm crit}$ = 23$\pm4$\,nT, while for the 10 ROIs with nominal HRO trends, $B_{\rm crit}$ ranges over 5--27\,nT.  This places Region 9 at a higher-$N_{\rm crit}$ and -$B_{\rm crit}$ level compared to these other clouds, when mapped at sub-pc scales.  The equivalent $B_{\rm crit}$ result for the BYF\,73 ALMA data is $B_{\rm crit}$ = 42$\pm$7\,nT \citep{b23}, which is commensurate with the higher density levels on the 0.1\,pc scale.  

%%%%%%%%%
%                      %
%   Section 4    %
%                      %
%%%%%%%%%
\section{Discussion}\label{disc}

The results of the DCF analysis can be summarised as: the mapped $B_\perp$ and log$\lambda$ values show a generally slightly subcritical group of molecular clumps, at somewhat higher than typical column densities $N$.  The results of the HRO analysis can be summarised as showing a clear trend towards criticality as $N$ rises, but without clearly crossing into criticality over most of the mapped area.  Encouragingly, we view these results as mutually compatible, and speak to the higher than typical $B_\perp$ environment in these high-density clumps with only moderately vigorous SF.  Exceptions to these statements do occur, but only in a few locations (most obviously, BYF\,73).

The overall impression from the comparisons in \S\ref{Bcomps} is that ``environment matters.''  In other words, since gravity is a weak force, and since the many $B$ field studies summarised by $B$-$n$ diagrams \citep[e.g.,][]{ppvii} show that magnetic forces only seem to be of $\sim$similar strength to gravity in star-forming (i.e., dense) gas, any other effects that might buffet a cold, dense molecular cloud need to be at a fairly minimal level in order for the transition to criticality to occur.  Specifically, strong feedback from a nearby classical HII region can potentially disrupt this process, even several pc beyond its own ionisation front.  This is made visually intuitive in the LIC image of {\color{red}Appendix \ref{LICimg}}.

More generally, the other HRO studies cited above show that the actual transition to criticality can happen at different column densities which seem to be location-dependent.  Region 9 seems to be at the higher end of column densities $N$ generally, and transition columns $X$ in particular (e.g, 1 vs 3$\times$10$^{26}$\,m$^{-2}$), than in most of the local clouds in the prior studies.  At the same time, when we zoom in to core scales (i.e., $<$0.1\,pc) like the MIR\,2+Streamer results from \cite{b23}, we find both $N$ generally and $X$ specifically are higher still (7$\times$10$^{26}$\,m$^{-2}$).

This general trend holds also for the derived $B_{\perp}$ values seen in Figs.\,\ref{northB}a--\ref{southB}a.  Overall, they range from $\sim$10--200\,nT, with a few excursions higher, up to 470\,nT (or in old-fashioned cgs units, $\sim$0.1--2\,mG, up to almost 5\,mG).  As recently as a decade ago, mG fields had been rarely observed in star-forming clouds, according to the data of \cite{cru12}.  More recent work has started to fill in this high-$B$ space (up to $\mu$T/10\,mG fields) in massive star forming clouds, e.g., \cite{cgs24}; the compilations of \cite{ppvii} and \cite{wh24} provide another $\sim$20 measurements with $B$ $>$ 100\,nT.

Yet here, in just a fraction of one GMC, we have 455 independent pixels with $B$ $>$ 100\,nT, or 5\% of our high-S/N data, mostly in and around some of the clump peaks (see Figs.\,\ref{northB}a--\ref{southB}a).\footnote{This amplification of analytical capability, 9000 independent data points with polarisation S/N $>$ 3, compared to the more modest number of clumps targeted (17), was enabled by the gradient technique, Eq.\,(3), to estimate $n$ at each pixel.}  Perhaps the $\eta$ Car clouds are unusual in their $B$ field strengths, limiting SF to areas where the column density (and gravity) build up to a level sufficient to overwhelm such strong fields only where log$\lambda$ is high, yielding massive SF in only select locations.

To illustrate this, we place our data in the $B$-$n$ diagram ({\color{red}Fig.\,\ref{bnd}a}).  There, the line of criticality coincides approximately with the \cite{cru12} relationship, with supercritical areas (gravity dominant) below that line and subcritical areas ($B$ fields dominant) above it.  Most of Region 9 is subcritical, trending towards criticality at only the highest densities ($\sim$10$^{11}$m$^{-3}$), where the $B$ fields are more modest ($\sim$30\,nT).  By contrast, the strongest $B$ fields (150--500\,nT) occur at modest densities (10$^{9-10}$m$^{-3}$).

Using this diagram to obtain a slope $\kappa$ in the $Bn$ relationship for Region 9, we can support values of either $\sim$$\tfrac{1}{2}$ or $\sim$$\tfrac{2}{3}$ from our DCF-derived data, depending on the analysis method.  Apparently due to sensitivity limitations, we also cannot find evidence for a density threshold $n_{0}$ where $\kappa$ transitions to a lower value.  But the strong turnover of the piecewise $\kappa$ from $\tfrac{2}{3}$ over most of the observed range of $n$, to a value $<$0 where $n$ $>$ 2.5$\times$10$^{10}$m$^{-3}$, suggests a possible breakdown at high density of the main assumption of DCF, that it arises from incompressible Alfv\'en waves, as discussed by \cite{ppvii}.

Variants of this analysis give very similar results.  For example, retaining a correlation length of 0.5\,pc from the DCF procedure instead of the assumed 0.33\,pc (Appx.\,\ref{DCFdetails}) raises slightly the overall dispersion $s$ in the distribution of polarisation angles, correspondingly lowering the resulting $B_{\perp}$ values to some extent.  A single, whole-sample power-law fit then gives $\kappa$ = 0.499$\pm$0.003, intercept = --3.72$\pm$0.03, and $r$ = 0.55, virtually identical to the result in Figure \ref{bnd}a.  Fitting the binned data, we obtain $\kappa$ = 0.462$\pm$0.025, intercept = --3.44$\pm$0.24, $r$ = 0.96, and $\chi^{2}$ = 0.40, again very close.  However, for log\,$n$ $<$ 10.5, the trend does soften somewhat, from $\kappa$ $\approx$ $\tfrac{2}{3}$ to 0.566$\pm$0.013, intercept = --4.37$\pm$0.12, $r$ = 0.99, and $\chi^{2}$ = 0.99, mainly because the pixels most affected by the different choice of correlation scale tend to be the ones where $s$ was small and $B$ large to begin with.  In this case, the maximum $B_{\perp}$ = 158\,nT and we only wind up with 79 independent pixels above 100\,nT, or about 1\% of our data.  But otherwise, the overall data distribution looks very similar to Figure \ref{bnd}a, including the displacement of the mean trend by a factor of $\sim$2 above the \cite{cru12} line, and the turnover to $\kappa$$<$0 for log\,$n$ $>$ 10.5.

However, when we recast the $Bn$ diagram with Eq.\,(2) from \cite{st21}, we find a somewhat more noticeable effect on the data distribution ({\color{red}Fig.\,\ref{bnd}b}).  Since the scatter in the ordinate of the $Bn$ diagram is mostly due to the scatter in $s$, Eq.\,(2)'s weaker dependence on $s$ produces a more compressed distribution, including having no data points with $B$ $>$ 100\,nT.  Interestingly, the turnover to smaller $\kappa$ at high $n$ is much softer, suggesting that \cite{st21}'s formalism may not have the same high-$n$ issues as has been suggested for classical DCF.  However, the plots in Figure \ref{bnd} do not indicate a clear preference for classical DCF vs.\ the \cite{st21} analysis.

\begin{figure}[t]
\centerline{\includegraphics[angle=0,scale=0.22]{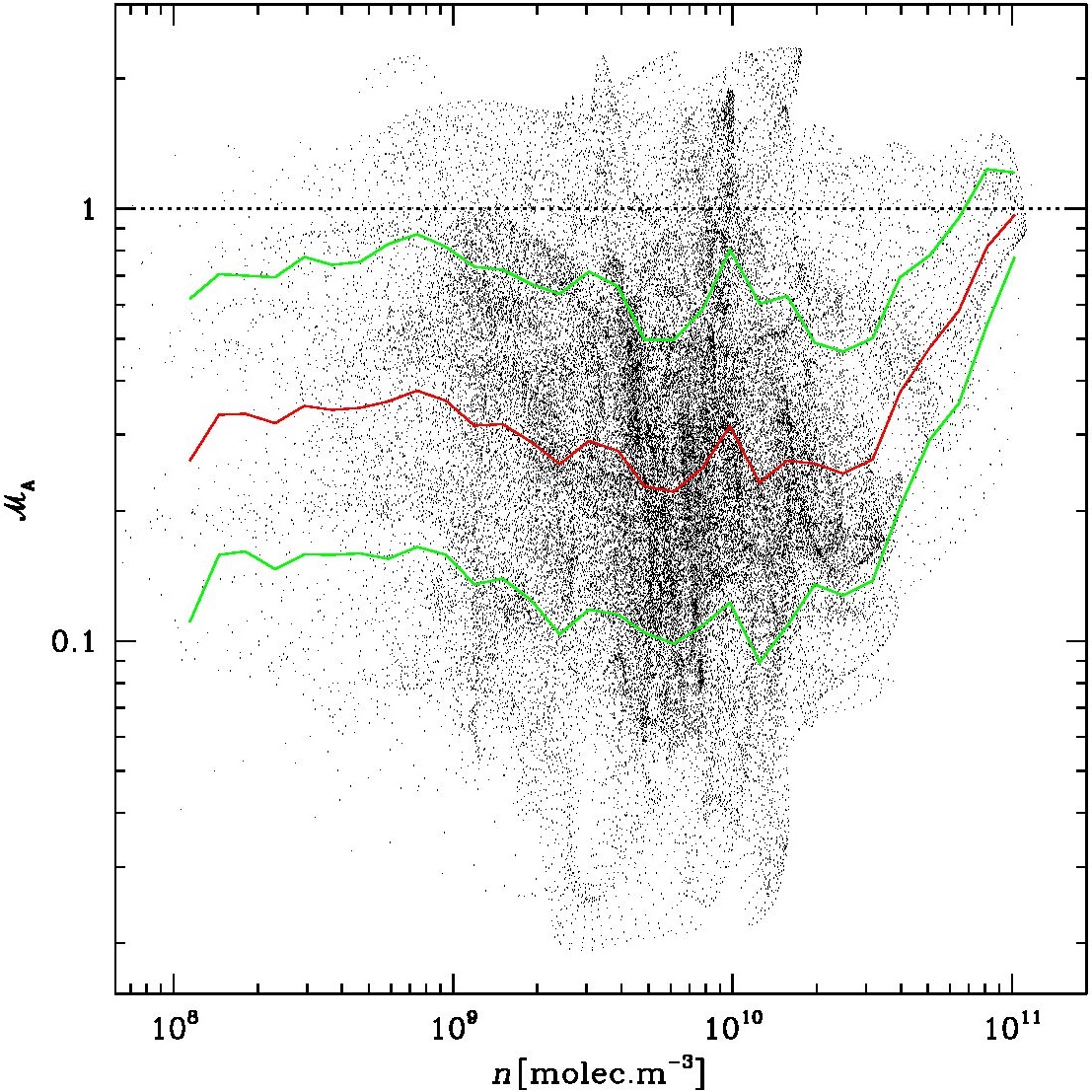}}
\vspace{0mm}\caption{ %\footnotesize % uncomment for aasj
Alfv\'en number as a function of density in Region 9.  The bin means$\pm$1$\sigma$ are shown merely to indicate trends.
}\label{MAcl}\vspace{0mm}
\end{figure}

\cite{ppvii} have also pointed out that the trends in the $Bn$ diagram are somewhat built-in, since with DCF measurements, $B$ and $n$ are not independent (see Eq.\,(1)).  Absent other factors, we would expect $\kappa$ = $\tfrac{1}{2}$, and so any structure in this diagram is really measuring the Alfv\'en number, $M_{A}$.  In the classical DCF analysis, this reduces to
\begin{equation} % EQUATION SEVEN
	M_{A} = s/Q~~,
\end{equation}
where $s$ is now in radians and $Q\approx0.5$ usually: we show this for the Region 9 data in {\color{red}Figure \ref{MAcl}}.  We recover much the same pattern as do \cite{ppvii}, where most clouds are sub- or trans-Alfv\'enic.  There, $M_{A}$ is typically $\sim$0.2--2 and flat with $n$, but at $n$ \gapp\ 10$^{12}$\,m$^{-3}$, $M_{A}$ rises significantly to 10 or more, suggesting a change to criticality.  In Region 9, this turn-up seems to begin at 10$^{10.5}$\,m$^{-3}$, but only rises to $M_{A}$ $\approx$ 1.

%%%%%%%
%   Fig.17   %   all-pixel MA-n plot
%%%%%%%

We can also examine criticality more directly through log$\lambda$.  From \S\ref{DCFstuff}, the overall log$\lambda$ distribution across Region 9 is rather close to gaussian at a mean of --0.75 (this is when we use the classical DCF $B_{\perp}$ values).  That is, the average $\lambda$ is 0.35, meaning somewhat strong support against gravity in most areas.  Only a rather small fraction of pixels (2\%) have this reversed, $\lambda$$>$3.

\begin{figure*}[ht]
\centerline{\includegraphics[angle=0,scale=0.226]{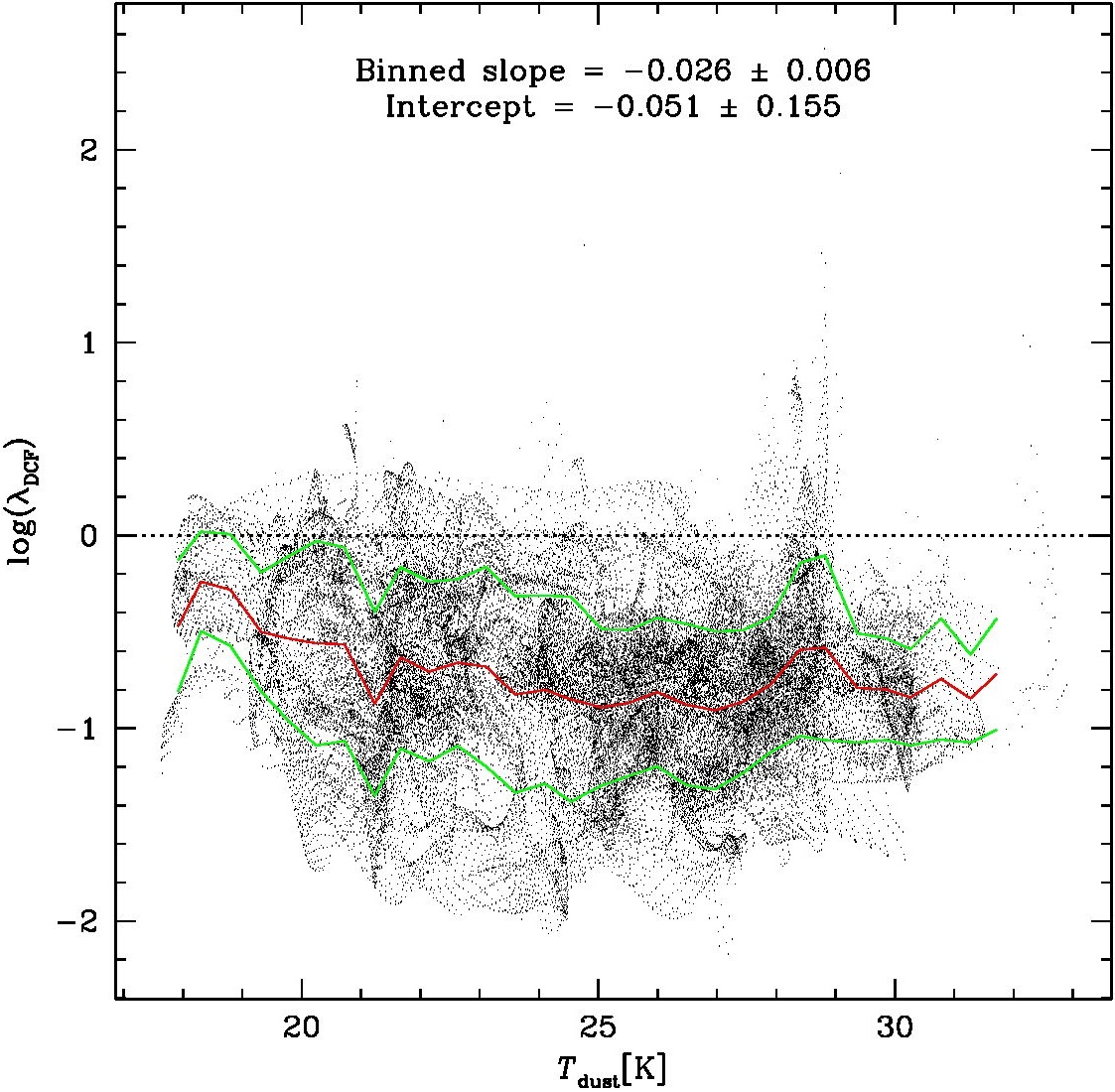}~\includegraphics[angle=0,scale=0.226]{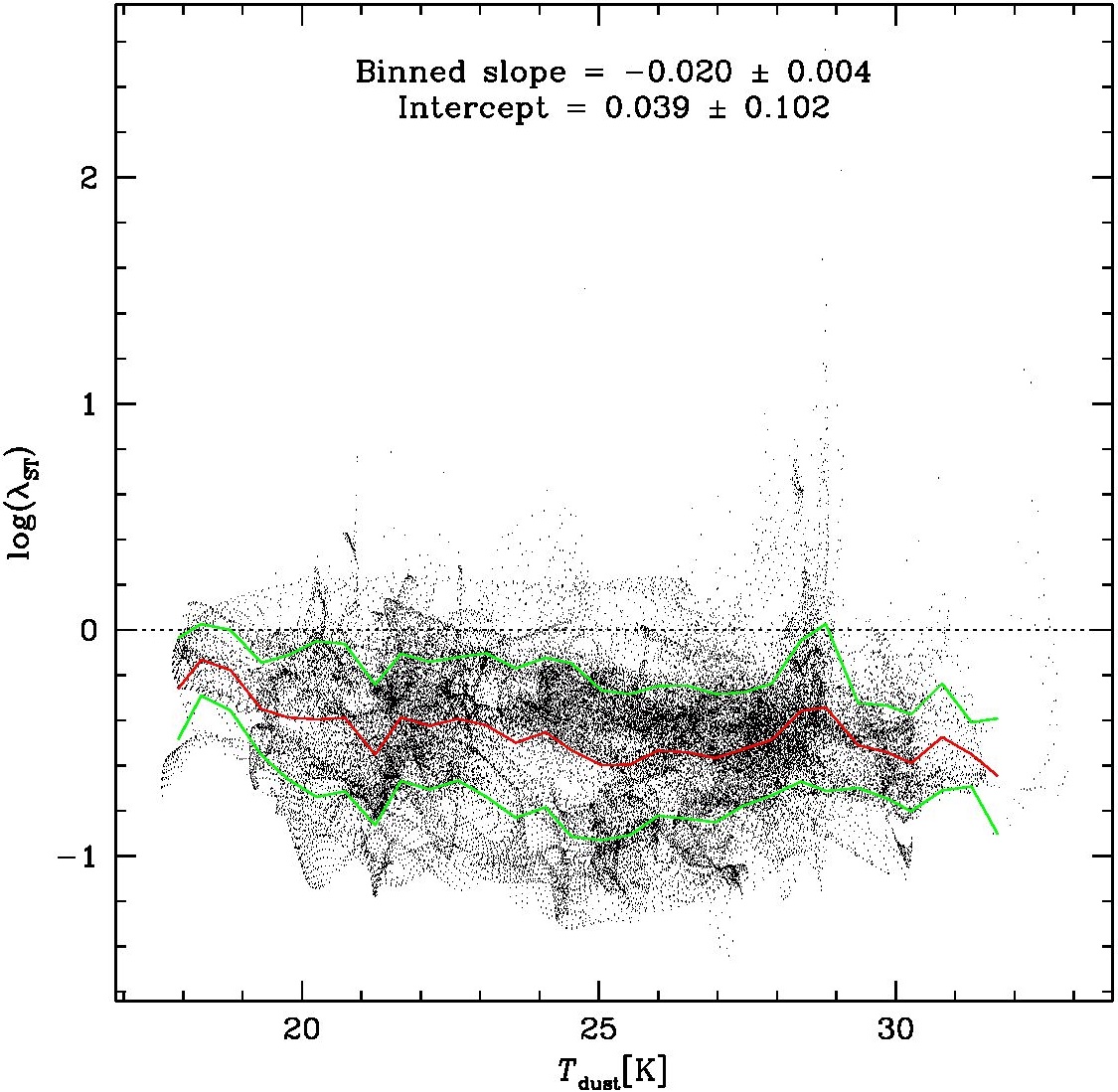}\hspace{-181mm}(a)\hspace{81mm}(b)\hspace{91mm}}
\vspace{-1mm}\caption{ %\footnotesize % uncomment for aasj
({\em a}) Comparison of pixel values for log$\lambda$ and \td\ maps across all of Region 9 using the DCF formula for $B_{\perp}$, Eq.\,(1).  The $\lambda$ values were also averaged in each of 30 bins in \td, and the bins' mean ($\pm$1$\sigma$) log$\lambda$ values are overlaid in red (green).  The label at the top describes a linear fit to the binned means, with correlation coefficient $r$ = --0.63 and $\chi^{2}$ = 0.55.
({\em b}) Same as (a) but using the \cite{st21} formula for $B_{\perp}$, Eq.\,(2).  Here the fit to the binned means has $r$ = --0.68 and $\chi^{2}$ = 0.24.
}\label{llTd}\vspace{0mm}
\end{figure*}

But if environment matters, then perhaps the gas or dust temperature is as important as the column density, as far as the transition to criticality is concerned.  We therefore reconsider the \td\ evidence first raised in \S\ref{Bcomps}, but in a more structure-independent way, i.e., pixel by pixel as shown in {\color{red}Figure \ref{llTd}a}.  The much larger $|r|$ than in Figure \ref{HRO-CTd} suggests a significant trend despite the large scatter in log$\lambda$ within each bin.  An (unbinned) simple linear fit gives similar slope and intercept but $r$ = --0.14, suggesting a weaker overall correlation.  However, what matters to SF is where the gas has positive log$\lambda$.

%%%%%%%
%   Fig.18   %   all-pixel log(lambda)-Td plot
%%%%%%%

For example, if we count the fraction of pixels within each \td\ bin with log$\lambda$$>$0 (i.e., $\lambda$$>$1) as representative of gas more likely to be gravitationally dominated, then accumulate these statistics into \td\ quintiles in order to reduce the bin-to-bin fluctuations, the fractions are 12.4\% for \td=17.6\---20.5\,K (1st quintile), 7.0\% for 20.5---23.4\,K (2nd quintile), 2.3\% for 23.4---26.7\,K (3rd quintile), 3.7\% for 26.7---29.6\,K (4th quintile), and 0.08\% for 29.6---32\,K (5th quintile).  This seems to demonstrate a real trend towards significantly more high-$\lambda$ gas as \td\ falls.

This impression is more formally supported with a $t$-statistic, of the form
\begin{equation} % EQUATION EIGHT
	t = \frac{\bar{x}-\mu}{(\sigma/\sqrt{n})}~,
\end{equation}
where $\bar{x}$ is the mean of a (small) test sample, $\mu$ is the mean of a parent population, $\sigma$ is the dispersion in the population, and $n$ is the test sample size.  We take the 23 highest-\td\ bins (i.e., 21\,K $<$ \td\ $<$ 32\,K) to define the hypothetical ``normal'' sample of molecular clouds, based on the appearance of the mean trend (red) in Figure \ref{llTd}a.  This population has a mean log$\lambda$ = --0.78$\pm$0.44 (1$\sigma$).  We then compute the $t$ statistic for the 7 lowest-\td\ bins (i.e., 17.7\,K $<$ \td\ $<$ 20.9\,K), both individually and collectively, and these are listed in column 4 of {\color{red}Table \ref{lamDCFstats}}.  In column 5 of each case is listed the probability of the null hypothesis (that the bin is drawn from the same population as the ``main sample'') being satisfied.

%%%%%%%%%%
%       Table 1       %
%%%%%%%%%%
\begin{deluxetable}{cccccccc}
\tabletypesize{} %\footnotesize % uncomment for aasj
\tablecaption{ %\footnotesize % uncomment for aasj
Region 9 log$\lambda$ $t$-statistics and probabilities for low-\td\ bins using DCF formulae, Eqs.\,(1) and (2)
\vspace{-1mm}\label{lamDCFstats}}
\tablewidth{0pt}
\tablehead{
\colhead{Bin} & \colhead{Sample} & \colhead{Mean} & \multicolumn{2}{c}{DCF} & & \multicolumn{2}{c}{ST} \vspace{0mm} \\
\cline{4-5} \cline{7-8}
\colhead{} & \colhead{Size\tablenotemark{a}} & \colhead{\td\ [K]} & \colhead{$t$} & \colhead{Prob.} & & \colhead{$t$} & \colhead{Prob.} \vspace{-3mm} \\
}
\startdata
1 &   368 & 17.9 & 5.49 & $<$10$^{-4}$ & & 5.83 & $<$10$^{-4}$ \\
2 &   455 & 18.3 & 10.54 & $<$10$^{-4}$ & & 10.02 & $<$10$^{-4}$ \\
3 &   325 & 18.8 & 8.21 & $<$10$^{-4}$ & & 7.40 & $<$10$^{-4}$ \\
4 &   929 & 19.3 & 7.86 & $<$10$^{-4}$ & & 5.69 & $<$10$^{-4}$ \\
5 & 1255 & 19.8 & 8.07 & $<$10$^{-4}$ & & 4.90 & $<$10$^{-4}$ \\
6 & 1695 & 20.3 & 8.42 & $<$10$^{-4}$ & & 5.22 & $<$10$^{-4}$ \\
7 & 1795 & 20.7 & 8.43 & $<$10$^{-4}$ & & 5.75 & $<$10$^{-4}$ \\
 & & & \vspace{-2mm} \\
1--7  &  6822 & 19.8 & 20.7 & $<$10$^{-4}$ & & 15.30 & $<$10$^{-4}$ \\
8--30 & 45518 & 25.5 & 0 & --- & & 0 & --- \vspace{-2mm} \\
\enddata
\tablenotetext{a}{The sample size refers to the number of pixels in the pipeline-rendered SOFIA data.  These images oversample the HAWC+ beam by a factor of 2.5 compared to Nyquist.  Therefore, we have divided the sample sizes by 2.5$^{2}$ for the number of degrees of freedom (i.e., independent measurements) in the probability calculation.}
\vspace{-1mm}
\end{deluxetable}

We compare again how the \cite{st21} analysis changes these results ({\color{red}Fig.\,\ref{llTd}b}).  Superficially, the log$\lambda$ distribution (still very close to gaussian) is systematically shifted to higher values compared to classical DCF, the mean increasing by 0.3 in the log (2$\times$ higher in $\lambda$).  But the distribution around this mean is also compressed, with the dispersion dropping from 0.44 to 0.31 in the log (25\% smaller in $\lambda$).  The end result is that the overall statistical trends are very similar to the classical DCF approach.  The fraction of points with log$\lambda$$>$0 in the same \td\ quintiles are now 8.8\%, 6.7\%, 3.2\%, 5.0\%, and 0.8\% respectively, and the $t$-statistics are listed in columns 6 \& 7 of Table \ref{lamDCFstats}.  While the quintile trend is now more muted, the $t$-statistics still strongly reject the null hypothesis.

Thus, with either approach, each low-\td\ bin shows a highly significant difference between its log$\lambda$ distribution and that of the warmer bins.  In other words, colder clouds are statistically more likely to be dominated by gravity, and not supported by magnetic fields.  Such a result comports well with the relationship between \td\ and \nhtwo\ in massive clumps \citep{p19,p21}.  They found a strong trend of falling \td\ as \nhtwo\ rises towards the centres of around 85--90\% of all CHaMP clumps on the sub-pc scale, as determined by SED fitting to far-IR/submm data.  This trend fails to appear in only about 10--15\% of clumps, accompanied by either strong internal heating from massive SF at clump centres or strong external heating from nearby massive SF, changing such clumps' original pre- or protostellar internal state.  Taken together, the SED fits and results here suggest that we can add a trend of rising $\lambda$ to the internal prestellar state of massive clumps, as a systematic feature of massive SF.

%%%%%%%%%
%                      %
%   Section 5    %
%                      %
%%%%%%%%%
\section{Conclusions}\label{concl}

Based on new SOFIA/HAWC+ far-IR continuum polarisation data, we have presented a systematic study of the 0.16--30\,pc scale magnetic field in 17 of 21 massive molecular clumps comprising the western end of the $\eta$ Carinae GMC (Region 9 of CHaMP, $d$ = 2.5\,kpc).  The polarisation data were analysed in order to learn about the structure, strength, role, and significance of the $B$ field in this cloud.  Our results include the following.

\textbullet\ The HAWC+ polarisation data are widely distributed but do not cover all the known far-IR emission features of these clumps at the sensitivity levels we had originally planned to reach.

\textbullet\ DCF analysis of the polarisation directions (i.e., $B_{\perp}$ magnetic field orientations) shows correlation lengths (areas within which the field is well-aligned) of $\sim$0.5--1.5\,pc in the various clumps.  We adopted a scale 0.33\,pc for subsequent calculations, in order to avoid numerical instabilities in mapping the orientation dispersion.  We also mapped the representative gas density across the clumps, estimated as the gradient of the column density.  Combining these with a published velocity dispersion map, we mapped $B_{\perp}$ and the mass-to-flux ratio $\lambda$ across the HAWC+ data, using both classical and modern DCF formulae.

\textbullet\ HRO analysis of all the HAWC+ data show a strong transition to criticality at log($N_{\rm crit}$/(molec\,m$^{-2}$)) = 26.56$\pm$0.07, reflecting the nominal HRO pattern of $B$ field alignments with gas structures.  For individual clumps, only 10/17 display the same nominal HRO pattern, with ($\mu$$\pm$$\sigma$) log$N_{\rm crit}$ = 26.27$\pm$0.25.  The other 7 clumps show atypical HRO patterns with $B$ fields affected by external forces, most notably the radiation or overpressure from the nearby HII region NGC\,3324.

\textbullet\ Region 9 has both stronger $B$ fields and higher column density $N$ relative to similar studies of more local clouds.  The distribution of data in the $Bn$ diagram is mostly parallel to but slightly above the \cite{cru12} relation, then dropping below it at $n$\gapp10$^{10.5}$\,m$^{-3}$.  The log$\lambda$ distribution is close to gaussian, with $\mu\pm$$\sigma$ = --0.75$\pm$0.45 (classical DCF) or --0.49$\pm$0.31 \citep{st21}, i.e., most areas are subcritical ($B$ dominant).  With either approach, there is a small but significant excess of areas above this distribution with log$\lambda$ $>$ 0 (supercritical), but the Region as a whole appears to have sufficient magnetic support to explain the low star formation efficiency.

\textbullet\ In addition to the existence of critical transition densities from magnetic to gravitational domination, lower dust temperatures also seem to trace a significantly higher likelihood of criticality in such gas.  This is consistent with previous studies of \td\ dropping towards high-$N$ peaks at the centres of massive clumps \citep{p19,p21}.

Despite molecular clouds' wide observational diversity, the physical keys to star formation in massive clumps may boil down to a balance between external heating, internal radiative cooling, and gravity's slow but inexorable force overcoming the support of magnetic fields, as clouds' central densities rise and gas temperatures drop.

%% No more than seven \figcaption commands are allowed per page,
%% so if you have more than seven captions, insert a \clearpage
%% after every seventh one.

\acknowledgments
We lament the grounding of SOFIA, and thank the aircraft crew and instruments' scientific staff for their outstanding support of this pioneering facility.  We also thank Prof.\ Enrique V\'azquez-Semadeni for suggesting the \cite{st21} analysis, and the anonymous referee for helpful comments which improved the presentation of several points in the paper.  PJB gratefully acknowledges financial support for this work provided by NASA through awards SOF 07-0089 and 09-0048 issued by USRA.  This paper is based in part on observations made with the NASA/DLR Stratospheric Observatory for Infrared Astronomy (SOFIA).  SOFIA was jointly operated by the Universities Space Research Association, Inc. (USRA), under NASA contract NNA17BF53C, and the Deutsches SOFIA Institut (DSI) under DLR contract 50 OK 2002 to the University of Stuttgart.

Facilities: \facility{SOFIA(HAWC+)}.

Software: {Jupyter}, {karma}, {Miriad}, {SuperMongo}.

\clearpage

\appendix
%%%%%%%%
%   Appx.  A   %
%%%%%%%%
\section{Details of DCF Angular Dispersion Analysis}\label{DCFdetails}
Here we describe in detail the steps in the DCF analysis procedure for each of the clumps in the South, North, and West portions of Region 9 %similarly to the results for the sample clumps in 
as desired in \S\ref{DCFstuff}.  We closely follow the procedure of \cite{b23}, which requires defining an appropriate correlation length over which $s$ is measured, as explained by \cite{mg91}.

%%%%%%%%
%     Fig.A1    %   South DCF regions
%%%%%%%%
\begin{figure*}[b]
\centerline{\includegraphics[angle=0,scale=0.66]{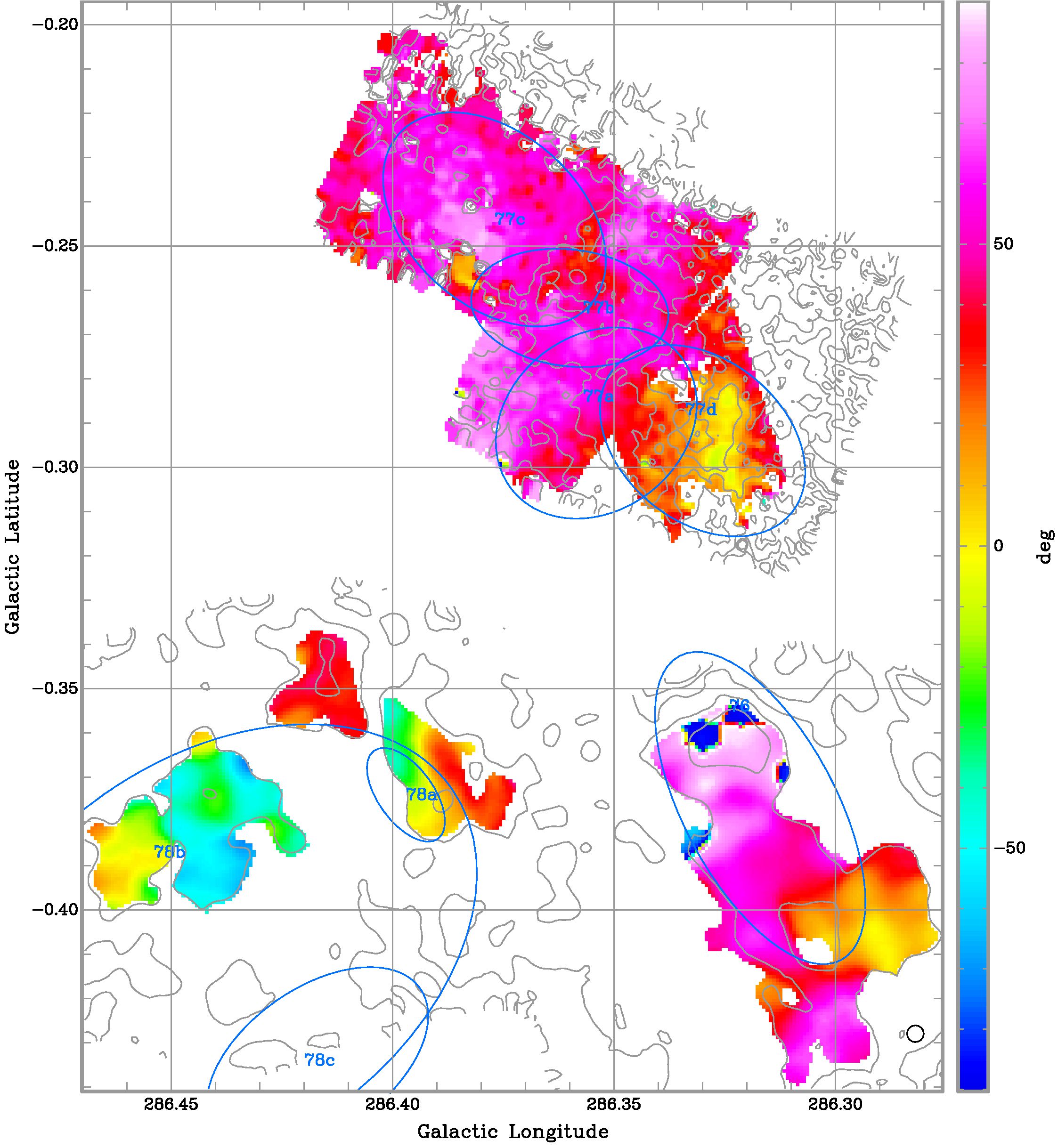}}
\vspace{-1.5mm}\caption{ %\footnotesize  % uncomment for aasj
Cutouts of the $\theta_{B_{\perp}}$ distribution in the Region 9 South ROIs (BYF\,76, 77a--d, 78a--b) with S/N($P'$) \gapp 5, overlaid by grey $P'$ contours at 0.2(0.4)1.0 Jy/bm and blue Mopra \tco\ ellipses.  The DCF analysis for these ROIs appears in Fig.\,\ref{DCFstats}.  For clump BYF\,78c, the data were of insufficient S/N to be included in the DCF analysis.
}\label{Sregs}\vspace{0mm}
\end{figure*}

The correlation length can be intuitively understood by inspection of Figures \ref{r9north}--\ref{r9south}.  The polarisation vectors are seen to be closely correlated in direction over small patches of each HAWC+ field, typically of size 3--10 beamwidths ($\sim$0.5--1.5\,pc).  Beyond that scale, the vectors are less correlated, presumably signifying the scale within which the $B$ field is also physically correlated (i.e., sampling the MHD wave in one domain).  We need to quantify this behaviour more precisely, and especially, to map it.

We first fit the distributions of the rotated polarisation position angle $\theta_B$ across a given map area with a simple gaussian $e^{-\theta_B^2/2s^2}$ to obtain a best-fit value for the dispersion $s$ in $\theta_B$ (measured in radians) over that area.  The problem then reduces to deciding what areas are appropriate to collect these statistics.  To bracket the scales described above, we collect $s$ data from a HAWC+ beam scale to complete patches above the S/N limit for each clump (or part-clumps with distinct $\theta_B$ patterns). %This can be tricky, however.  If we choose an area that is too small, we risk underestimating $s$ by missing the true variance of $\theta_B$ in that location.  If the area is too large, we could overestimate $s$ by mixing together dynamically-unrelated $\theta_B$ across multiple $B$ field domains.

%%%%%%%%
%     Fig.A2    %   South thetaBs and DCF dispersions
%%%%%%%%
\begin{figure}[b]
\centerline{\includegraphics[angle=0,scale=0.22]{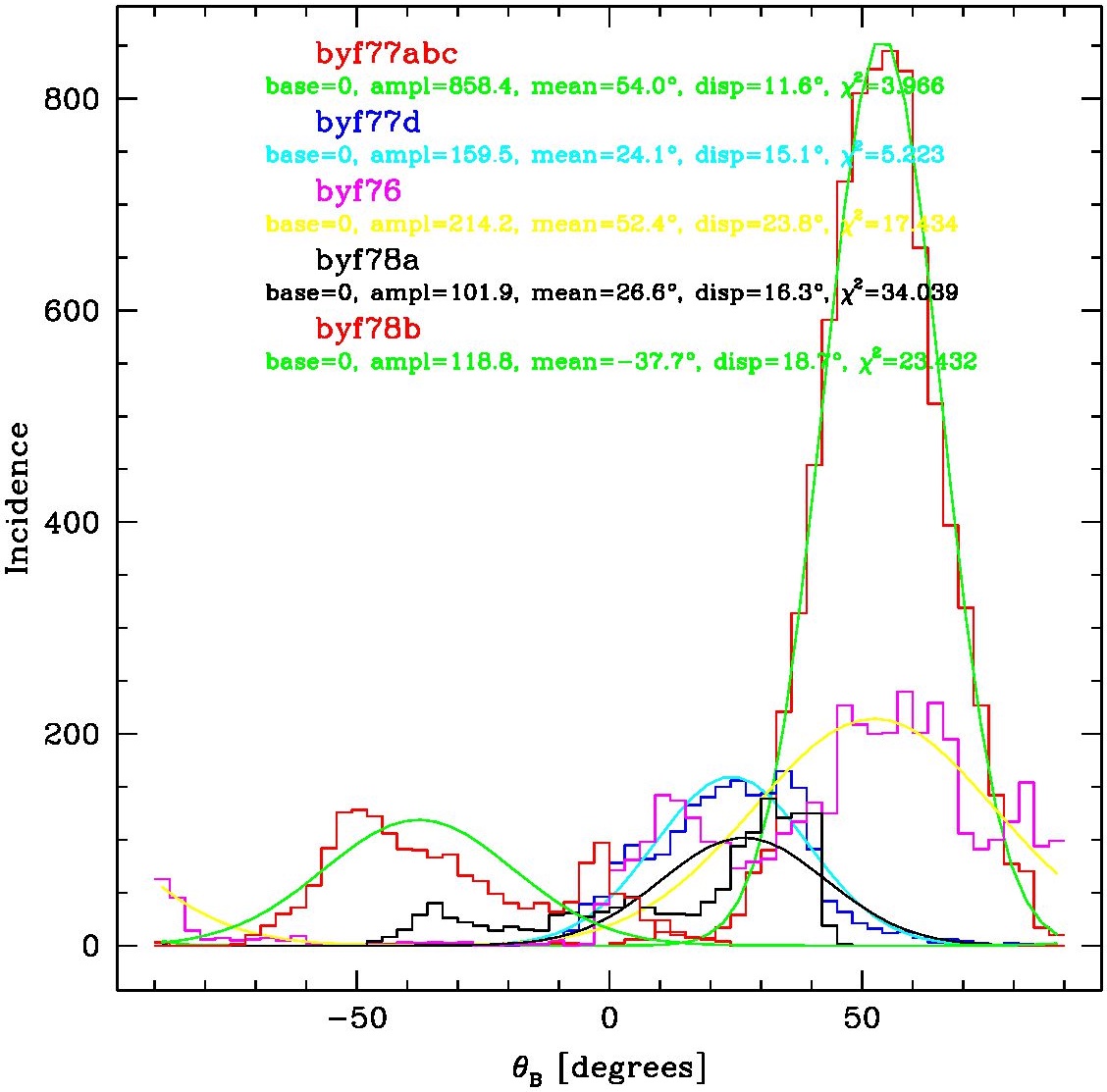}
\hspace{2mm}\includegraphics[angle=0,scale=0.22]{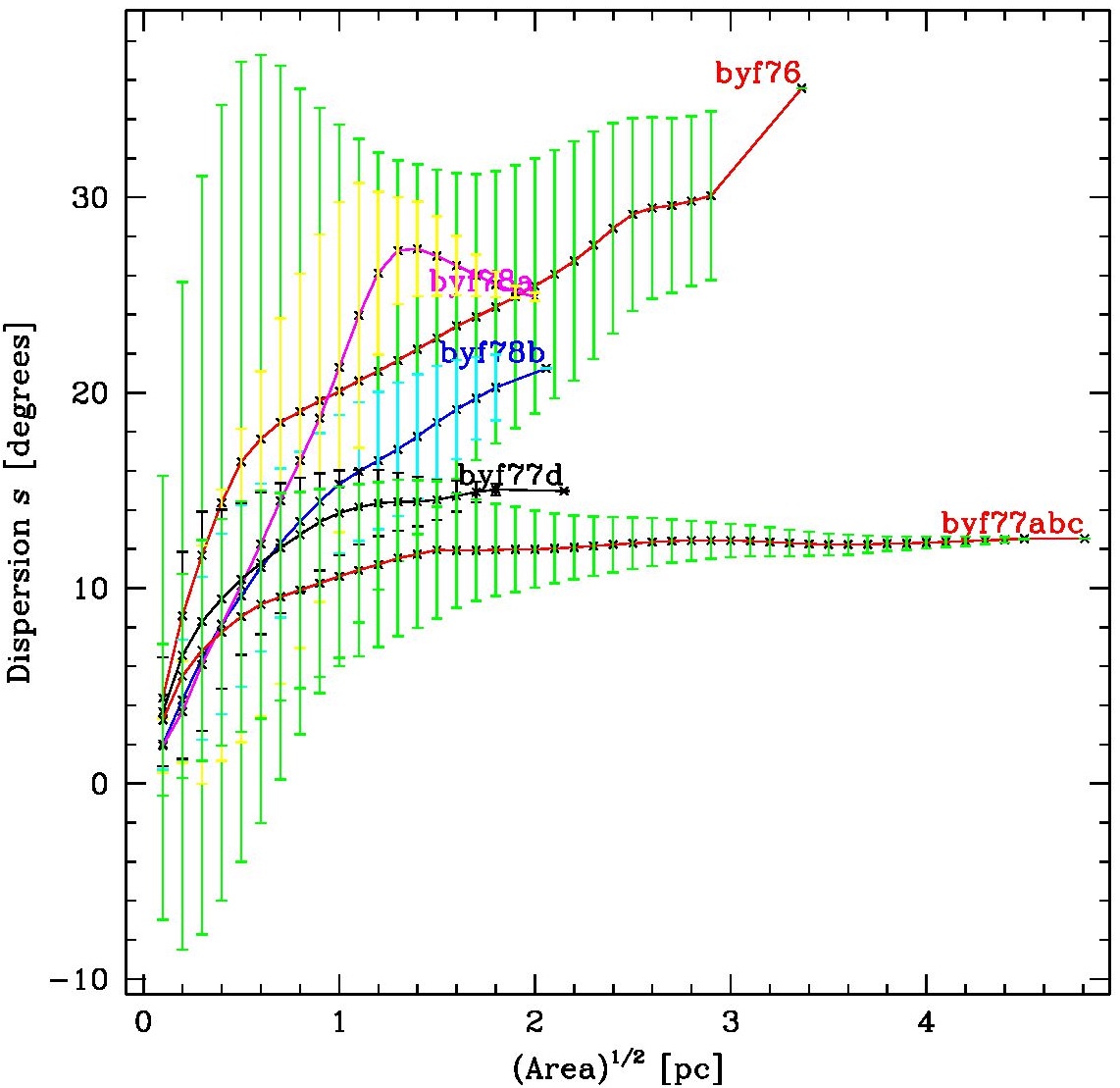}}
\vspace{-2mm}\caption{ \footnotesize  % uncomment for aasj
{\em Left:} Histograms of $\theta_{B_{\perp}}$ at all pixels within each of the ROI cutouts from Fig.\,\ref{Sregs}, as labelled.  Also shown are gaussian fits to each $\theta_{B_{\perp}}$ distribution, along with each fit's parameters.  %TO DO?--> Make yellows darker in eps.
%}\label{Sthetas}
%\caption{ \footnotesize  % uncomment for aasj
{\em Right:} Mean dispersion $s$ of polarization position angles $\theta_{B_{\perp}}$, with the 1$\sigma$ variation in $s$ shown as error bars, as a function of box size within each ROI shown in Fig.\,\ref{Sregs}.
}\label{DCFstats}\vspace{0mm}
\end{figure}

In the HAWC+ data, the measured $\theta_{B_{\perp}}$ in any given region with S/N \gapp 5 will have an uncertainty in orientation dominated by instrumental noise, $\Delta\theta_{B_{\perp}}$ \lapp 3\degree.  This occurs at a level around $P'$ = 0.3\,Jy/beam in Figures \ref{r9north}--\ref{r9south}, but even down to a formal S/N $\sim$ 2--3 around such areas ($P'$ $\sim$ 0.2\,Jy/beam), the polarisation signal will be correlated over small areas (a few beams) while the noise will not.  Therefore, such data also contains useful information, and we show the corresponding $\theta_{B_{\perp}}$ vectors in Figures \ref{r9north}--\ref{r9south} down to this lower level. 

In many of the Cycle 9 fields, this meaningful polarisation signal is detected over a number of irregular patches, but most are smaller than the individual fields.  In the Cycle 7 fields, the brighter emission generally results in wider $P'$ detections across each field.  In either case, we can define regions of interest (hereafter ROI) where the $\theta_{B_{\perp}}$ pattern appears reasonably well-correlated, as a first cut for areas over which we want to evaluate $s$.  %Making the ROIs too big is an easier-to-diagnose problem than making them too small, but with images of $\theta_B$, it is relatively easy to visually avoid either problem.  
Practically, the ROIs are usually clump- or sub-clump sized areas with contiguous $\theta_{B_{\perp}}$ data.  This is illustrated for Region 9 South in {\color{red}Figure \ref{Sregs}} (details of the same treatment for Region 9 North \& West follow). %appear in {\color{red}Appendix \ref{DCFdetails}}).
These ROIs typically have S/N$>$5 over most of their interiors, with some edges having S/N$\sim$2--3.

Having defined the ROIs, we can see in {\color{red}Figure \ref{DCFstats}, left panel} the $\theta_{B_{\perp}}$ distribution in all pixels within each.  While some of these are only approximately gaussian, we show these fits in order to illustrate effective dispersions in $\theta_{B_{\perp}}$ for the full ROIs.  The point is that the distributions are relatively limited, meaning we have roughly enclosed the maximum correlation scale within each ROI.

We next construct histograms for subsets (comprised of square boxes of area $A$) of each ROI, computing a dispersion $s$($A$) in $\theta_{B_{\perp}}$ for each subset.  The smaller the boxes, the more choice we have of where to fit them inside each ROI.  We then compute a mean dispersion $<$$s_{\theta_{B_{\perp}}}(A)$$>$ ($\pm$ a standard deviation) in the field orientation for all small boxes of a given area $A$, no matter where they are placed within the ROI.  Finally, in {\color{red}Figure \ref{DCFstats}, right panel} we plot all such results as a function of box size $A^{1/2}$, ranging from a minimal useful size of $A$ = 3$\times$3 pixels (roughly one Nyquist sample given the HAWC+ band D beam) to the full size of each ROI.  

In each ROI, the mean dispersion $<$$s$$>$ within all boxes of area $A$ rises as the box size increases, meaning that $\theta_{B_{\perp}}$ is more correlated on small scales (e.g., $s$\lapp5\degr\ within $\sim$0.3\,pc), and becomes less correlated over longer distances ($s$\gapp10\degr\ beyond $\sim$1.5\,pc).  The scale at which $s$ first starts to plateau is where we identify the correlation length as per \cite{mg91}.  Thus, the HAWC+ polarisation data suggest that, as far as the $B$ field is concerned, the correlation length across all of Region 9 is $\sim$0.5--1\,pc.

In what follows though, we use a correlation scale length of 0.33\,pc for computing $s$ as a practical matter.  This is because we wish to make a rough map of $s$ for purposes of computing $B_{\rm \perp}$, but the actual variations in $\theta_{B_{\perp}}$ are such that, on scales \gapp0.5\,pc we start to overlap more and more areas where the $\theta_{B_{\perp}}$ wrap across $\pm$90\degr\ distorts the averaging, in a way that is difficult to compensate for.  In other words, we choose to slightly underestimate $s$ on these smaller scales, rather than grossly overestimating it on larger scales.

Thus, we can map the rms dispersion $s$ in $\theta_{B_{\perp}}$ at each pixel, even if it is systematically a slight underestimate.  But with a single correlation length, we are not limited by ROIs and can compute $s$ at every pixel in the HAWC+ maps where the polarisation S/N $>$ 3.  This is shown in {\color{red}Appendix \ref{parmaps}} ({\color{red}Fig.\,\ref{smap}}). 

\begin{figure*}[h]
\vspace{-1mm}
\centerline{\includegraphics[angle=-90,scale=0.63]{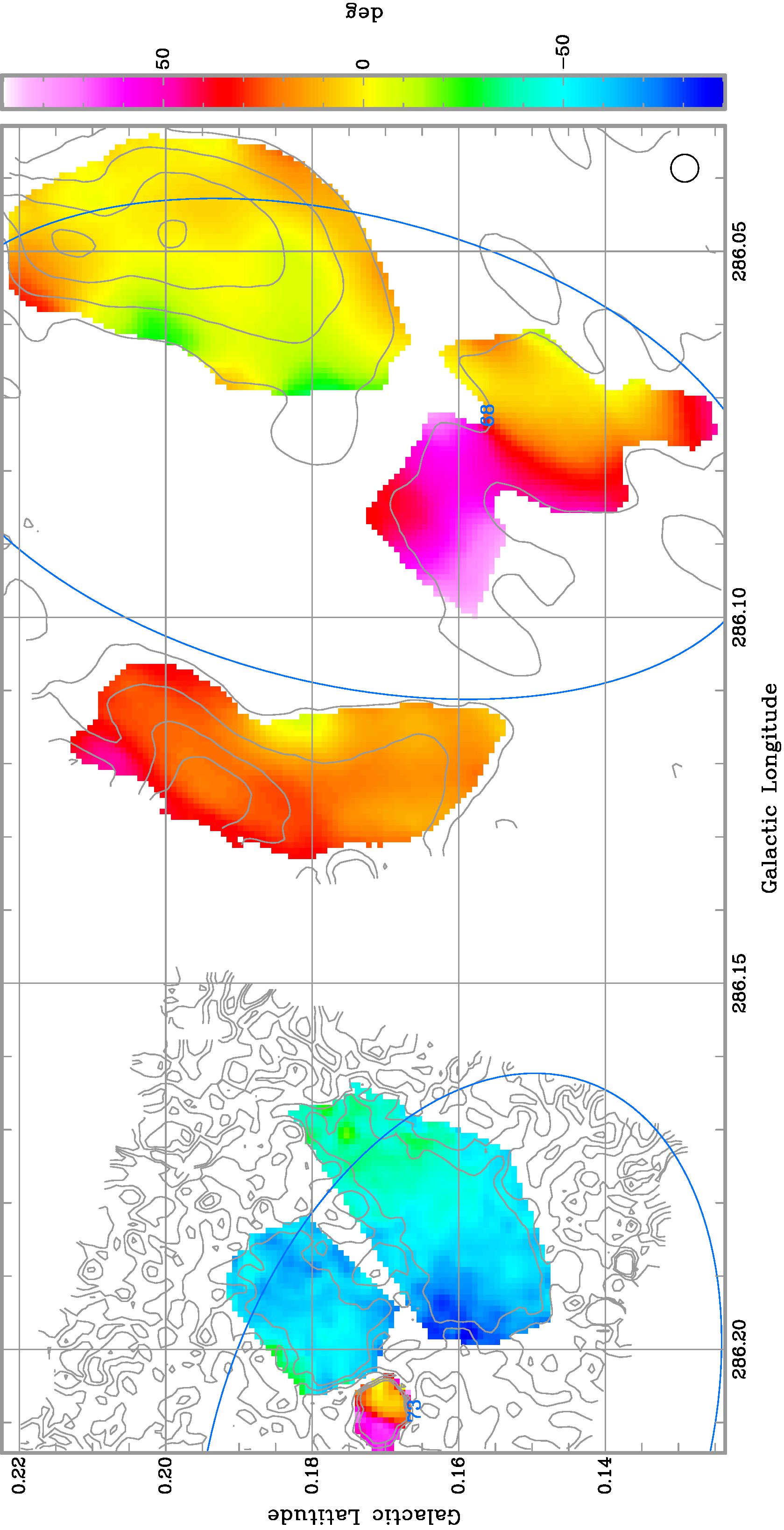}}
\vspace{-1mm}\caption{ %\footnotesize  % uncomment for aasj
Cutouts of the $\theta_{B_{\perp}}$ distribution in the Region 9 North ROIs (three for BYF\,68 --- centre, north, and east --- and four for BYF\,73 --- north and south portions of the HII region, and the MIR\,2 and EPL portions of the massive core) with S/N($P'$) \gapp 5, overlaid by grey $P'$ contours at 0.2(0.4)1.4 Jy/bm and blue Mopra \tco\ ellipses.  The DCF analysis for these ROIs appears in Fig.\,\ref{Nthetas}.
}\label{Nregs}\vspace{-2mm}
\end{figure*}

\begin{figure}[b]
\centerline{\includegraphics[angle=0,scale=0.21]{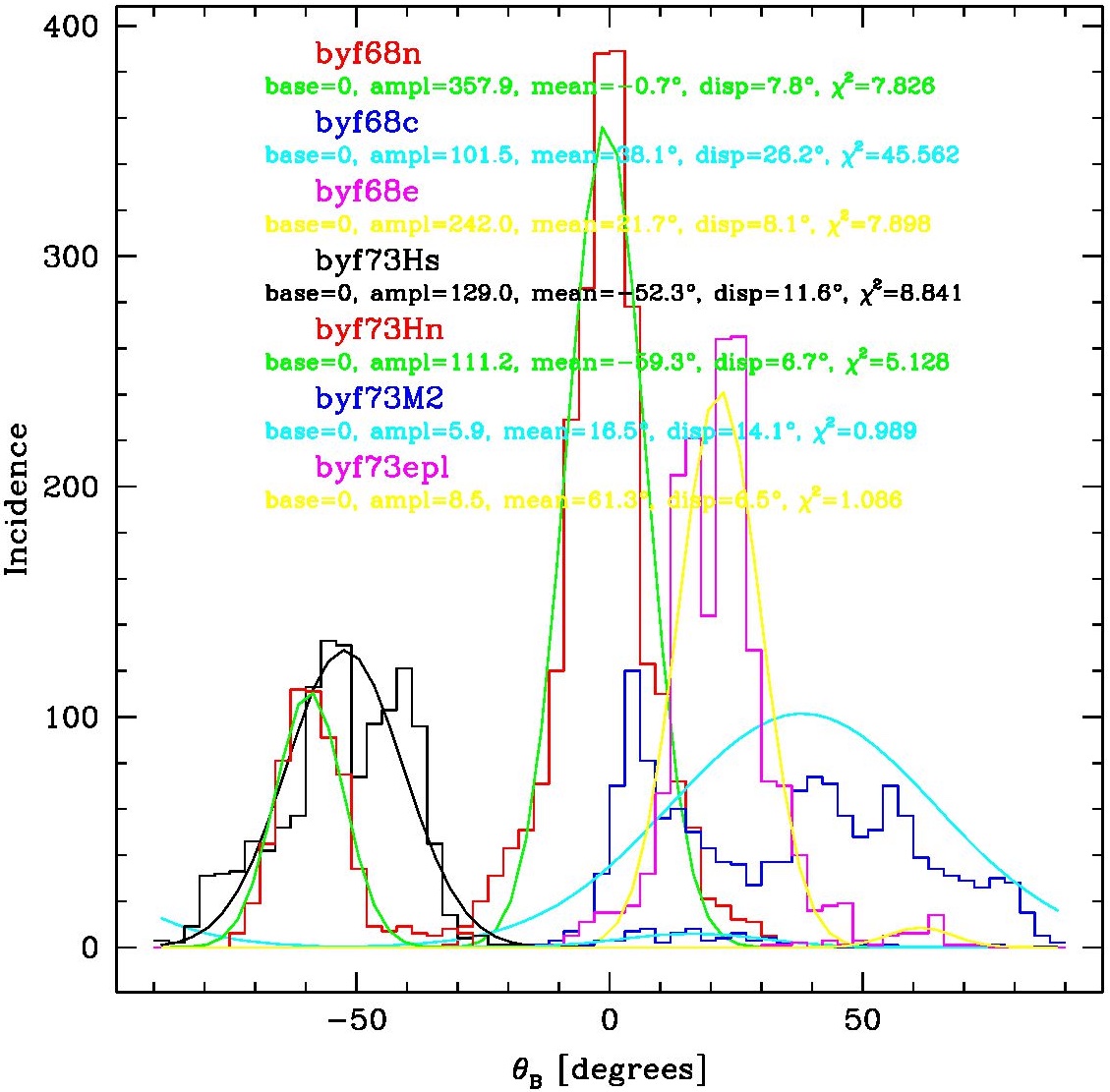}
\hspace{2mm}\includegraphics[angle=0,scale=0.21]{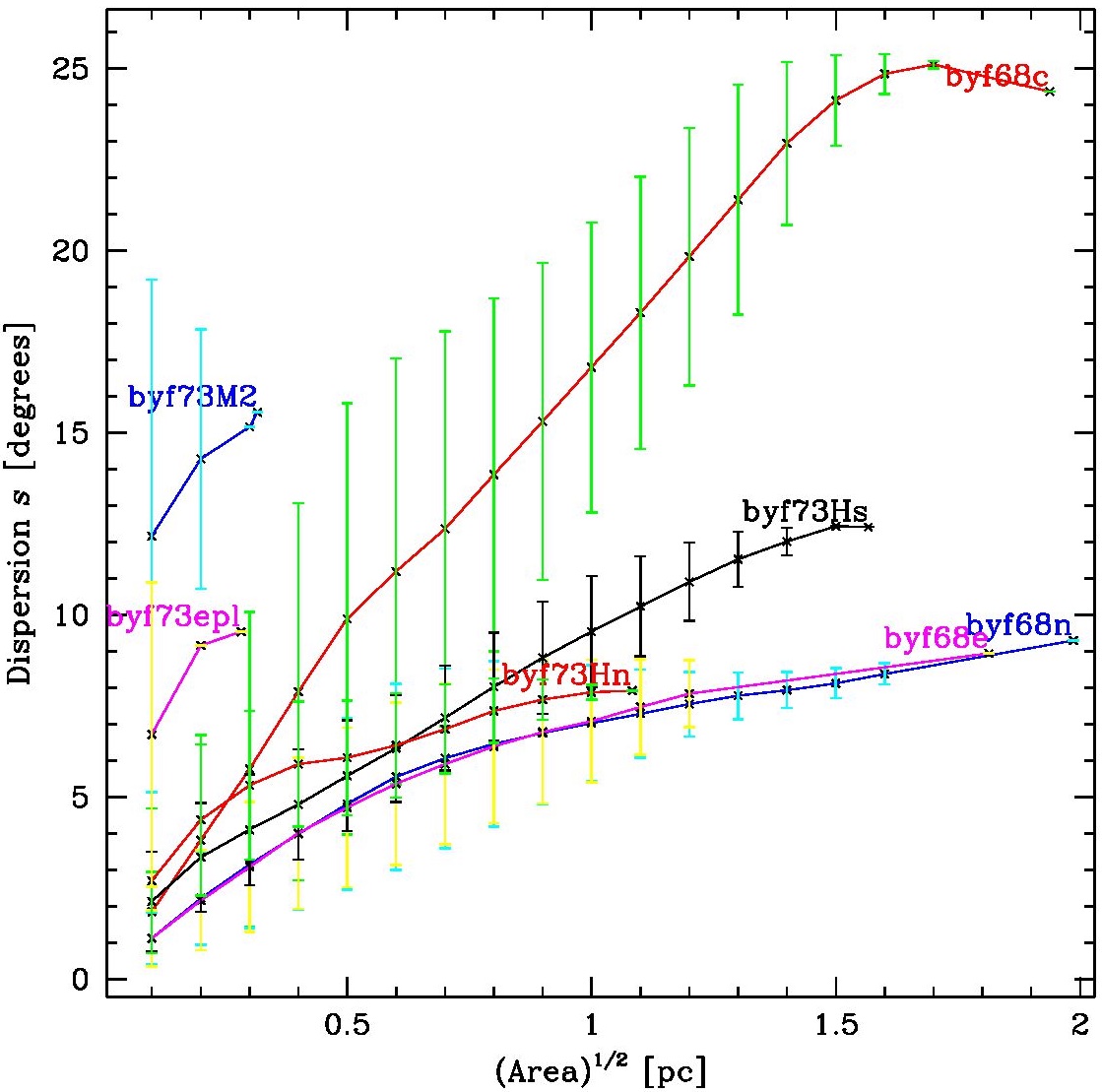}}
\vspace{-1mm}\caption{ %\footnotesize  % uncomment for aasj
{\em Left:} Histograms of $\theta_{B_{\perp}}$ at all pixels within each of the ROI cutouts from Fig.\,\ref{Nregs}, as labelled.  Also shown are gaussian fits to each $\theta_{B_{\perp}}$ distribution, along with each fit's parameters.
{\em Right:} Mean dispersion $s$ of polarization position angles $\theta_{B_{\perp}}$, with the dispersion in the dispersion shown as error bars, as a function of box size within each ROI shown in Fig.\,\ref{Nregs}.
}\label{Nthetas}\vspace{-1mm}
\end{figure}

\begin{figure*}[t]
\centerline{\includegraphics[angle=0,scale=0.47]{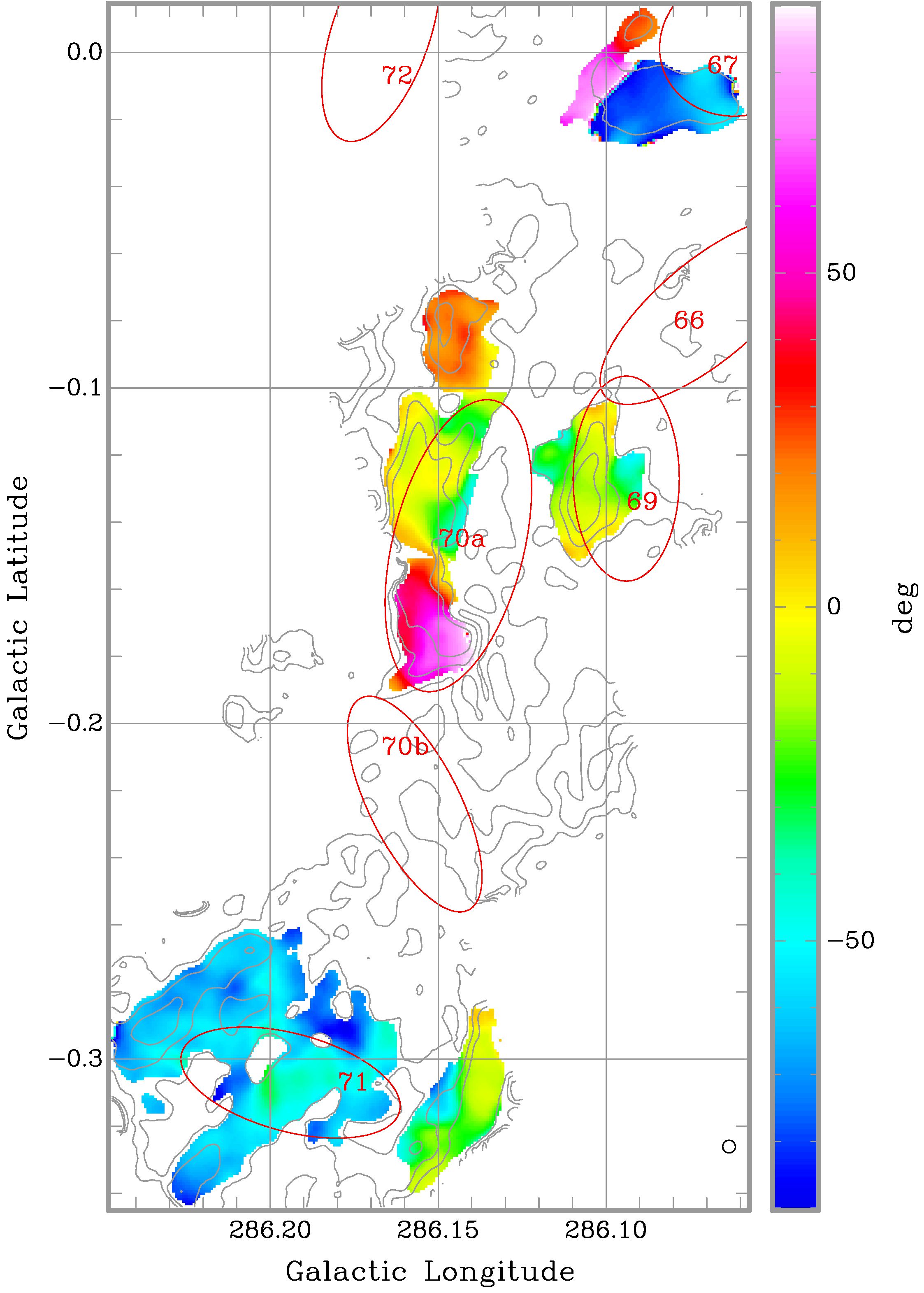}}
\vspace{-1mm}\caption{ %\footnotesize  % uncomment for aasj
Cutouts of the $\theta_{B_{\perp}}$ distribution in the Region 9 West ROIs (BYF\,67, 69, three for BYF\,70a --- centre, north, and south --- and two for BYF\,71 --- main and west) with S/N($P'$) \gapp 5, overlaid by grey $P'$ contours at 0.2(0.4)1.0 Jy/bm and red Mopra \tco\ ellipses.  The DCF analysis for these ROIs appears in Fig.\,\ref{Wthetas}.  For clumps BYF\,66 and 70b, the data were of insufficient S/N to be included in the DCF analysis.
}\label{Wregs}\vspace{0mm}
\end{figure*}

\begin{figure}[b]
\centerline{\includegraphics[angle=0,scale=0.21]{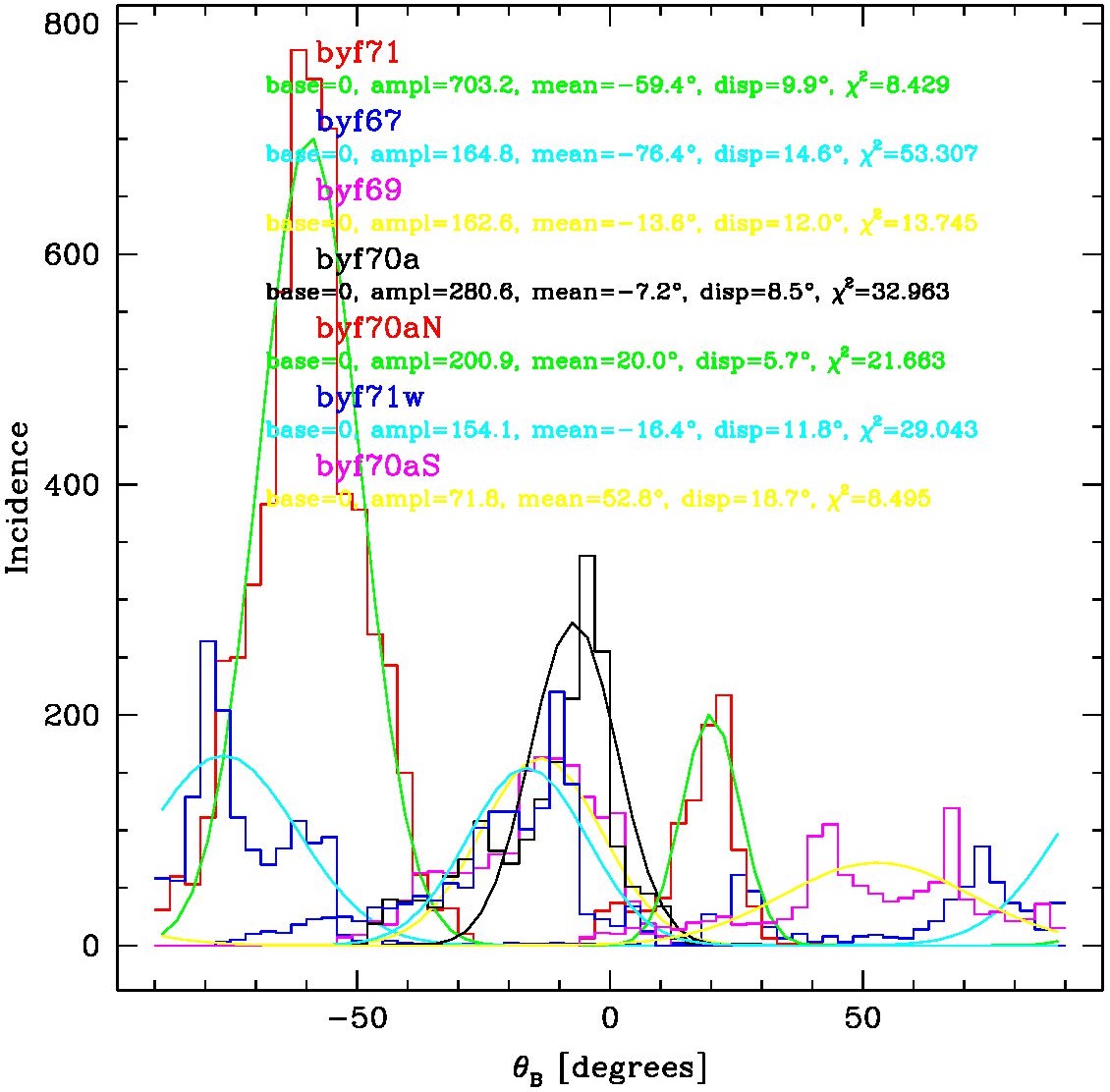}
\hspace{2mm}\includegraphics[angle=0,scale=0.21]{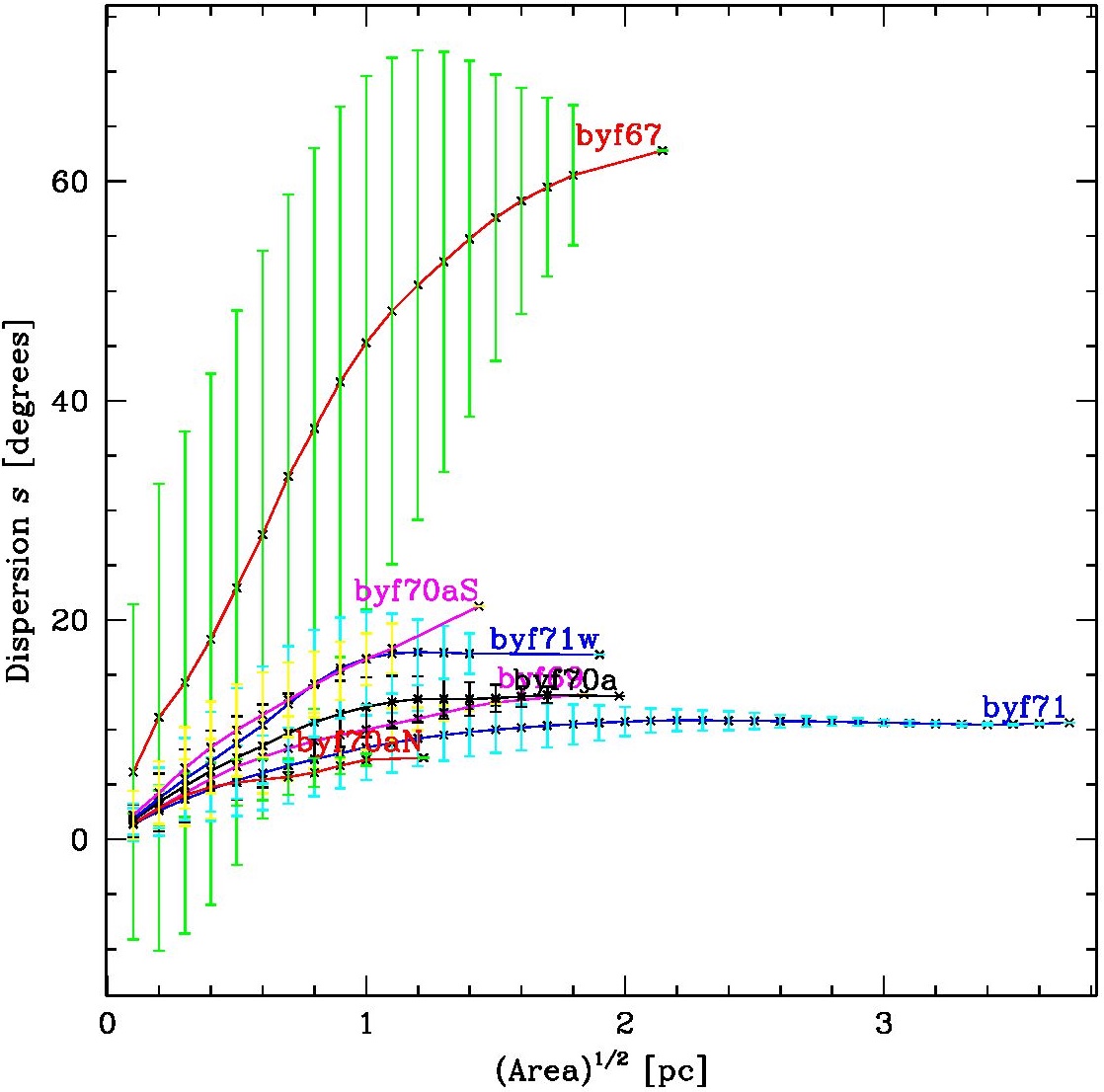}}
\vspace{-1mm}\caption{ %\footnotesize  % uncomment for aasj
{\em Left:} Histograms of $\theta_{B_{\perp}}$ at all pixels within each of the ROI cutouts from Fig.\,\ref{Wregs}, as labelled.  Also shown are gaussian fits to each $\theta_{B_{\perp}}$ distribution, along with each fit's parameters.
{\em Right:} Mean dispersion $s$ of polarization position angles $\theta_{B_{\perp}}$, with the dispersion in the dispersion shown as error bars, as a function of box size within each ROI shown in Fig.\,\ref{Wregs}.
}\label{Wthetas}\vspace{-1mm}
\end{figure}

In the North, {\color{red}Figure \ref{Nregs}} shows the ROI cutouts with sufficient S/N in the $P'$ signal to perform the analysis, while the two panels of {\color{red}Figure \ref{Nthetas}} show the statistics of the $\theta_{B_{\perp}}$ distributions.  {\color{red}Figures \ref{Wregs}} and {\color{red}\ref{Wthetas}} reflect the same information for the West cutouts.

%%%%%%%
%   Fig.A3   %   North DCF regions
%%%%%%%

%%%%%%%
%   Fig.A4   %   North thetaBs and DCF dispersions
%%%%%%%

%%%%%%%
%   Fig.A5   %   West DCF regions
%%%%%%%

%%%%%%%
%   Fig.A6   %   West thetaBs and DCF dispersions
%%%%%%%

\clearpage

%%%%%%%%
%   Appx.  B   %
%%%%%%%%
\section{Global Parameter Inputs to $B_{\perp}$ Calculation}\label{parmaps}

Here we show the derived parameter maps for $s$, $R$, $n$, and $\Delta$$V$, which are combined in Eq.\,(1) to produce the $B_{\perp}$ maps shown in Figures \ref{northB}--\ref{southB}.

\begin{figure*}[b]
\vspace{0mm}
\centerline{\includegraphics[angle=0,scale=0.86]{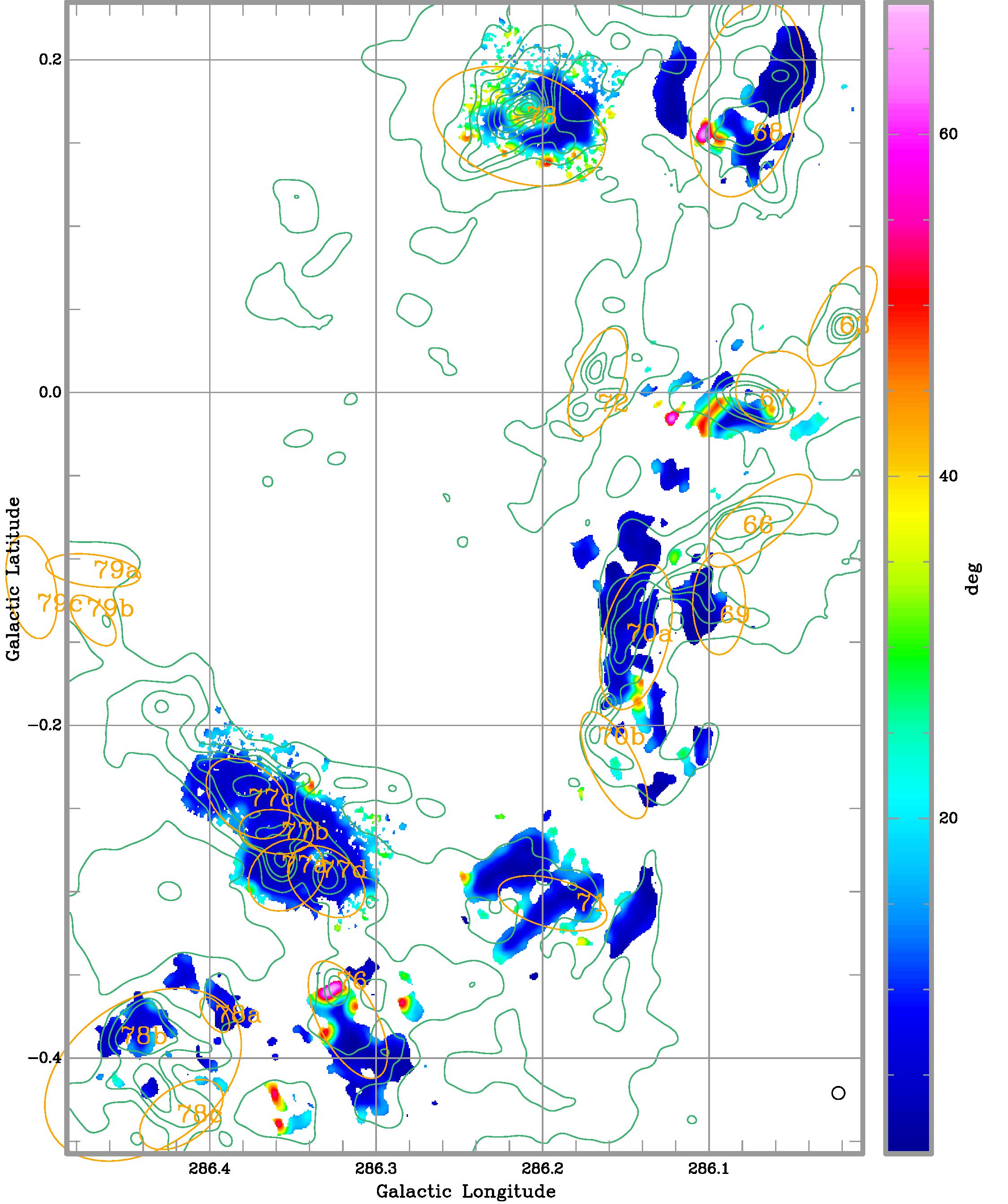}}
\vspace{0mm}\caption{ %\footnotesize  % uncomment for aasj
Complete Region 9 map of computed angular dispersion $s$ from DCF analysis described in the text.  Also overlaid, in this and subsequent figures, are green \nhtwo\ contours at levels 0.5(0.5)2 and 4(2)14 $\times$10$^{26}$ molecules\,m$^{-2}$ and orange Mopra \tco\ half-power ellipses.  The convolved beamsize of the DCF analysis (twice the HAWC+ beam) is shown in the bottom-right corner.
}\label{smap}\vspace{0mm}
\end{figure*}

We calculate $s$ as follows.  Starting with the HAWC+ $\theta_{B_{\perp}}$ maps, we make maps of (1) the local average of $\theta_{B_{\perp}}$ within the chosen correlation scale (0.33\,pc), (2) the difference between each pixel's $\theta_{B_{\perp}}$ and this local mean, (3) the square of these differences, (4) the local mean of these squared differences on the same correlation scale as (1), and (5) the square root of this local mean.  This is shown in {\color{red}Figure \ref{smap}}.

%%%%%%%
%   Fig.B1   %   s map
%%%%%%%

Note that in most areas the $s$ values are consistently small, \lapp10\degr, on scales of $\sim$1--5\,pc across most of Region 9, and are more strictly $<$20\degr\ over $\sim$95\% of the pixels; the modal range of values is flat-topped over $\sim$2--6\degr.  The values become noticeably larger, however, only in small patches where the $\pm$90\degr\ polarisation wrap occurs within the correlation scale, numerically distorting the $s$ calculation.  The only exception(s) where $s$ is truly large ($\sim$30\degr) on these small scales is in the MIR\,2 core of BYF\,73, and in a few lower-S/N spots in BYF\,67, 68, and 76.

%%%%%%%
%   Fig.B2   %   R map
%%%%%%%
\begin{figure*}[b]
\vspace{0mm}
\centerline{\includegraphics[angle=0,scale=0.86]{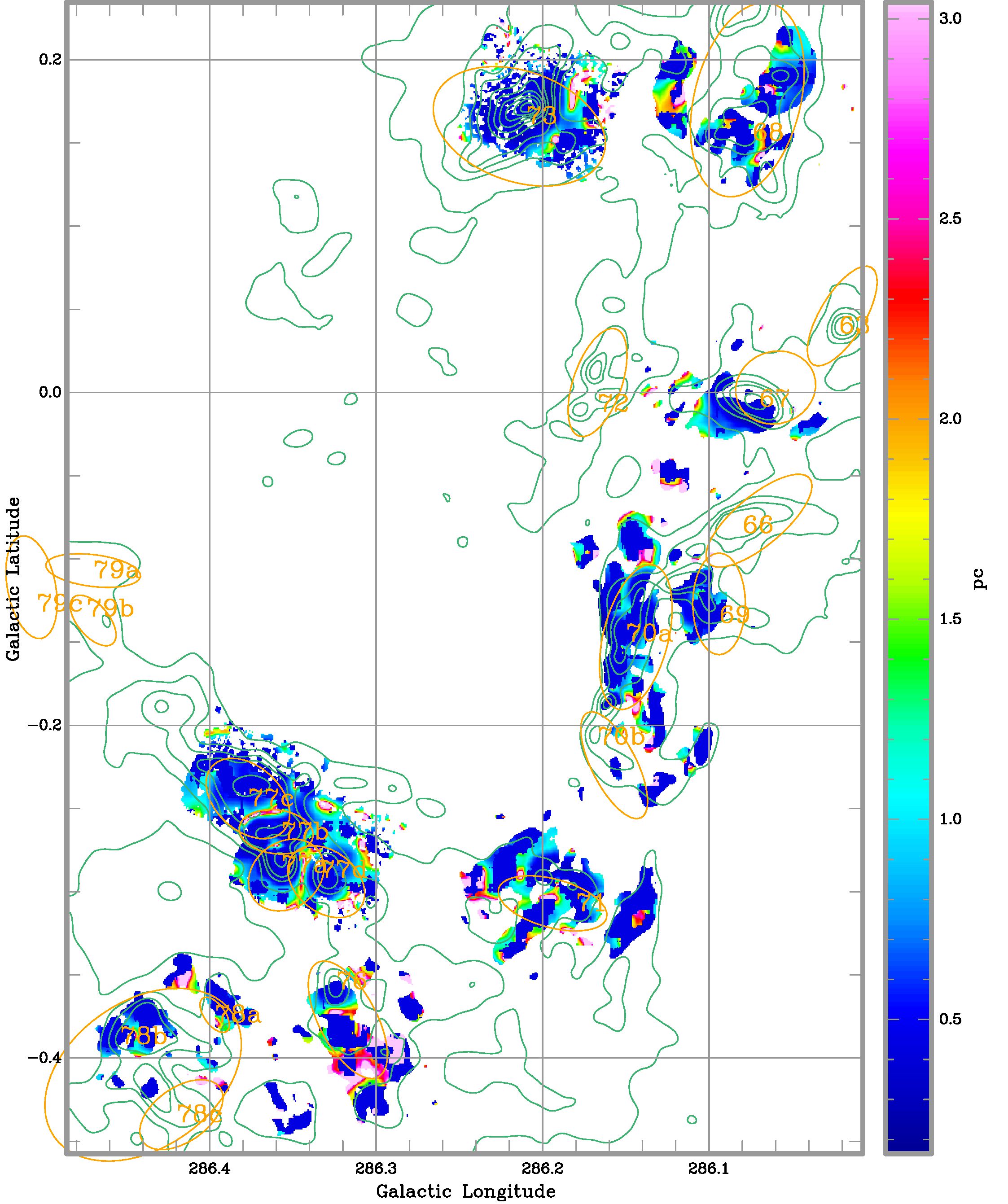}}
\vspace{0.3mm}\caption{ %\footnotesize  % uncomment for aasj
Complete Region 9 map of clump size/depth scale $R$ as inferred from gradient method described in the text.  The $R$ values are computed on every pixel with available $N$ data (e.g., Fig.\,\ref{density}), but masked to the $s$ map in Fig.\,\ref{smap} to avoid visual clutter away from the polarisation signal.
}\label{Rscale}\vspace{0mm}
\end{figure*}

We also show the derived $R$ scale map as described in the main text ({\color{red}Fig.\,\ref{Rscale}}), mainly to demonstrate that the gradient method gives reasonable results for the HAWC+ data analysis.  Most $R$ values are well under 1.5\,pc, although small patches have $R$ $>$ 5\,pc.

\begin{figure*}[b]
\centerline{\includegraphics[angle=0,scale=0.86]{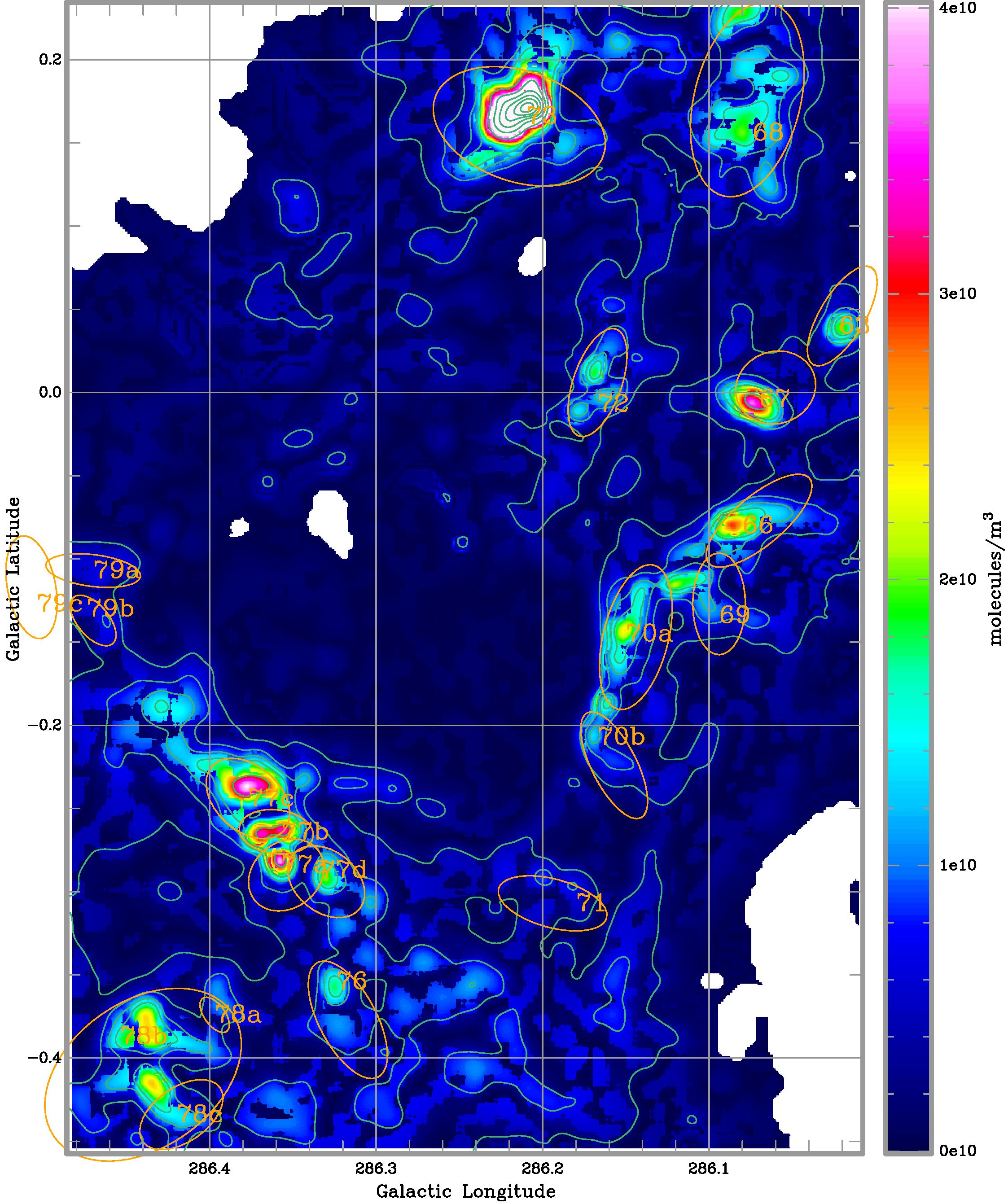}}
\vspace{-0.5mm}\caption{ %\footnotesize  % uncomment for aasj
Complete Region 9 map of gas-phase molecular density $n_{\rm H_2}$, derived from column density \nhtwo\ map \citep{p19} and gradient method to estimate $R$ (Fig.\,\ref{Rscale}) as described in the text.  The colour scale displays most areas well, except for BYF\,73 which is heavily saturated at a peak $n$ = 1.23$\times$10$^{11}$molecules\,m$^{-3}$.
}\label{density}\vspace{0mm}
\end{figure*}

The $n$ map ({\color{red}Fig.\,\ref{density}}) obtained with our gradient method also appears satisfactory as a rough, structure-agnostic way to compute $B_{\perp}$.

\begin{figure*}[b]
\centerline{\includegraphics[angle=0,scale=0.95]{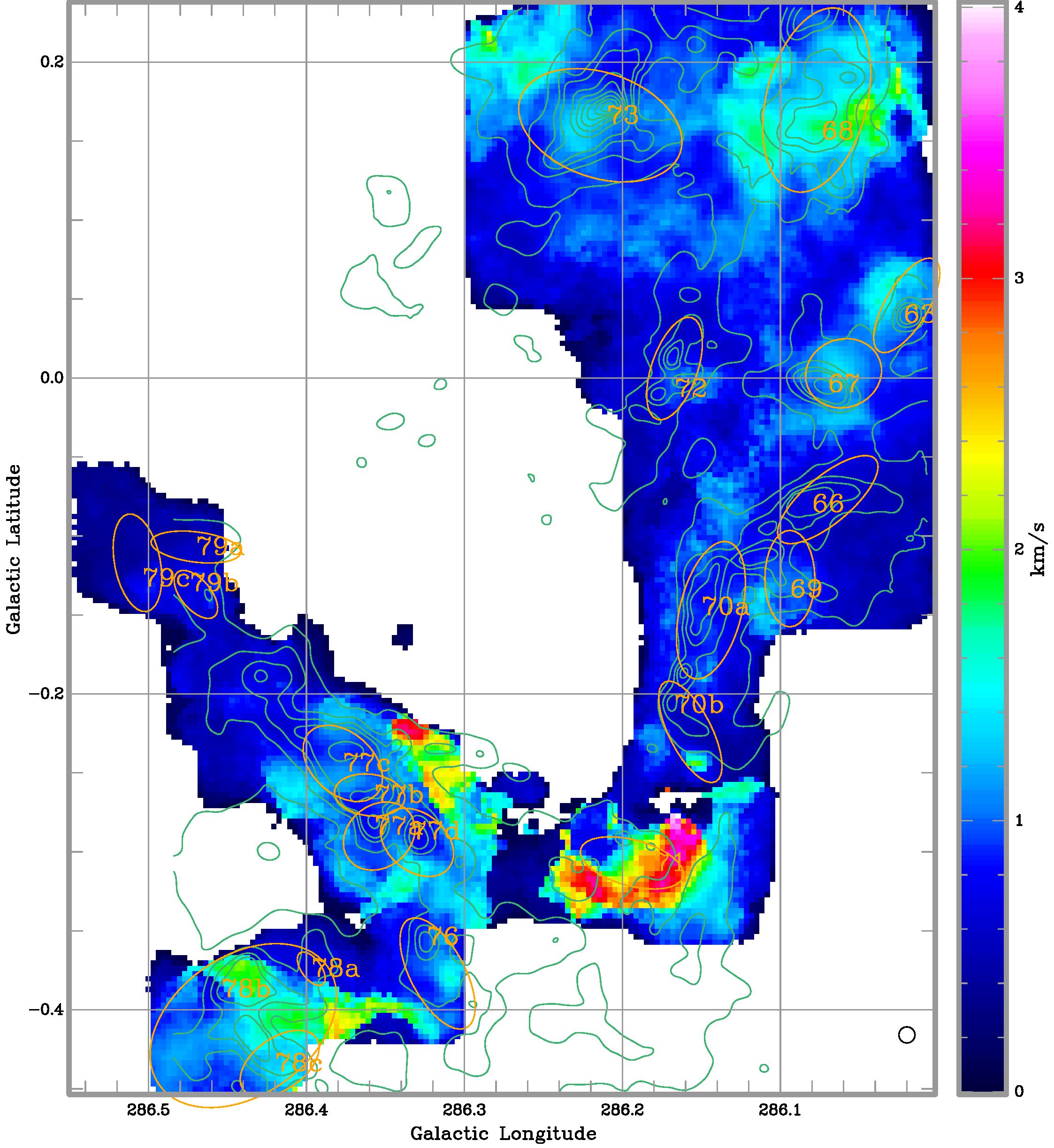}}
\vspace{-1mm}\caption{ %\footnotesize  % uncomment for aasj
Complete Region 9 map of velocity dispersion from \nco\ data \citep{b18}.  The Mopra beam (37$''$) is shown in the bottom right corner.
}\label{sigmaV}\vspace{0mm}
\end{figure*}

Finally we show in {\color{red}Fig.\,\ref{sigmaV}} a $\sigma_{V}$ map for the \nco\ cube from \cite{b18}, derived from a radiative transfer treatment of \tco, \ttco, and \ceto\ data cubes that fully samples the opacity and column density in each line.

%%%%%%%
%   Fig.B3   %   n map
%%%%%%%

%%%%%%%
%   Fig.B4   %   sigmaV map
%%%%%%%

\clearpage
%%%%%%%%
%   Appx.  C   %
%%%%%%%%
\section{HRO Analysis of Individual Cutouts}\label{HROdetails}

\begin{figure*}[h]
\centerline{\includegraphics[angle=0,scale=0.148]{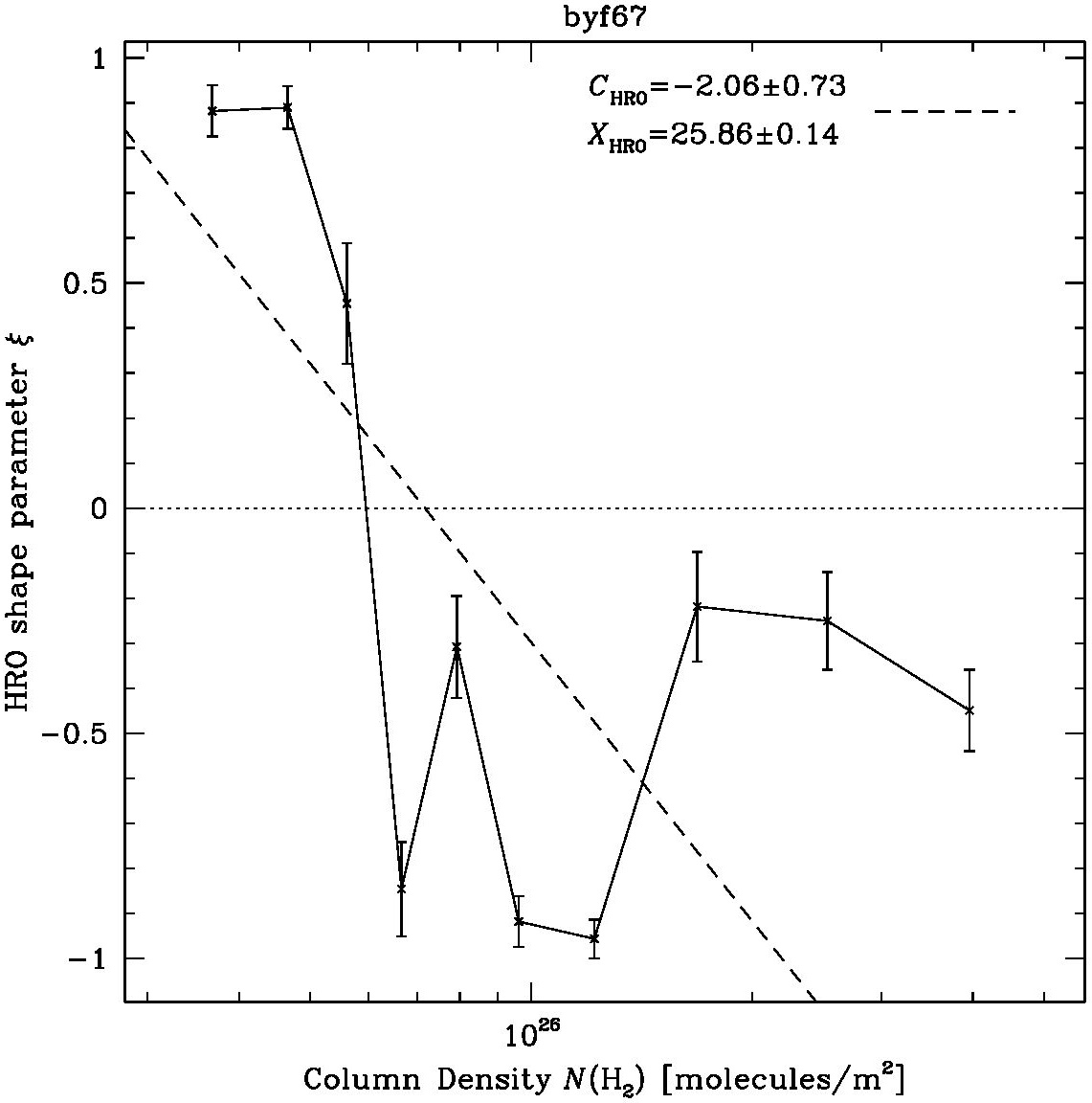}
		\includegraphics[angle=0,scale=0.148]{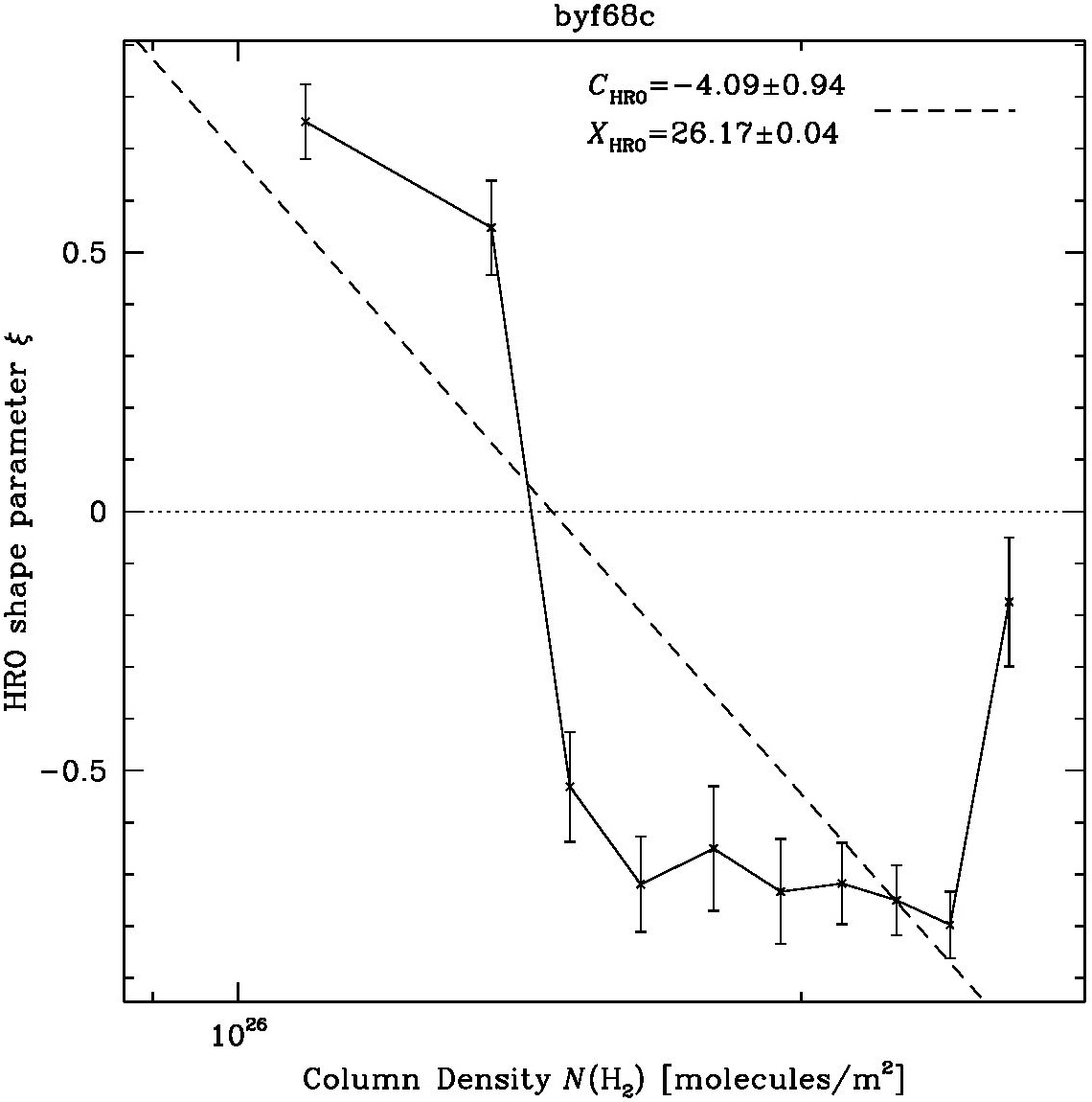}
		\includegraphics[angle=0,scale=0.148]{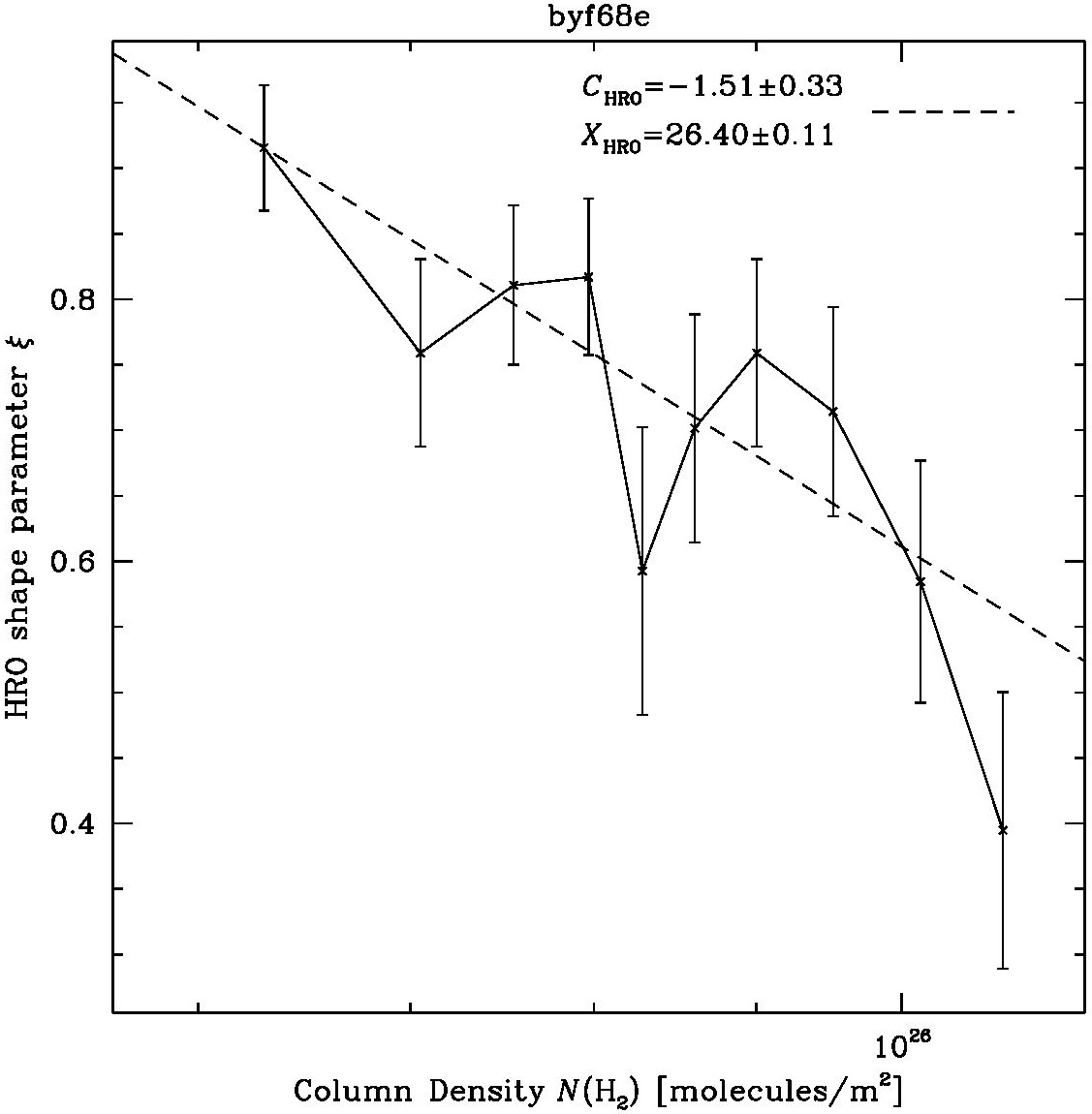}}
\centerline{\includegraphics[angle=0,scale=0.148]{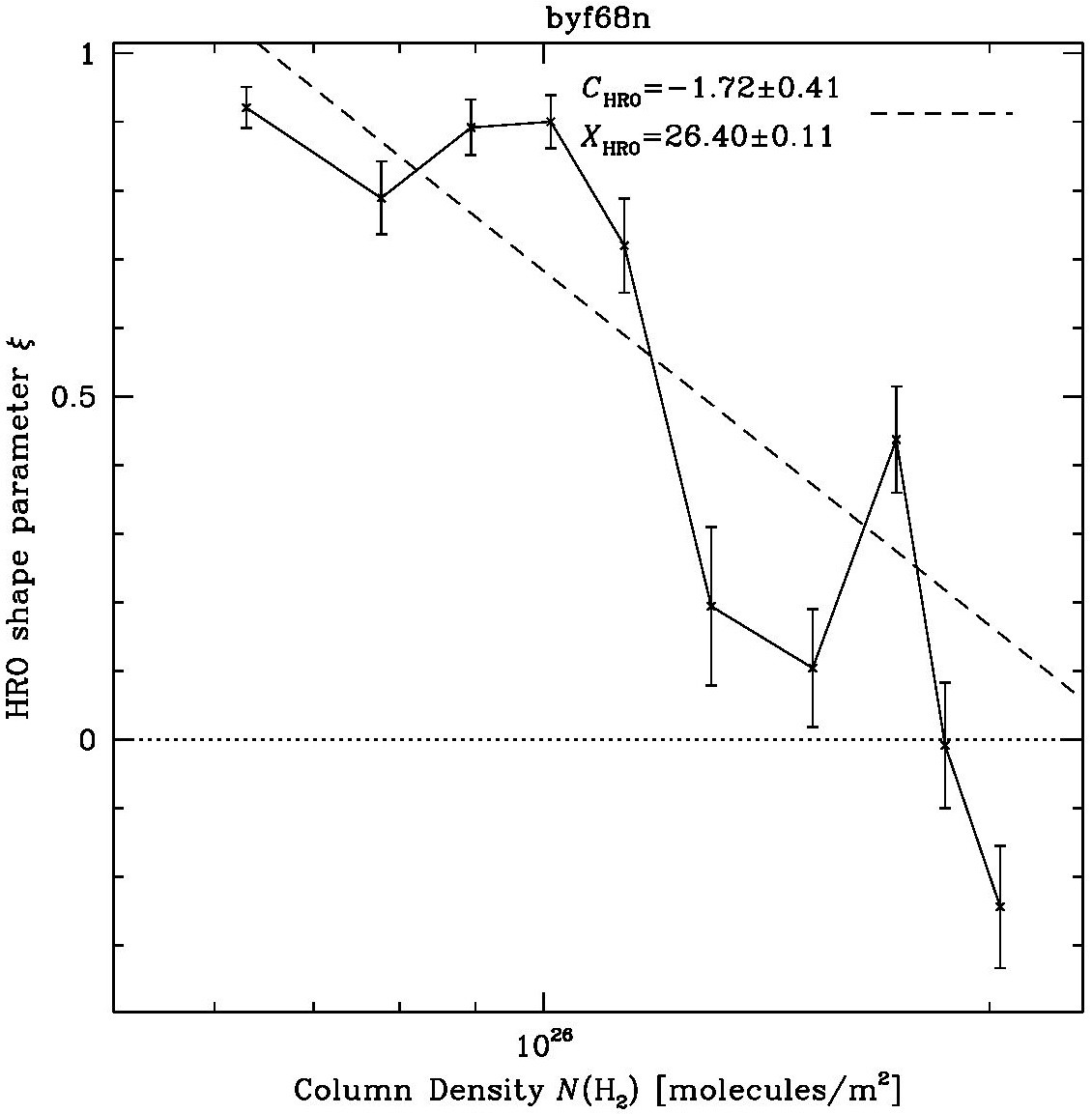}
		\includegraphics[angle=0,scale=0.148]{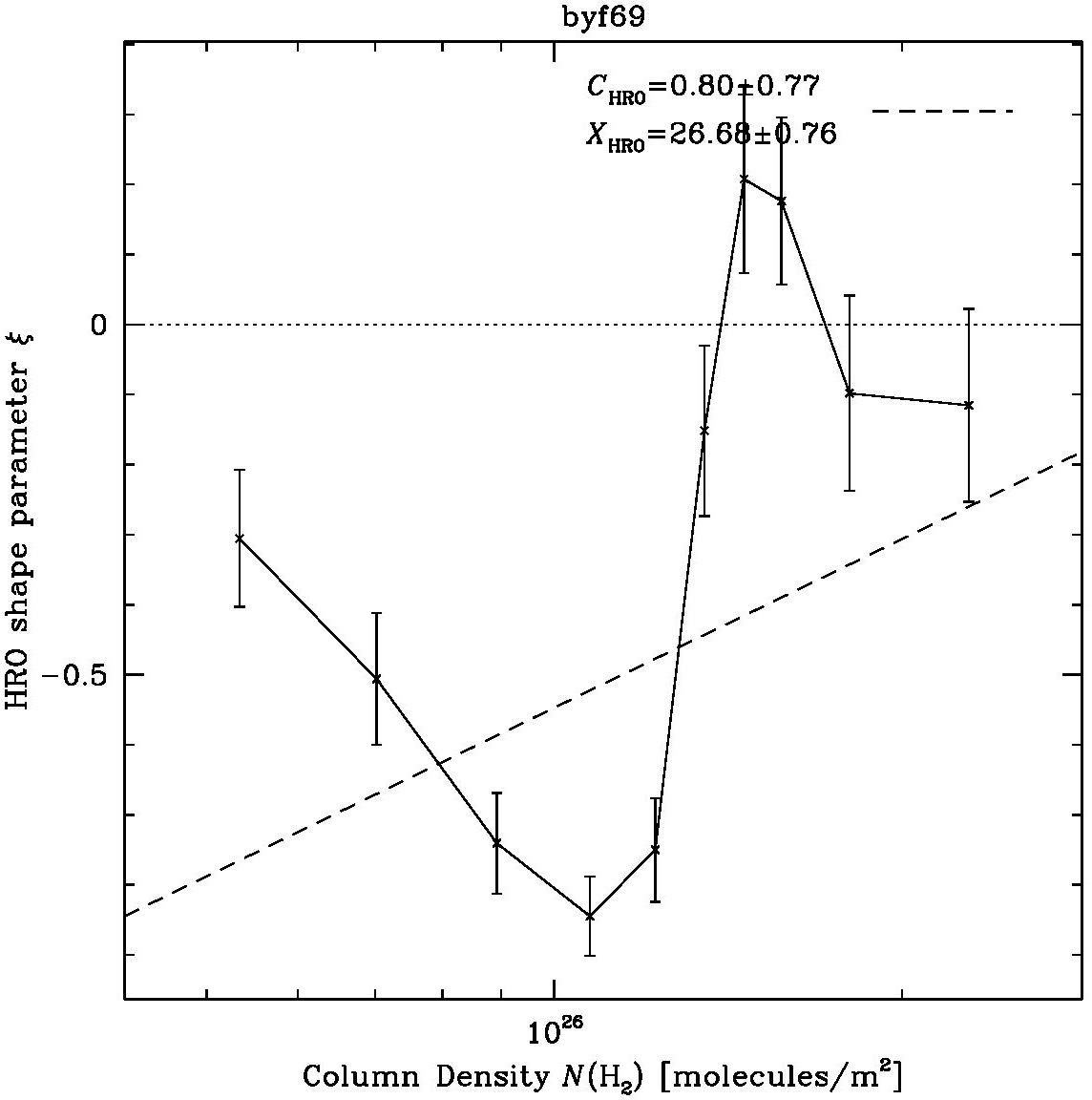}
		\includegraphics[angle=0,scale=0.148]{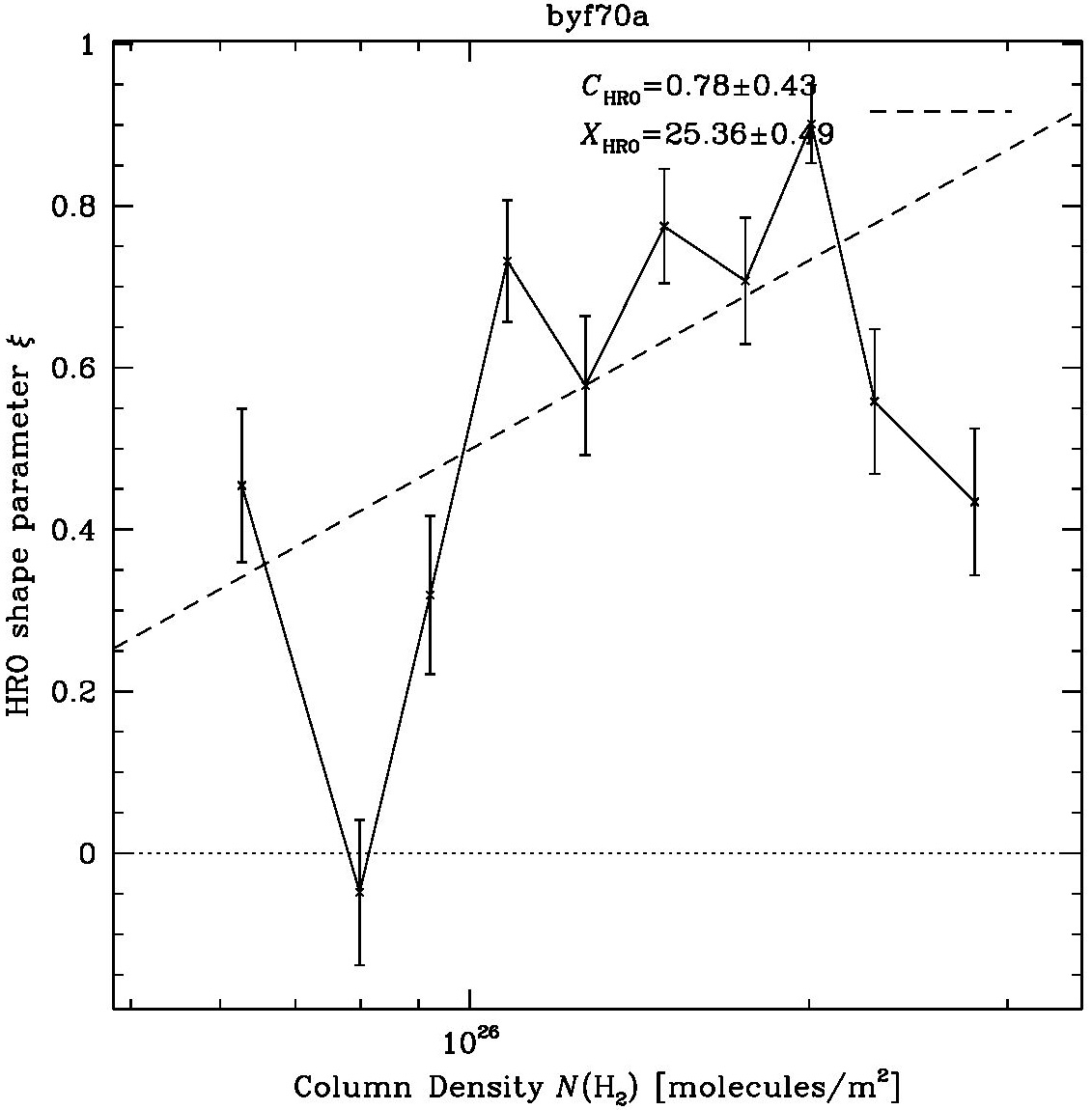}}
\centerline{\includegraphics[angle=0,scale=0.148]{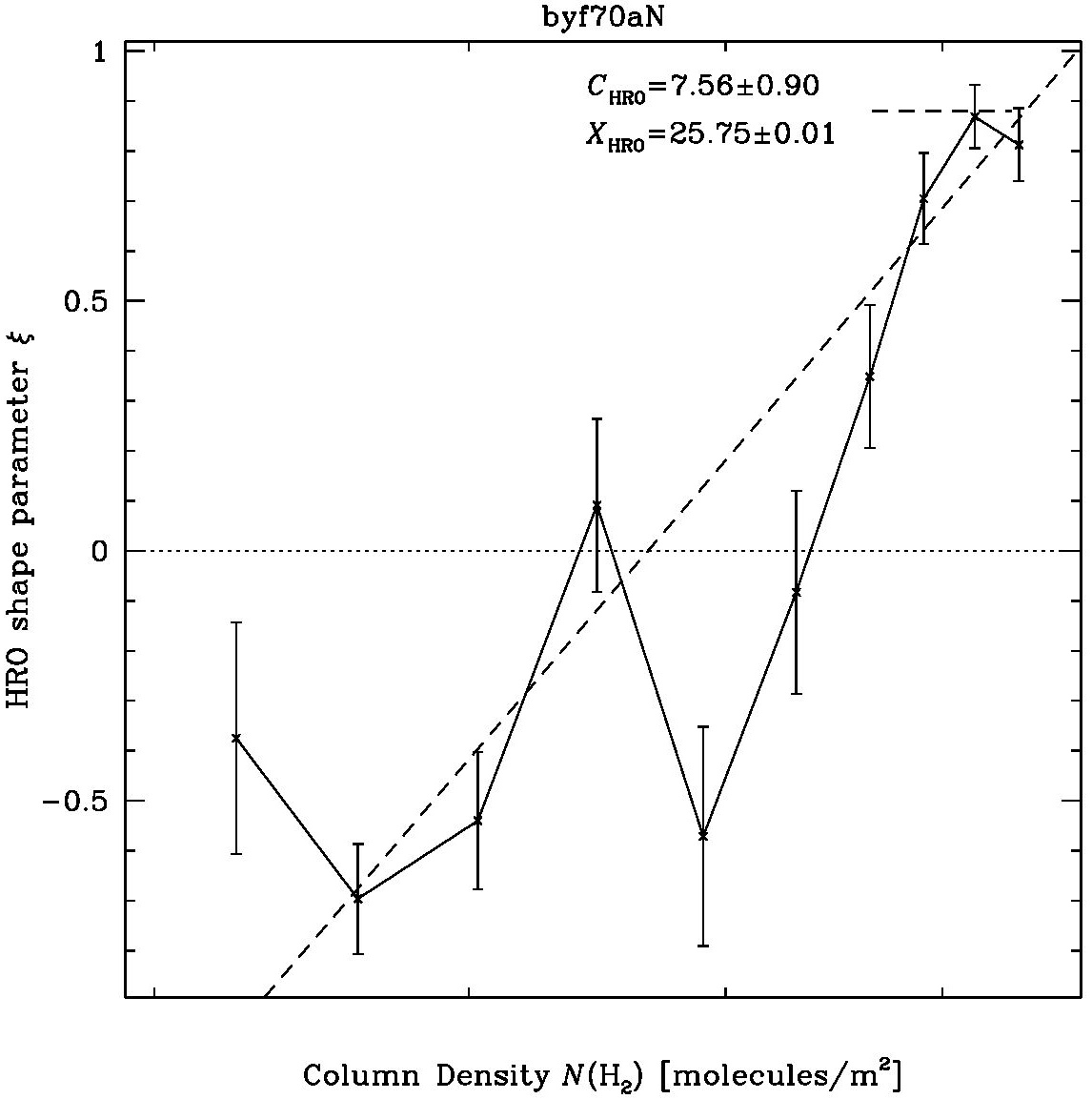}
		\includegraphics[angle=0,scale=0.148]{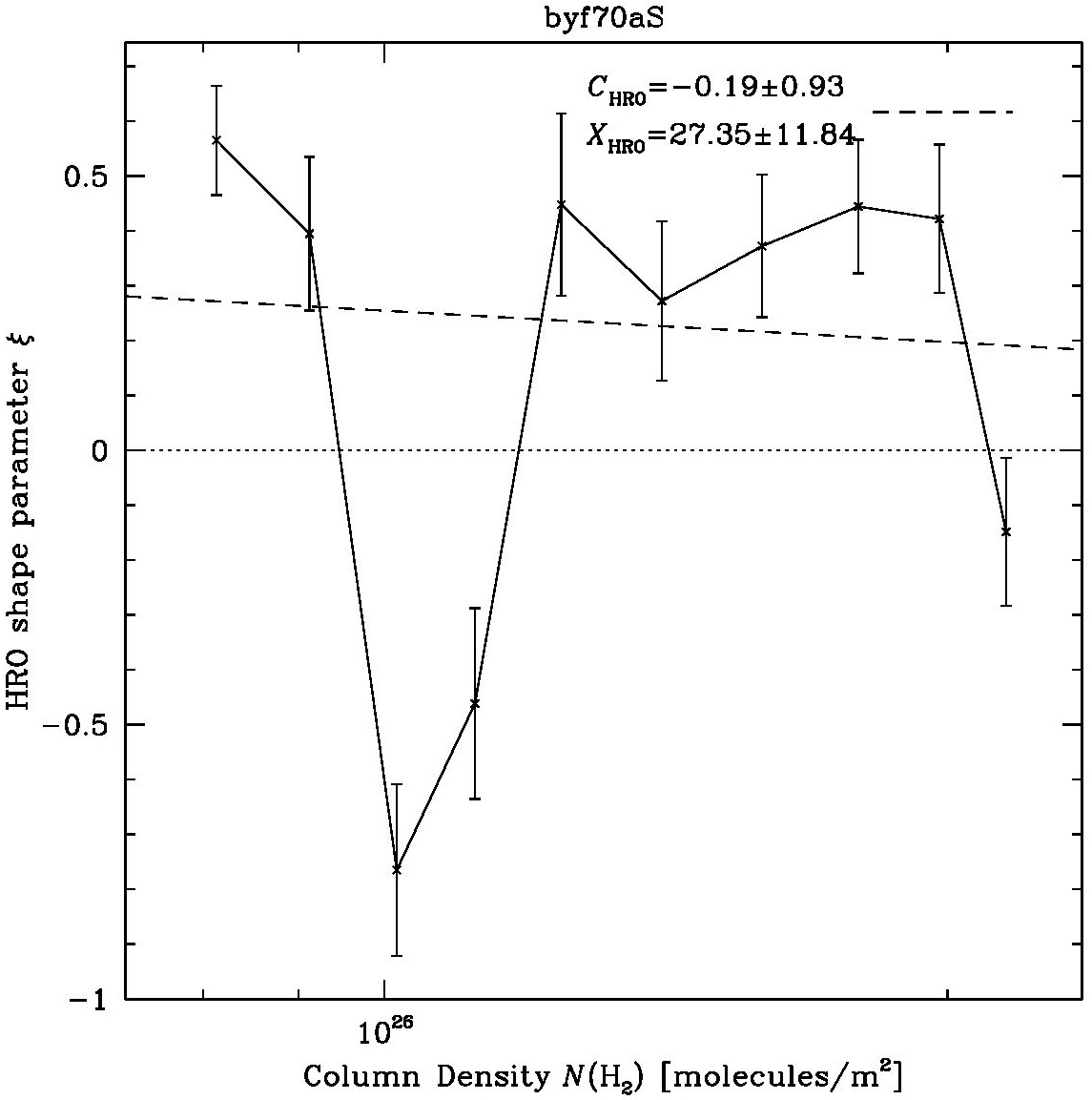}
		\includegraphics[angle=0,scale=0.148]{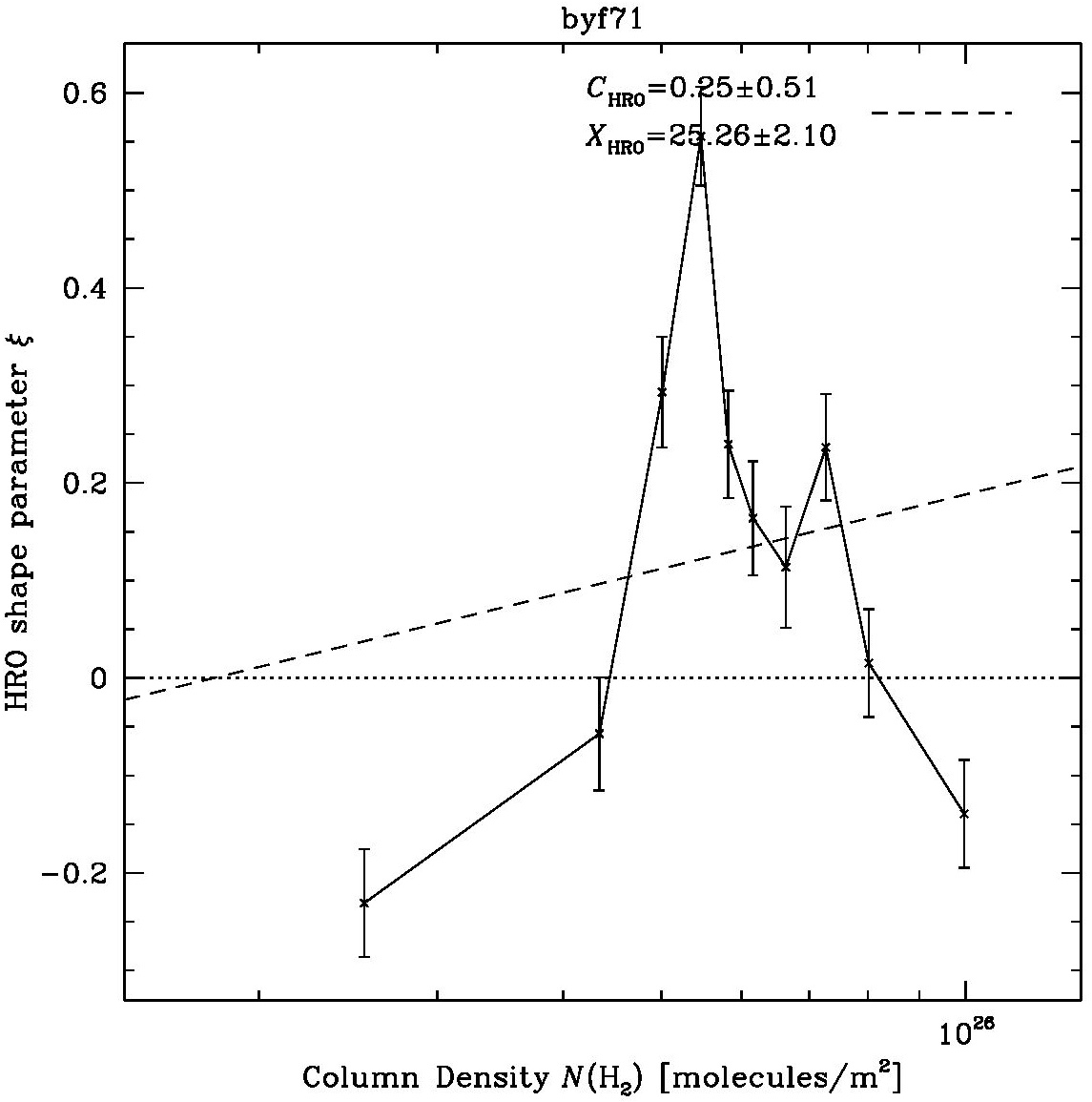}}
\vspace{0mm}\caption{ %\footnotesize  % uncomment for aasj
HRO analysis for 9 of the individual ROI cutouts shown in Figs.\,\ref{Sregs}, \ref{Nregs}, and \ref{Wregs}, as labelled.  The other panels are shown in Fig.\,\ref{xiplots2}.
}\label{xiplots1}\vspace{0mm}
\end{figure*}

\begin{figure*}[h]
\centerline{\includegraphics[angle=0,scale=0.148]{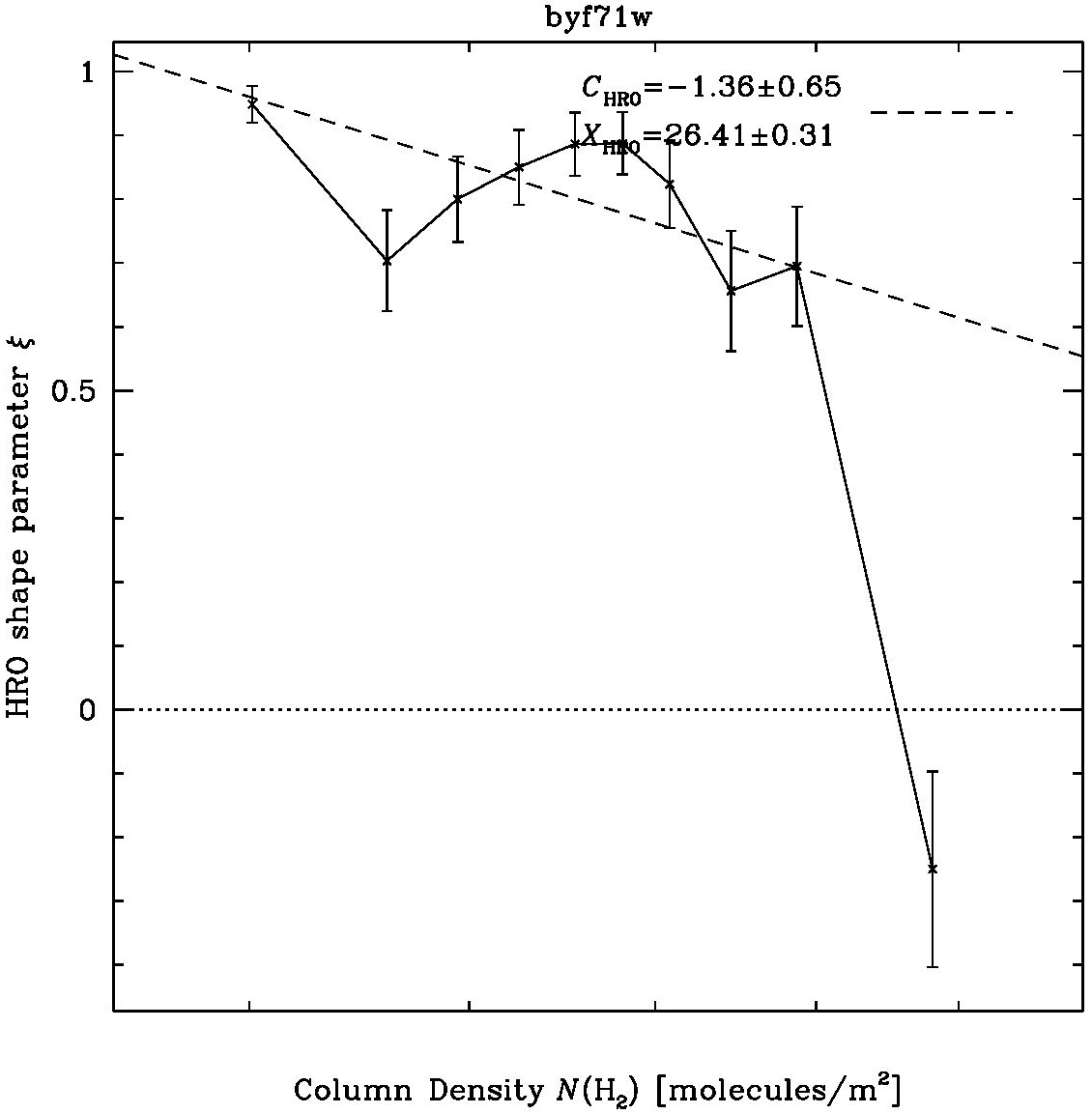}
		\includegraphics[angle=0,scale=0.148]{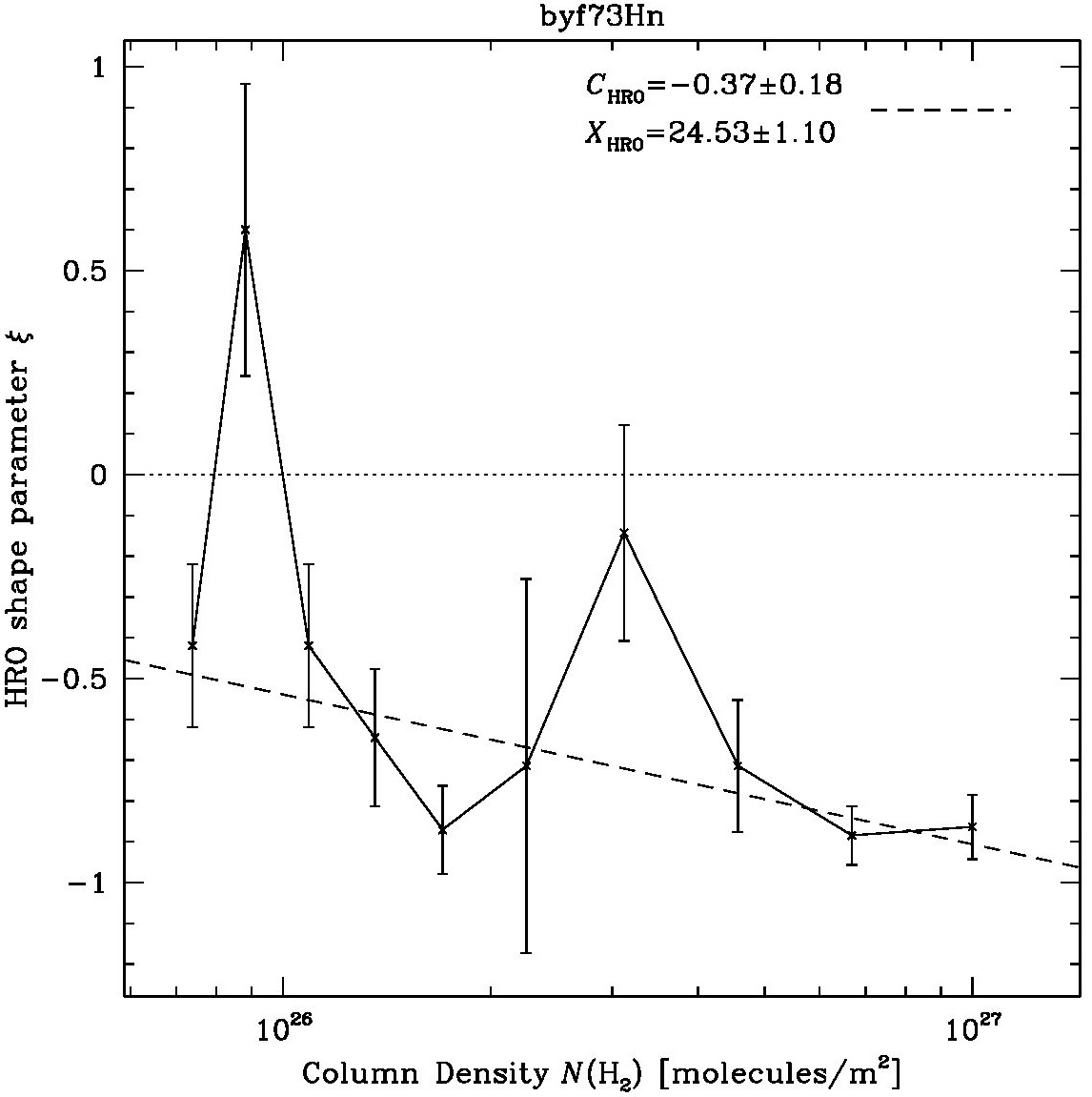}
		\includegraphics[angle=0,scale=0.148]{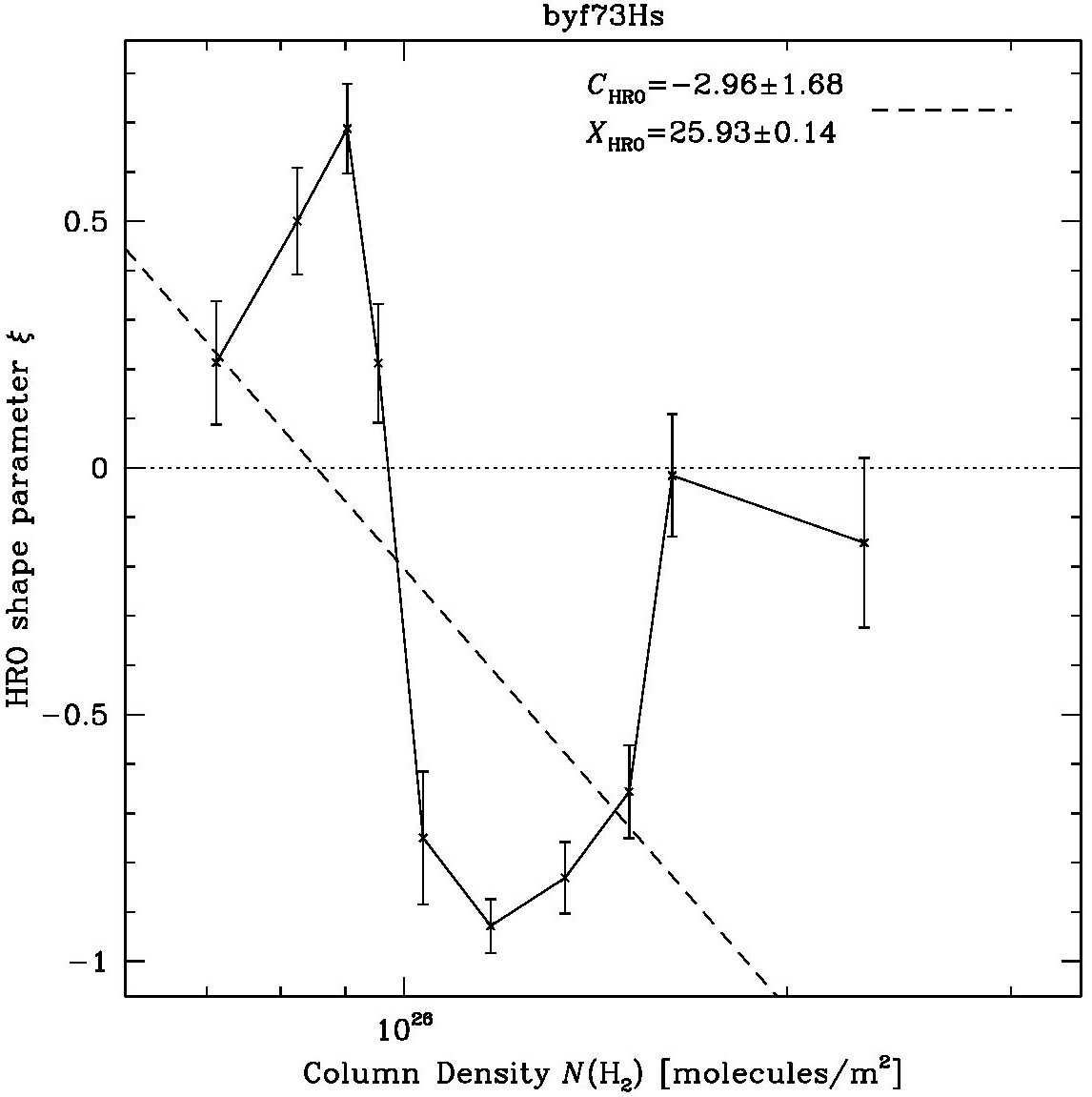}}
\centerline{\includegraphics[angle=0,scale=0.148]{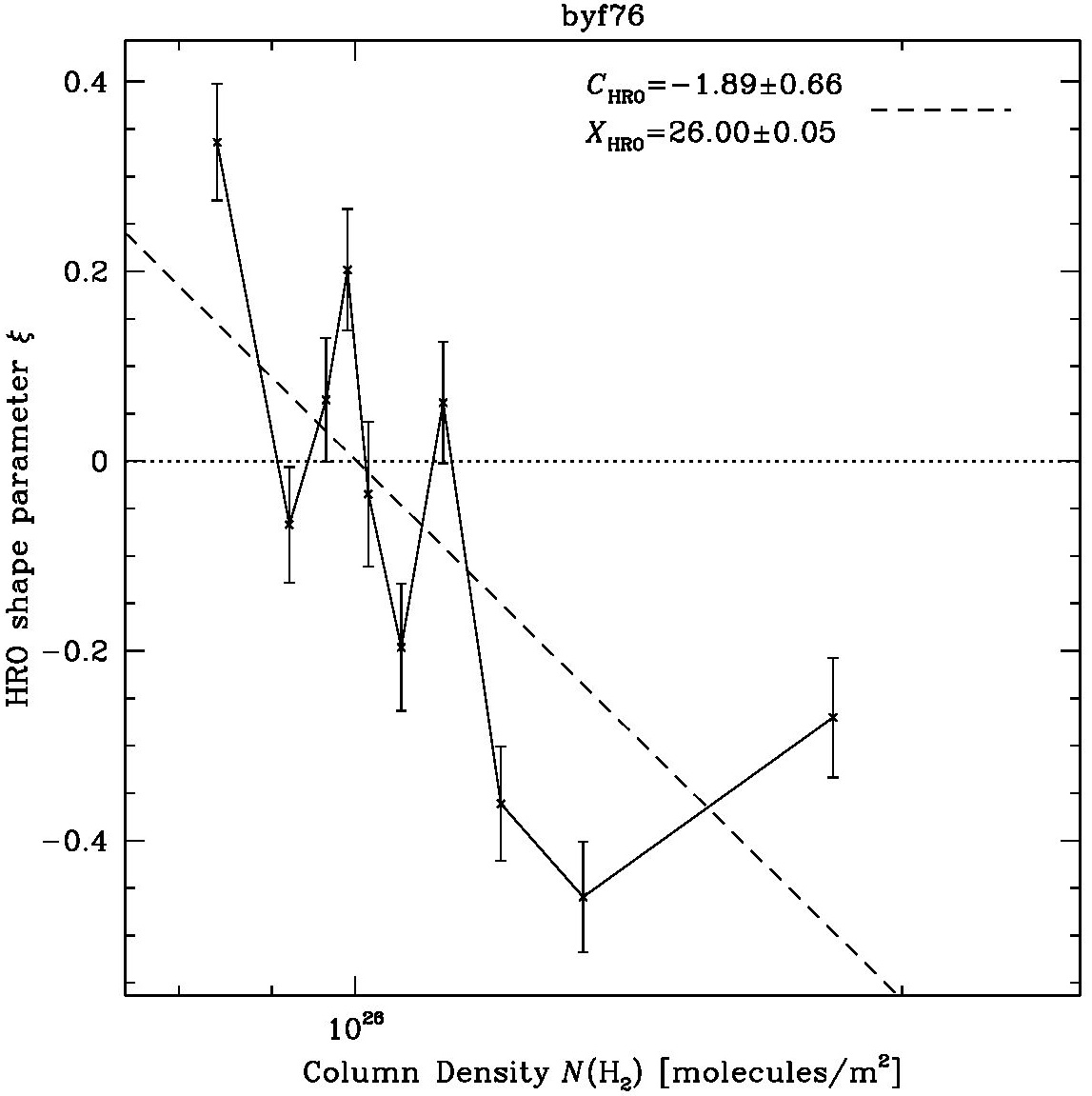}
		\includegraphics[angle=0,scale=0.148]{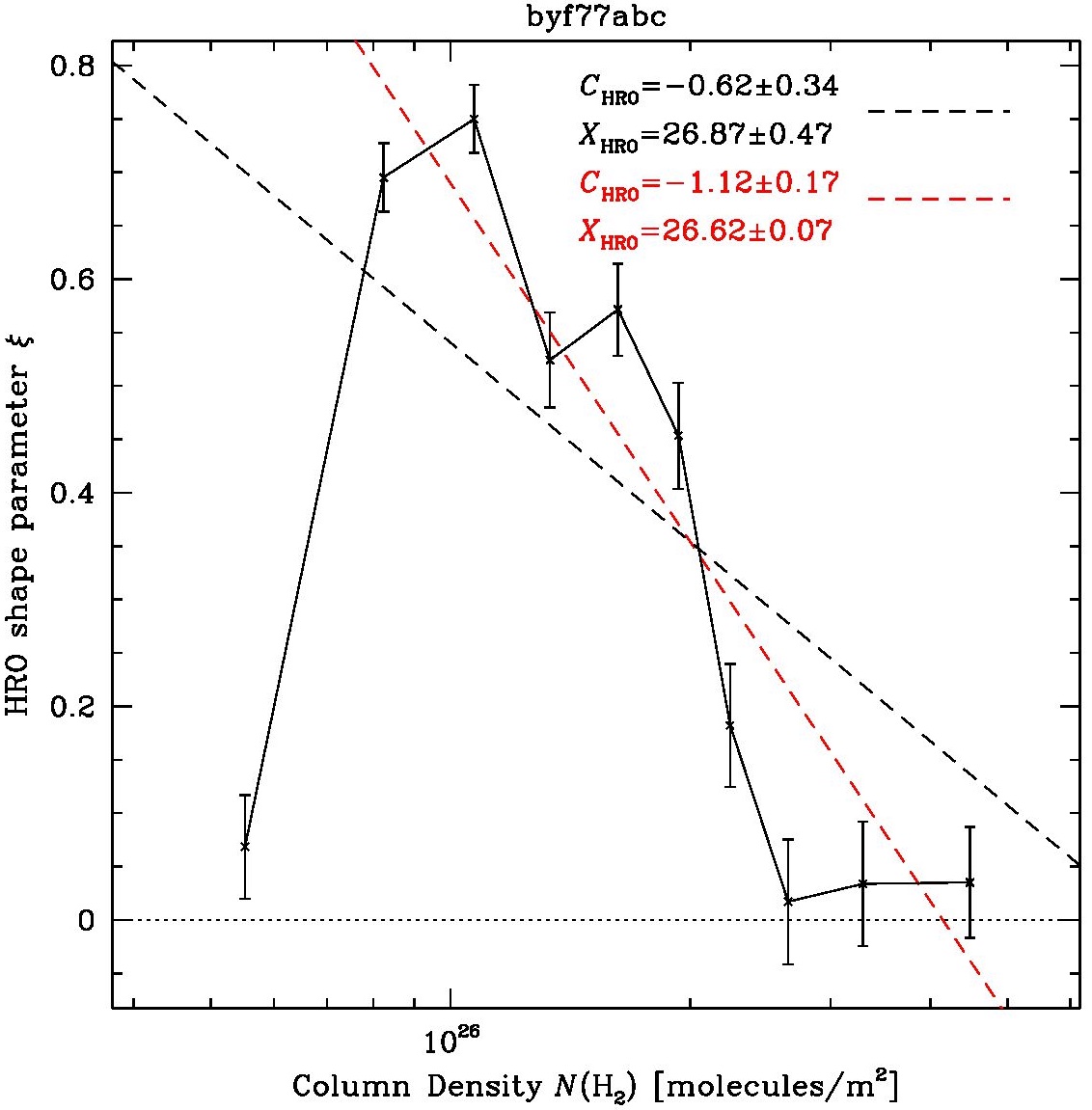}
		\includegraphics[angle=0,scale=0.148]{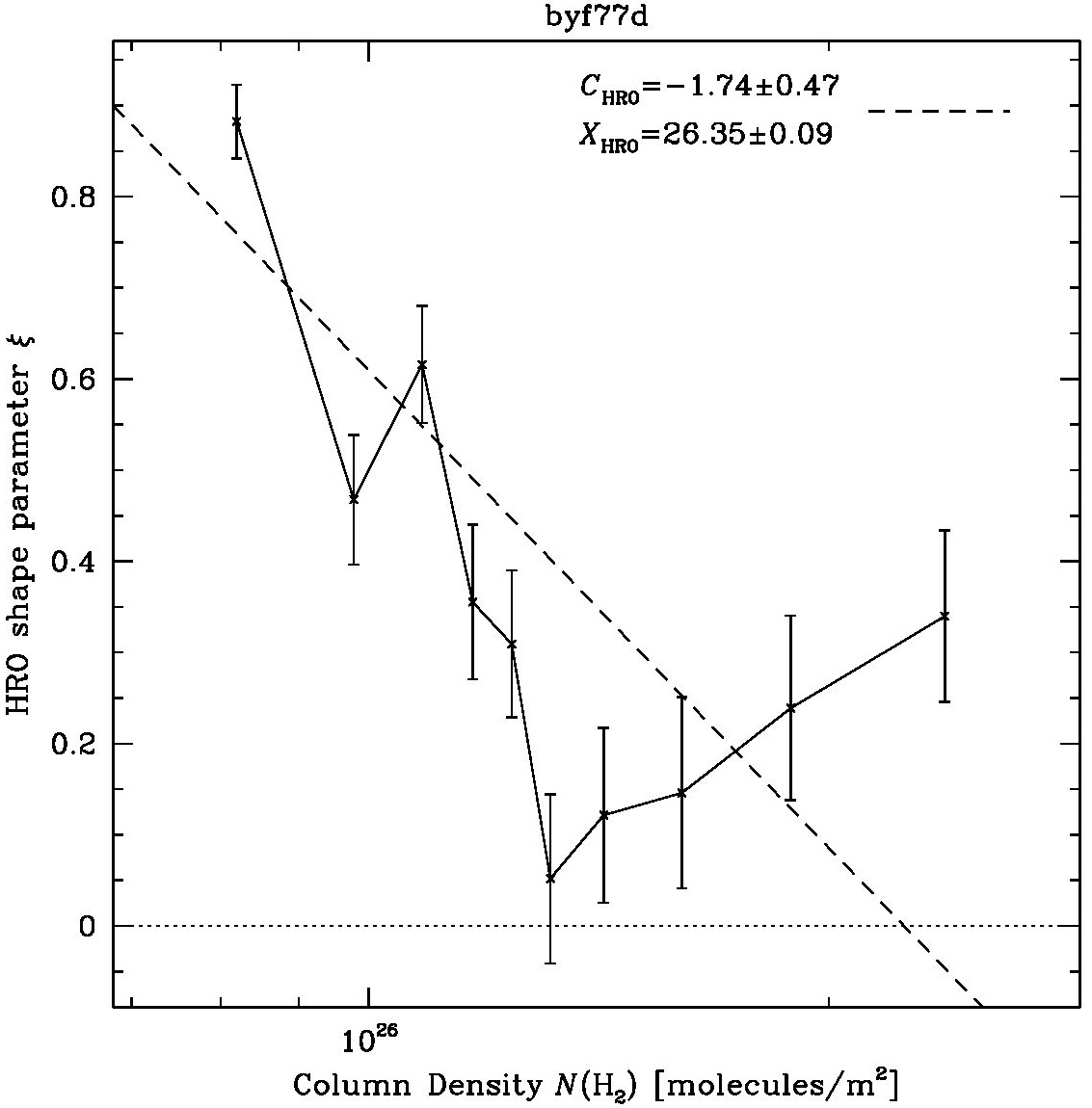}}
\centerline{\includegraphics[angle=0,scale=0.148]{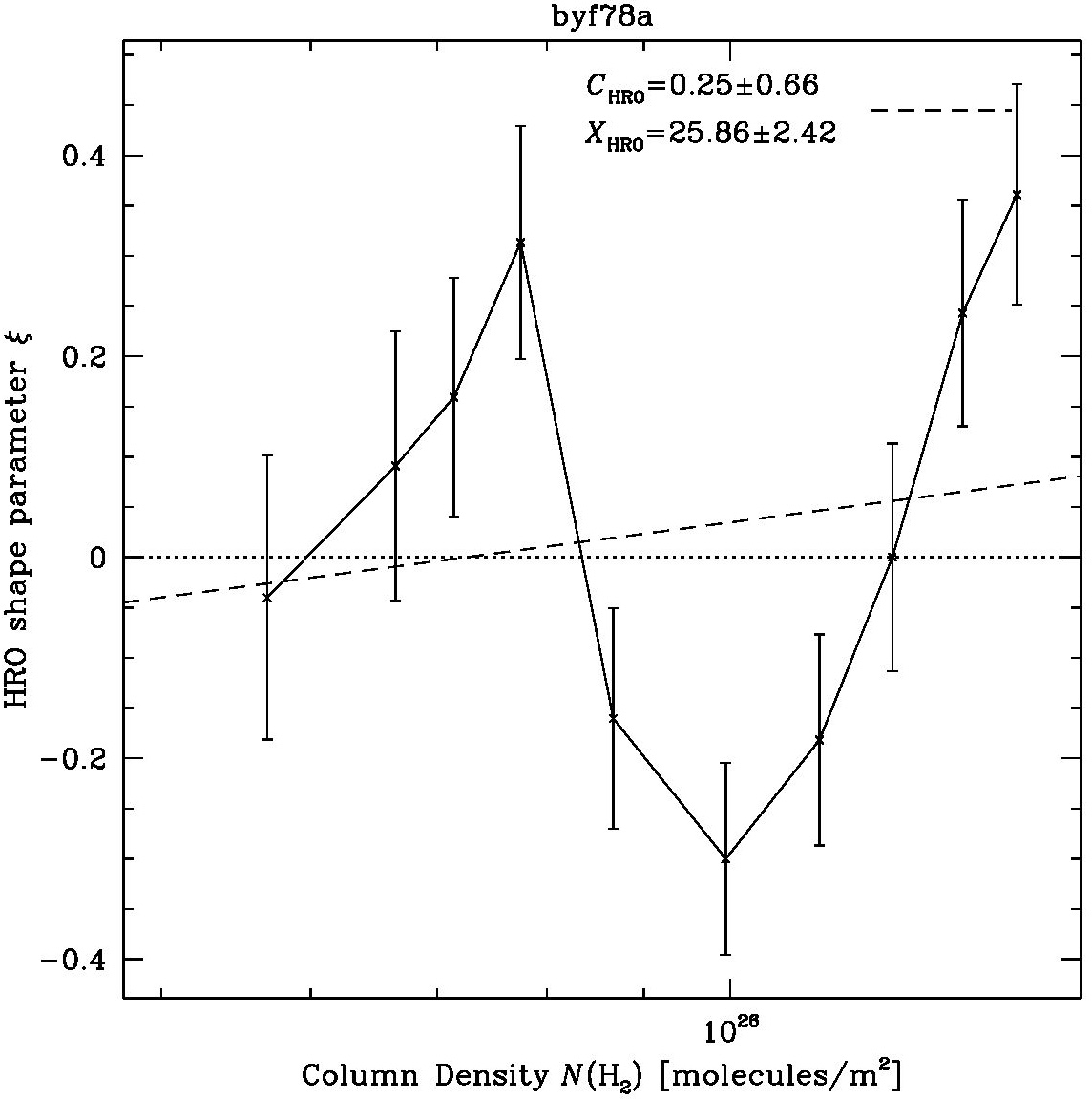}
		\includegraphics[angle=0,scale=0.148]{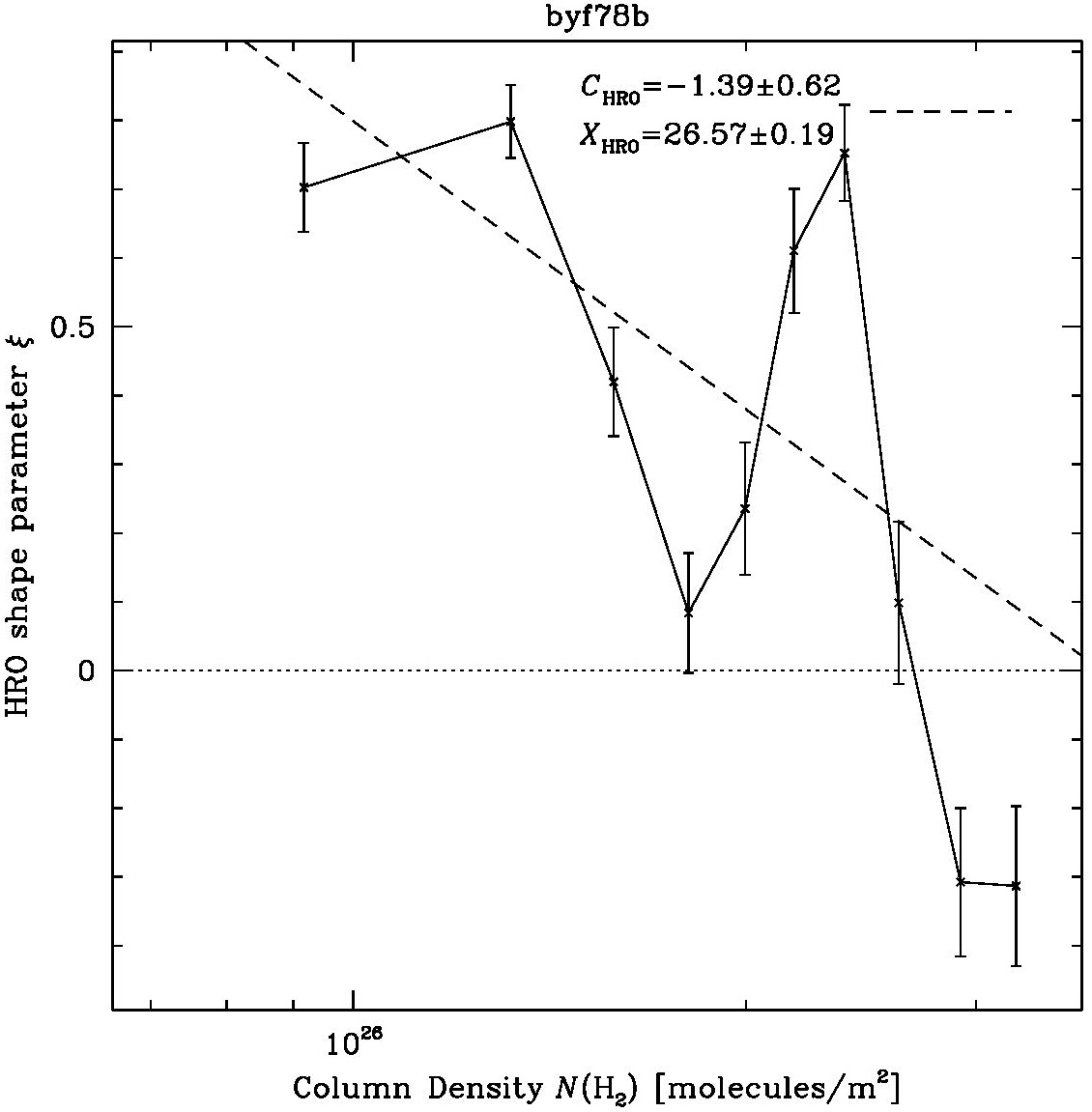}}
\vspace{0mm}\caption{ %\footnotesize  % uncomment for aasj
(Continuation of Fig.\,\ref{xiplots1}.)  HRO analysis for 8 of the other 10 individual ROI cutouts shown in Figs.\,\ref{Sregs}, \ref{Nregs}, and \ref{Wregs}, as labelled.  In the panel for BYF\,77abc only, we show two fits in red and black, where the red fit omits the first bin at low $N$ and should be considered the more reliable fit, as in Fig.\,\ref{hawcHROxi} and for the same reasons.  None of the other panels required the same treatment.
}\label{xiplots2}\vspace{0mm}
\end{figure*}

{\color{red}Figures \ref{xiplots1} and \ref{xiplots2}} show 17 panels of the HRO analysis for individual ROI cutouts, out of 19 total ROIs.  Each panel shows the relative orientation between the $B$ field and gas structures (via the shape parameter $\xi$) as a function of column density $N$ in each ROI, as in Fig.\,\ref{hawcHROxi}.  In each ROI/panel we use only 10 bins in log$N$ for the analysis, since the number of points available is (obviously) much smaller than for Region 9 as a whole (Fig.\,\ref{hawcHROxi}).  Excluded are the plots for the 2 ROIs at the centre of BYF\,73 (the MIR\,2 core and EPL feature), since those ROIs are too small to be analysed by HAWC+ data alone.  They were fully analysed with HAWC+ALMA data by \cite{b23}.

%%%%%%%
%   Fig.C1   %   single-ROI HROs
%%%%%%%

%%%%%%%
%   Fig.C2   %   single-ROI HROs
%%%%%%%

\clearpage
%%%%%%%%
%   Appx.  D   %
%%%%%%%%
\section{Line Integral Convolution Image}\label{LICimg}

\begin{figure*}[b]
\centerline{\includegraphics[angle=0,scale=0.84]{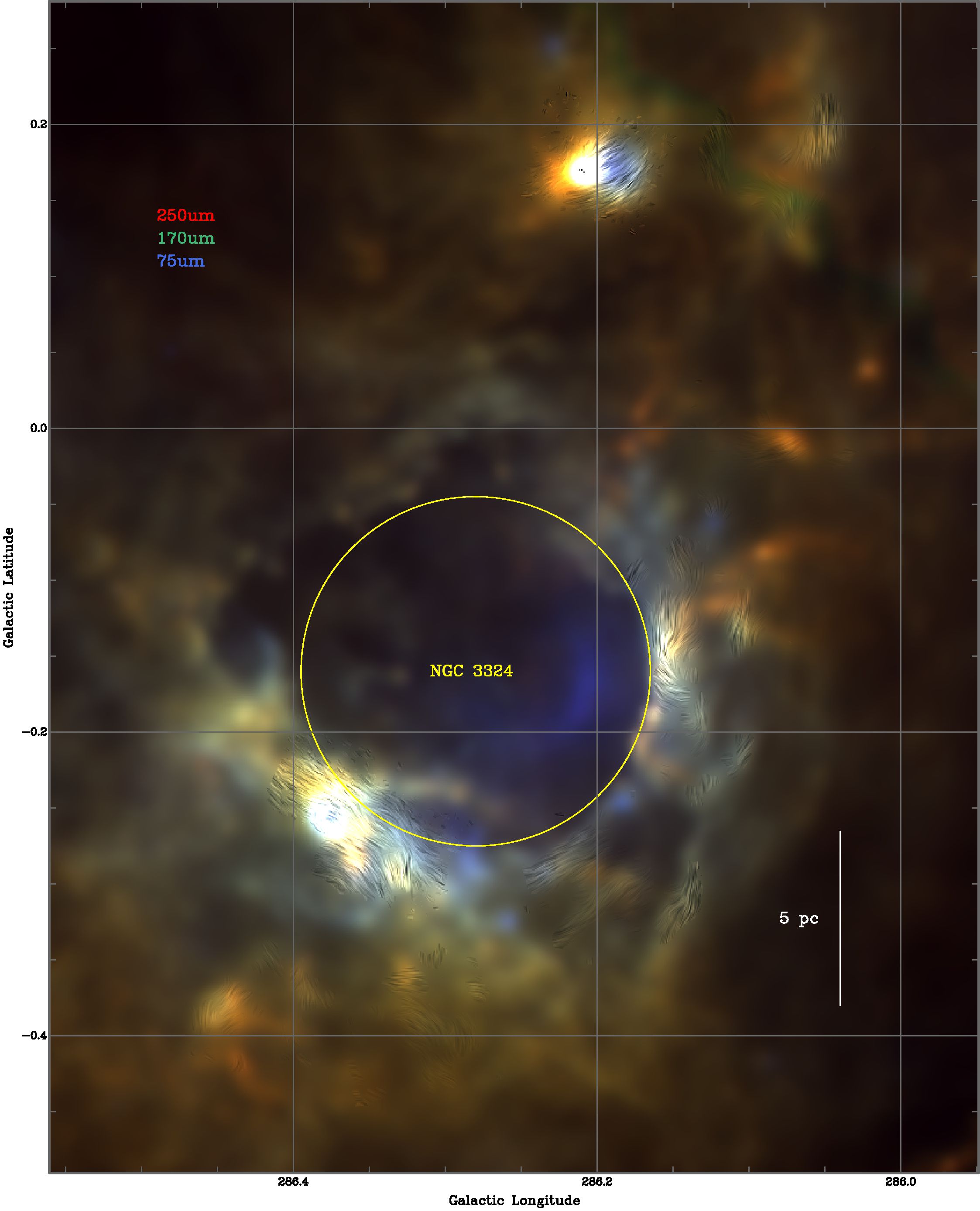}}
\vspace{0mm}\caption{Line Integral Convolution (LIC) image of Region 9, obtained as follows.  We started with a uniform noise map of the same size as the image shown; performed a one-dimensional 10-pixel convolution (the Line Integral) along the polarisation direction at each pixel in the noise map; multiplied the result by the polarisation amplitude at each pixel, normalised around unit brightness; and finally multiplied each colour channel in this image by the LICed noise map.
}\label{lic}\vspace{0mm}
\end{figure*}

We computed a LIC image \citep{CL93} of the {\em Herschel} data shown in Figure \ref{higal} based on the HAWC+ polarisation vectors of Figures \ref{r9north}--\ref{r9south}; see {\color{red}Figure \ref{lic}}.  This is useful as a visual confirmation of the preferentially circumferential $B_{\perp}$ orientation around the NGC 3324 HII region and more bluish (= warmer) molecular clouds there, whereas the more reddish (= colder) molecular clouds that are further from the HII region's influence seem to be unaffected by the HII region, magnetically speaking.  This also contextualises the \td-log$\lambda$ trend seen in Figure \ref{llTd}.

%%%%%%%
%   Fig.D1   %   single-ROI HROs (or Fig.30)
%%%%%%%


\begin{thebibliography}{}
%\bibitem[Andersen et al.(2017)]{a17}Andersen, M., Barnes, P. J., Tan, J. C., Kainulainen, J., \& G. de Marchi, G. 2017, \apj, 850, 12
%\bibitem[Barnes et al.(2010)]{b10}Barnes, P. J., Yonekura, Y., Ryder, S. D., et al. 2010, \mnras, 402, 73 %Hopkins, A. M., Miyamoto, Y., Furukawa, N., \& Fukui, Y. %(CHaMP Paper 0)
\bibitem[Barnes et al.(2011)]{b11}Barnes, P. J., Yonekura, Y., Fukui, Y. et al. 2011, \apjs, 196, 12 %(CHaMP Paper I)
%\bibitem[Barnes et al.(2013)]{b13}Barnes, P. J., Ryder, S. D., O'Dougherty, S. N., et al. 2013, \mnras, 432, 2231 %(CHaMP Paper I.5)
\bibitem[Barnes et al.(2015)]{BL15}Barnes, P. J., Li, D., Telesco, C., et al. 2015, \mnras, 453, 2622
%\bibitem[Barnes et al.(2015)]{bm15}Barnes, P. J., Muller, E., Indermuehle, B., et al. 2015,\apj,812,6 %(ThrUMMS Paper I)
\bibitem[Barnes et al.(2016)]{b16}Barnes, P. J., Hernandez, A. K., O'Dougherty, S. D., Schap, W. J. III, \& Muller, E. 2016, \apj, 831, 67 %(CHaMP Paper III)
\bibitem[Barnes et al.(2018)]{b18}Barnes, P. J., Hernandez, A. K., Muller, E., \& Pitts, R. L. 2018, \apj, 866, 19 %(CHaMP Paper IV)
\bibitem[Barnes et al.(2023)]{b23}Barnes, P. J., Ryder, S. D., Novak, G., Crutcher, R. M., Fissel, L. M., Pitts, R. L., \& Schap, W. J. III 2023, \apj, 945, 34
\bibitem[Cabral \& Leedom(1993)]{CL93} Cabral, B. \& Leedom, L. C., 1993, ``Imaging Vector Fields Using Line Integral Convolution,'' Proc. 20th ann. conf. {\em Computer graphics and interactive techniques} (SIGGRAPH '93: Anaheim, CA), pp.\ 263--270
\bibitem[Chandrasekhar \& Fermi(1953)]{cf53}Chandrasekhar, S., \& Fermi, E. 1953, \apj, 118, 113
%\bibitem[Chen et al.(2016)]{ck16}Chen, C.-Y., King, P. K., \& Li, Z.-Y. 2016, \apj, 829, 84
\bibitem[Cortes et al.(2024)]{cgs24}Cortes, P. C., Girart, J. M., Sanhueza, P., et al 2024, \apj\ submitted, arXiv:2406.14663
\bibitem[Crutcher et al.(2004)]{cn04}Crutcher, R. M., Nutter, D. J., Ward-Thompson, D. \& Kirk, J. M. 2004, ApJ, 600, 279
\bibitem[Crutcher(2012)]{cru12}Crutcher, R. M. 2012, \araa, 50, 29
\bibitem[Davis(1951)]{d51}Davis, L. 1951, Phys.Rev., 81, 890
%\bibitem[Falgarone et al.(2008)]{ftc08}Falgarone, E., Troland, T. H., Crutcher, R. M., \& Paubert, G. 2008, A\&A, 487, 247
\bibitem[Fissel et al.(2016)]{faa16}Fissel, L. M., Ade, P. A. R., Angil\`e, F. E., et al. 2016, \apj, 824, 134
\bibitem[Fissel et al.(2019)]{faa19}Fissel, L. M., Ade, P. A. R., Angil\`e, F. E., et al. 2019, \apj, 878, 110
%\bibitem[Franco et al.(1990)]{ftb90}Franco, J., Tenorio-Tagle, G., Bodenheimer, P. 1990, \apj, 349, 126
%\bibitem[Girart et al.(2004)]{gc04}Girart, J. M., Greaves, J. S., Crutcher, R. M., \& Lai, S.-P. 2004, Ap.SpaceSci. 292, 119
%\bibitem[Goldreich \& Kylafis(1981)]{gk81}Goldreich P. \& Kylafis N. 1981, \apjl, 243, L75
\bibitem[Gooch(1997)]{g97} Gooch, R.\ E. 1997, Proc. Astr. Soc. Aust., 14, 106
%\bibitem[Green et al.(1999)]{mgps}Green, A. J., Cram, L. E., Large, M. I., \& Ye, T. 1999, \apjs, 122, 207
%\bibitem[Habing \& Israel(1979)]{hi79}Habing, H. J. \& Israel, F. P. 1979, \araa, 17, 345
%\bibitem[Hakobian \& Crutcher(2011)]{hc11}Hakobian, N. \& Crutcher, R. 2011, \apj, 733, 6
\bibitem[Harper et al.(2018)]{hrd18}Harper, D. A., Runyan, M. C., Dowell, C. D., et al. 2018, {\em J.\,Astr.\,Instrum.}, 7(4), 184008
%\bibitem[Howard et al.(1997)]{h97}Howard E. M., Koerner D. W., Pipher J. L., 1997, \apj, 477, 738
%\bibitem[Jiang et al.(2021)]{j21}Jiang, C., et al. 2021, Nat.Astron., doi:10.1038/s41550-021-01414z
%\bibitem[Lada(2015)]{L15}Lada, C. J. 2015, in Proc IAU Symposium 309, {\em Galaxies in 3D across the Universe}, B Ziegler et al. eds, p31
\bibitem[Lazarian(2007)]{L07}Lazarian, A. 2007, JQSRT, 106, 225
%\bibitem[Lazarian \& Hoang(2014)]{LH14}Lazarian, A. \& Hoang, T. 2014, 40th COSPAR Scientific Assembly, Abstract F3.2-4-14, http://adsabs.harvard.edu/abs/ 2014cosp...40E1761L
\bibitem[Lazarian et al.(2018)]{LY18}Lazarian A., Yuen K. H., Ho, K. W., et al. 2018, \apj, 865, 46
%\bibitem[Lee et al.(2000)]{L00}Lee, C-F., Mundy, L. G., Reipurth, B., Ostriker, E. C., \& Stone, J. M. 2000, \apj, 542, 925
\bibitem[Lee et al.(2021)]{lbc21}Lee, D., Berthoud, M., Chen, C-Y., et al. 2021, \apj, 918, 39
\bibitem[Liu et al.(2022)]{liu22}Liu, J., Zhang, Q., Qiu, K., 2022, Front.\ Ast.\ Space Sci., 9, 943556
\bibitem[Lupton \& Monger(2000)]{lm00}Lupton, R. \& Monger, P. 2000, SuperMongo Manual, unpublished
\bibitem[McKee \& Ostriker(2007)]{mo07}McKee, C. \& Ostriker, E. 2007, \araa, 45, 565
%\bibitem[Mezger \& Henderson(1967)]{mh67}Mezger, P. G. \& Henderson, A. P. 1967, \apj, 147, 471
%\bibitem[Mezger et al.(1967)]{mst67}Mezger, P. G., Schraml, J., \& Terzian, Y. 1967, \apj, 150, 807
%\bibitem[Mouschovias \& Ciolek(1999)]{mc99}Mouschovias, T. Ch., \& Ciolek, G. E. 1999, in {\em The Origin of Stars and Planetary Systems}, ed. C.J. Lada, N.D. Kylafis, p. 305. Dordrecht: Kluwer
\bibitem[Myers \& Goodman(1991)]{mg91}Myers, P. C., \& Goodman, A. A. 1991, \apj, 373, 509
%\bibitem[Novak et al.(1997)]{ndd97}Novak, G., Dotson, J. L., Dowell, C. D., Goldsmith, P. F., Hildebrand, R. H., Platt, S. R., \& Schleuning, D. A. 1997, \apj, 487, 320
\bibitem[Ostriker et al.(2001)]{osg01}Ostriker, E. C., Stone, J. M., \& Gammie, C. F., 2001, \apj, 546, 980
%\bibitem[Ouyed \& Pudritz(1997)]{op97}Ouyed, R., \& Pudritz, R. E. 1997, \apj, 482, 712
%\bibitem[Panagia(1973)]{p73}Panagia, N. 1973, \aj, 78, 929
\bibitem[Pattle et al.(2022)]{ppvii}Pattle, K., Fissel, L., Tahani, M., Liu, T., \& Ntormousi, E.\ 2022, to be published in {\em Protostars and Planets VII}, S.-I.\ Inutsuka et al., eds. (arXiv:2203.11179)
\bibitem[Pitts et al.(2018)]{p18}Pitts, R. L., Barnes, P. J., Ryder, S. D, \& Li, D. 2018, \apjl, 867, L7
\bibitem[Pitts et al.(2019)]{p19}Pitts, R. L., Barnes, P. J., \& Varosi, F. 2019, \mnras, 484, 305
\bibitem[Pitts \& Barnes(2021)]{p21}Pitts, R. L., \& Barnes, P. J. 2021, \apjs, 256, 3 %(CHaMP Paper V)
\bibitem[Planck Collaboration(2016)]{pc16}Planck Collaboration XXXV 2016, A\&A, 586, A138
%\bibitem[Pudritz \& Ray(2019)]{pr19}Pudritz, R. E., \& Ray, T. P. 2019, Front.Astron.SpaceSci., 6, 54
%\bibitem[Raga et al.(1993)]{r93}Raga, A. C., Canto, J., Calvet, N., Rodr\'iguez, L. F., \& Torrelles, J. M. 1993, A\&A, 276, 539
%\bibitem[Rots et al.(1990)]{rbv90}Rots, A. H., Bosma, A., van der Hulst, J. M., Athanassoula, E., \& Crane, P. C. 1990, \aj, 100, 387
%\bibitem[Rygl et al.(2013)]{r13}Rygl, K. L. J., Wyrowski, F., Schuller, F., \& Menten, K. M. 2013, A\&A, 549, 5
\bibitem[Samson(2021)]{s21}Samson, W. B. 2021, J.Br.Astron.Assoc., 131(2), 97
%\bibitem[Santos et al.(2019)]{scd19}Santos, F. P., Chuss, D. T., Dowell, C. D., %Martin Houde, Leslie W. Looney, Enrique Lopez Rodriguez, Giles Novak, Derek Ward-Thompson, Marc Berthoud, Daniel A. Dale, Jordan A. Guerra, Ryan T. Hamilton, Shaul Hanany, Doyal A. Harper, Thomas K. Henning, Terry Jay Jones, Alex Lazarian, Joseph M. Michail, Mark R. Morris, Johannes Staguhn, Ian W. Stephens, Konstantinos Tassis, Christopher Q. Trinh, Eric Van Camp, C. G. Volpert, and Edward J. Wollack
%et al 2019, \apj, 888, 2 %``The Far-infrared Polarization Spectrum of $\rho$ Ophiuchi A from HAWC+/SOFIA Observations,'' 
\bibitem[Sault et al.(1995)]{st95}Sault, R. J., Teuben, P. J., \& Wright, M. C. H., 1995. in ``Astronomical Data Analysis Software and Systems IV,'' eds. R. Shaw, H. E. Payne, J. J. E. Hayes, ASP Conference Series, 77, 433
%\bibitem[Sharpe(2019)]{sh19}Sharpe, M. 2019, M.Sc. Thesis, University of New England, unpublished
%\bibitem[Shu et al.(1994)]{sn94}Shu, F.H., Najita, J., Ostriker, E., Wilkin, F., Ruden, S., \& Lizano, S. 1994, \apj, 429, 781
\bibitem[Skalidis \& Tassis(2021)]{st21}Skalidis, R. \& Tassis, K., 2021, A\&A, 647, A186
\bibitem[Soler et al.(2017)]{saa17}Soler, J., Ade, P., Angil\`e, F., et al. 2017, A\&A, 603, A64
%\bibitem[Sullivan et al.(2021)]{sfk21}Sullivan, C., Fissel, L. M., King, P. K., Chen, C.-Y., Li, Z.-Y., \& Soler, J. D. 2021, \mnras, 503, 5006
%\bibitem[Tomisaka(2011)]{t11}Tomisaka, K. 2011, PASJ, 63, 147
\bibitem[Whitworth et al.(2024)]{wh24}Whitworth, D.\ J., Srinivasan, S., Pudritz, R.\ E., et al.\ 2024, \mnras, submitted (arXiv:2407.18293)
\bibitem[Zucker et al.(2020)]{zss20}Zucker, C., Speagle, J. S., Schlafly, E. F., Green, G. M., Finkbeiner, D. P., Goodman, A., \& Alves, J. 2021, A\&A, 633, 51
\end{thebibliography}
\end{document}